\documentclass[fleqn,usenatbib]{mnras}

\usepackage{newtxtext,newtxmath}
\usepackage{comment}
\usepackage{multirow}
\usepackage{subcaption}
\usepackage{longtable}
\usepackage{supertabular}
\usepackage{anyfontsize}
\usepackage{placeins}

\usepackage[T1]{fontenc}
\usepackage{booktabs}
\usepackage{siunitx}
\DeclareRobustCommand{\VAN}[3]{#2}
\let\VANthebibliography\thebibliography
\def\thebibliography{\DeclareRobustCommand{\VAN}[3]{##3}\VANthebibliography}

\usepackage{graphicx}	
\usepackage{amsmath}
\usepackage{amsbsy}
\usepackage{subcaption}

\title[Galaxy merger at high-$z$]{Galaxy Mergers in the Epoch of Reionization I: A JWST Study of Pair Fractions, Merger Rates, and Stellar Mass Accretion Rates at $z = 4.5 - 11.5$}

\author[Qiao Duan et al.]{
Qiao Duan,$^{1}$ Christopher J. Conselice,$^{1}$\thanks{E-mail: conselice@manchester.ac.uk} Qiong Li,$^{1}$\thanks{E-mail: qiong.li@manchester.ac.uk} Duncan Austin,$^{1}$\thanks{E-mail: duncan.austin@postgrad.manchester.ac.uk} Thomas Harvey,$^{1}$  Nathan J. Adams,$^{1}$ \newauthor Kenneth J. Duncan,$^{2}$ James Trussler,$^{1}$ Leonardo Ferreira,$^{3}$ Lewi Westcott,$^{1}$ Honor Harris,$^{1}$ Rogier A. Windhorst,$^{4}$ \newauthor Benne W. Holwerda,$^{5}$ Thomas J. Broadhurst,$^{6,7,8}$ Dan Coe,$^{9,10,11}$ Seth H. Cohen,$^{4}$ Xiaojing Du,$^{12}$ \newauthor Simon P. Driver,$^{13}$  Brenda Frye,$^{14}$  Norman A. Grogin,$^{9}$ Nimish P. Hathi,$^{9}$ Rolf A. Jansen,$^{4}$  Anton M. Koekemoer,$^{9}$  \newauthor Madeline A. Marshall,$^{15,16}$  Mario Nonino,$^{17}$ Rafael {Ortiz~III},$^{4}$ Nor Pirzkal,$^{9}$   Aaron Robotham,$^{13}$  \newauthor Russell E. Ryan, Jr.$^{9}$ Jake Summers,$^{4}$ Jordan C. J. D'Silva,$^{13, 16}$  
Christopher N. A. Willmer,$^{14}$ Haojing Yan$^{18}$ 
\\ \\
$^{1}$ Jodrell Bank Centre for Astrophysics, University of Manchester, Oxford Road, Manchester, UK \\
$^{2}$ Institute for Astronomy, University of Edinburgh Royal Observatory, Blackford Hill, Edinburgh, EH9 3HJ, UK \\
$^{3}$ Department of Physics \& Astronomy, University of Victoria, Finnerty Road, Victoria, British Columbia, V8P 1A1, Canada \\
$^{4}$ School of Earth and Space Exploration, Arizona State University, Tempe, AZ 85287-1404 \\
$^{5}$ University of Louisville, Department of Physics and Astronomy, 102 Natural Science Building, 40292 KY Louisville, USA \\
$^{6}$ Department of Physics, University of the Basque Country UPV/EHU, E-48080 Bilbao, Spain \\
$^{7}$ DIPC, Basque Country UPV/EHU, E-48080 San Sebastian, Spain \\ 
$^{8}$ Ikerbasque, Basque Foundation for Science, E-48011 Bilbao, Spain \\
$^{9}$ Space Telescope Science Institute, 3700 San Martin Drive, Baltimore, MD 21218, USA \\
$^{10}$ Association of Universities for Research in Astronomy (AURA) for the European Space Agency (ESA), STScI, Baltimore, MD 21218, USA \\
$^{11}$ Center for Astrophysical Sciences, Department of Physics and Astronomy, The Johns Hopkins University, 3400 N Charles St. Baltimore, MD 21218, USA \\
$^{12}$ Department of Statistical Science, Wake Forest University, Winston-Salem, NC, USA \\
$^{13}$ International Centre for Radio Astronomy Research (ICRAR) and the
International Space Centre (ISC), The University of Western Australia, M468, 35 Stirling Highway, \\ Crawley, WA 6009, Australia \\
$^{14}$ Department of Astronomy/Steward Observatory, University of Arizona, 933 N Cherry Ave,
Tucson, AZ, 85721-0009, USA \\
$^{15}$ National Research Council of Canada, Herzberg Astronomy \&
Astrophysics Research Centre, 5071 West Saanich Road, Victoria, BC V9E 2E7,
Canada \\
$^{16}$ ARC Centre of Excellence for All Sky Astrophysics in 3 Dimensions
(ASTRO 3D), Australia \\
$^{17}$ INAF-Osservatorio Astronomico di Trieste, Via Bazzoni 2, 34124
Trieste, Italy \\
$^{18}$ Department of Physics and Astronomy, University of Missouri,
Columbia, MO 65211, USA \\
}

\date{Accepted 2025 April 14. Received 2025 April 13; in original form 2024 July 25}

\pubyear{2025}

\newcommand{\fpp}{$\mathrm{f}_\mathrm{p}$ }

\begin{document}
\label{firstpage}
\pagerange{\pageref{firstpage}--\pageref{lastpage}}
\maketitle

\begin{abstract}
We present a full analysis of galaxy major merger pair fractions, merger rates, and mass accretion rates, thus uncovering the role of mergers in galaxy formation at the earliest previously unexplored epoch of $4.5<z<11.5$. We target galaxies with masses $\log_{10}(\mathrm{M}_*/\mathrm{M}_\odot) = 8.0 - 10.0$, utilizing data from eight JWST Cycle-1 fields (CEERS, JADES GOODS-S, NEP-TDF, NGDEEP, GLASS, El-Gordo, SMACS-0723, MACS-0416), covering an unmasked area of 189.36 $\mathrm{arcmin}^2$. We develop a new probabilistic pair-counting methodology that integrates full photometric redshift posteriors and corrects for detection incompleteness to quantify close pairs with physical projected separations between 20 and 50 kpc. Our analysis reveals an increase in pair fractions up to $z = 8$, reaching $0.211 \pm 0.065$, followed by a statistically flat evolution to $z = 11.5$. We find that the galaxy merger rate increases from the local Universe up to $z = 6$ and then stabilizes at a value of $\sim 6$ Gyr$^{-1}$ up to $z = 11.5$. The redshift evolution of both pair fractions and merger rates is well described by a power-law plus exponential model. In addition, we measure that the average galaxy increases its stellar mass due to mergers by a factor of $2.77 \pm 0.99$ from redshift $z = 10.5$ to $z = 5.0$. Lastly, we investigate the impact of mergers on galaxy stellar mass growth, revealing that mergers contribute as much as $71 \pm 25\%$ to galaxy stellar mass growth. This indicates that mergers drive about half of galaxy assembly at high redshift.

\end{abstract}



\clearpage
\section{Introduction}
The formation and evolution of galaxies are fundamentally driven by their merger histories. Within the hierarchical galaxy formation framework, galaxy mergers are directly triggered by the mergers of cold dark matter halos \citep{1984Natur.311..517B}. This process progressively assembles larger galaxies from smaller progenitors, influencing their subsequent growth and evolution \citep{1998ApJ...499...20J, maller2006galaxy}. Beyond the simple merging of galaxies, these activities profoundly impact various galaxy properties at various redshifts, affecting phenomena such as size growth \citep{Trujillo2007, Buitrago2008, naab2009minor, bluck2012structures}, star formation \citep{mihos1995gasdynamics, 2012MNRAS.426..549S, 2013MNRAS.433L..59P, 2019A&A...631A..51P,2020MNRAS.494.4969P, 2021MNRAS.501.2969G, ellison2022galaxy, 2022ApJ...940....4S, 2023ApJ...950...56H, 2023RNAAS...7..232W, 2023MNRAS.521..800M, 2024arXiv241104944D}, stellar-mass growth \citep{2017MNRAS.470.3507M, 2019ApJ...876..110D, 2022ApJ...940..168C, 2025MNRAS.538L..31F}, AGN activity \citep{satyapal2014galaxy, barrows2017observational, 2020A&A...637A..94G, 2023ApJ...944..168L, 2024MNRAS.527.9461S, 2024arXiv241104944D}, morphological transformations \citep{naab2006properties, bournaud2011hydrodynamics}, stellar mass growth \citep{2010ApJ...709.1018V, 2011MNRAS.411.2148K, 2014MNRAS.445.2198O} as well as possible neutrino emissions \citep{bouri2024search}. 

Recent advancements have also led to discoveries such as dual and triple AGN systems at redshifts around $z = [3,4]$ \citep{2023arXiv231003067P}, as well as evidence that mergers may catalyze the formation of extreme emission line galaxies in the early Universe \citep{2023ApJ...957L..35G}. Mergers also provide possible insights into magnetic fields and how they play a pivotal role in the dynamics of disk galaxy mergers through magnetohydrodynamic (MHD) simulations \citep{2023MNRAS.526..224W}. Galaxy mergers thus play a crucial role in shaping not only the gas content, stellar populations, and morphology of galaxies but also in the formation of central massive black holes and the triggering of star formation. Mergers are an essential process in galaxy formation, yet we do not have very much observational information on their influence across cosmic time. 

Quantifying the rate at which galaxies merge (merger rates) across the Universe’s history not only elucidates the merger process itself but also provides insights into broader cosmic environments and structure at different epochs. Two main methods are employed to study merger phenomena: morphology-based selection \citep[e.g.,][]{2003AJ....126.1183C, conselice2008structures, 2008MNRAS.391.1137L, 2008ApJ...672..177L,2009ApJS..181..233H, 2010MNRAS.404..575L, 2010MNRAS.404..590L, 2021ApJ...919..139W, 2022A&A...661A..52P, 2023ApJ...958...96R, 2023AAS...24145304R, 2023MNRAS.523.4381D, 2024MNRAS.528.5558W, 2024arXiv240311428D} and close pair analysis \citep[e.g.,][]{zepf1989close, burkey1994galaxy, carlberg1994survey, woods1995counting, yee1995statistics, 1997ApJ...475...29P, 2000MNRAS.311..565L, patton2000new, 2009A&A...498..379D, 2013A&A...553A..78L, 2015A&A...576A..53L, 2016MNRAS.461.2589P, 2016ApJ...830...89M, 2017MNRAS.470.3507M, 2017MNRAS.468..207S, 2017A&A...608A...9V, 2018MNRAS.475.1549M, 2019ApJ...876..110D, 2022ApJ...940..168C, 2024arXiv240700594I, 2025arXiv250201721P}. The morphological selection method can be both quantitative and qualitative. Quantitative methods employ criteria such as the CAS system \citep{1995ApJ...451L...1S, 1996ApJS..107....1A, 2000AJ....119.2645B, 2003ApJS..147....1C, 2003AJ....126.1183C}, as well as the \textit{Gini} coefficient and $M_{20}$ \citep{lotz2004new}, to identify merging galaxy pairs. Qualitative approaches include visual classification \citep[e.g.,][]{2015ApJS..221...11K, 2017MNRAS.464.4420S} and methods based on detecting tidal signatures indicative of mergers \citep[e.g.,][]{2019MNRAS.486.2643M, 2020AJ....159..103K}. The close-pair selection method, on the other hand, relies on redshifts and projected physical separations ($r_p$) to identify galaxies likely to merge in the near future. Different studies employ varying selection criteria for projected distances and redshift differences. For example, spectroscopic studies such as \cite{2013A&A...553A..78L, 2017A&A...608A...9V} use selection thresholds of $0 < r_p < 20$–$30$ kpc and $\Delta v < 500$ km/s to classify close-pair mergers. In contrast, photometric redshifts are required to analyze larger galaxy samples and study pair fractions and merger rates. Studies such as \cite{2015A&A...576A..53L, 2017MNRAS.470.3507M, 2019ApJ...876..110D, 2022ApJ...940..168C} employ a probabilistic pair-counting methodology, utilizing the full photometric redshift probability distribution to define redshift selection criteria, along with projected separation thresholds of $5 < r_p < 30$–$50$ kpc.

The measurement of pair fractions and merger rates up to $z < 3.5$ has been extensively explored \citep[e.g.,][]{2003ApJS..147....1C, conselice2008structures, 2011ApJ...742..103L, bluck2012structures, 2014MNRAS.445.2198O, 2015A&A...576A..53L,2016ApJ...830...89M, 2017MNRAS.470.3507M,2018MNRAS.475.1549M, 2019ApJ...876..110D, 2022ApJ...940..168C, 2024MNRAS.529.1493P}. However, studies analyzing these metrics at redshifts of $3\leq z \leq 6$ are less frequent \citep{2017A&A...608A...9V, 2019ApJ...876..110D}, primarily due to the limitations of earlier observational data. Previous observations lacked the necessary sensitivity and wavelength coverage to accurately determine the redshifts of high-$z$ galaxy samples, constraining our understanding of galaxy evolution during these epochs.

The high-redshift Universe is increasingly unveiled by JWST, highlighted by numerous recent studies on early galaxy discoveries \citep[e.g.,][]{2022ApJS..259...20H, 2022MNRAS.511.4464A, 2022ApJ...938L..15C, 2022NatAs...6..599D, 2022ApJ...940L..14N, curtis2022spectroscopic, 2023MNRAS.518.4755A, 2023arXiv230801230M,2023ApJ...952L...7A,donnan2023evolution, finkelstein2023ceers, 2023Natur.622..707A, 2023MNRAS.523.1036B, 2023arXiv230414469M, 2023arXiv230602468H, 2023arXiv230800751F, 2023arXiv230810932C, 2024Natur.627...59M,2024ApJ...965..169A, 2024arXiv240605306C}. These studies reveal a significant population of distant galaxies at $z > 6$, previously underrepresented in HST data, providing a new window to explore merger evolution at even high redshifts.

Given the challenges of identifying mergers through morphology at high redshifts, where galaxies are more compact and may appear as point-like sources \citep{2024MNRAS.527.6110O}, and where cosmological surface brightness dimming scales as $(1+z)^{-4}$ \citep{1930PNAS...16..511T, 1934rtc..book.....T}, making faint morphological signatures difficult to detect without extremely deep imaging, this paper pioneers the study of high-redshift merger evolution using the close-pair method, which is effective as it does not rely on such systematics. We utilize data from JWST's Near-InfraRed Camera \citep[NIRCam;][]{Rieke2005, 2023ApJS..269...16R} in eight deep Cycle-1 blank imaging fields (CEERS, JADES GOODS-S, NEP-TDF, NGDEEP, GLASS, El-Gordo, SMACS-0723, MACS-0416), covering unmasked area of $189.36 \, \mathrm{arcmin}^2$. We expand upon a novel close-pair methodology, building on foundational works \citep{2015A&A...576A..53L, 2017MNRAS.470.3507M, 2019ApJ...876..110D, 2022ApJ...940..168C} to compute pair fractions, merger rates, and Stellar-mass accretions from $z = 4.5$ to $z = 11.5$ ($\sim 350$\,–\,1350 Myr after the Big Bang) for the first time.  We then use these to derive the history of galaxy merging and how mergers contribute to the mass assembly of galaxies at this very early cosmic times. 

The structure of this paper is outlined as follows. In Section~\ref{sec: Observations and Data Reduction}, we detail the dataset sourced from 8 JWST fields. Our photometric redshift and stellar mass determinations are outlined in Sections  \ref{sec: Photometric Redshift and Sample Selection} and \ref{sec: Stellar Mass Estimation}. The close pair methodology is presented in Section \ref{sec: Galaxy Pair Fraction Methodology}. Our main findings and analysis are presnted in Section~\ref{sec: high redshift Merger Evolution}. A summary of our conclusions is provided in Section~\ref{sec: Conclusions}. Throughout this work, we adhere to the Planck 2018 Cosmology \citep{2020A&A...641A...6P} with $H_0=67.4 \pm 0.5\,\text{km}\,\text{s}^{-1}\,\text{Mpc}^{-1}$, $\Omega_{\rm M}=0.315 \pm 0.007$, and $\Omega_{\Lambda} = 0.685 \pm 0.007$ to facilitate comparison with other observational studies. All magnitudes reported are consistent with the AB magnitude system \citep{Oke1974,Oke1983}.

\section{Observations and Data Reduction}
\label{sec: Observations and Data Reduction}
The launch of the James Webb Space Telescope (JWST) in December 2021 \citep{2023PASP..135d8001R} marked the beginning of a new era in the exploration of the high-redshift Universe. We utilized a consistent approach for reducing and analyzing publicly available deep JWST NIRCam data from the PEARLS, CEERS, GLASS, JADES GOODS-S, NGDEEP, and SMACS0723 surveys. Specifically in this paper, for the PEARLS program, we included data from the NEP-TDF, MACS-0416, and El-Gordo fields, leading to a combined total of 189.36 arcmin$^2$ of unmasked sky when integrated with the other five surveys. From this comprehensive dataset, we have identified 3625 robust galaxy candidates at $z > 4.5$, which includes 1276 candidates at $z > 6.5$, forming our EPOCHS V1 sample. For the \(z = 4.5 - 6.5\) range, only data from the CEERS, JADES GOODS-S, and NEP-TDF surveys are used due to the availability of HST ACS/WFC F606W and F814W band data. 

In this paper, we use our own EPOCHS reduction, with a detailed description given in \cite{2024ApJ...965..169A, austin2024epochs, harvey2024epochs, 2024arXiv240714973C}.  We do not repeat this information and process here, but many details are available in these papers. It is important to keep in mind that we re-reduce all these fields using the same methods and furthermore measure and detect our galaxies through a homogeneous method on each of these fields.  We have uniformly reprocessed all lower-level JWST data products following our modified version of the official JWST pipeline. For detailed information on data reductions, source extractions, and catalogs, please refer to \cite{2024ApJ...965..169A,austin2024epochs, harvey2024epochs}. 

In the following subsections, we describe the fields analyzed in this study and their relevant properties. We outline the depth of our data across different fields and JWST filters, as well as ACS observations, along with the number of robustly detected sources in each field in \autoref{tab: fields parameters}.

\subsection{CEERS}

The Cosmic Evolution Early Release Science Survey \citep[CEERS; PI: S. Finkelstein, PID: 1345,][]{2023ApJ...946L..12B} is one of the 13 Director's Discretionary Early Release Science (DD-ERS) programs conducted during the first year of JWST's operation. CEERS features a $\sim$100 arcmin$^2$ NIRCam imaging mosaic covering wavelengths from 1 to 5 microns, across 10 NIRCam pointings. Concurrently, CEERS includes imaging with the mid-infrared instrument (MIRI) over four pointings, covering 5 to 21 microns, and multi-slit spectroscopy using the near-infrared spectrograph (NIRSpec) across six pointings. NIRCam imaging data is used in this work. This includes image across seven distinct filters: F115W, F150W, F200W, F277W, F356W, F410M, and F444W, with a $5\sigma$ depth of 28.6 AB magnitudes using 0.1 arcsec radius circular apertures.  The dataset we use encompasses observations collected during June 2022, accounting for 40\% of the total NIRCam area covered for CEERS with the remaining data taken in the latter half of the same year.

\subsection{JADES Deep GOODS-S}
The JWST Advanced Deep Extragalactic Survey \citep[JADES;][]{rieke2023jades, 2023arXiv230602465E, bunker2023jades, 2024arXiv240406531D, 2023AAS...24221202H} cover both the GOODS-S and GOODS-N fields. In this paper, we focus on the region covered by the JADES DR1 release which is in the GOODS-S field (PI: Eisenstein, N. Lützgendorf, ID:1180, 1210). The observations utilise nine filter bands: F090W, F115W, F150W, F200W, F277W, F335M, F356W, F410M, and F444W, encompassing a spatial extent of 24.4 - 25.8 arcmin\(^2\). A minimum of six dither points was used for each observation, with exposure times spanning 14-60 ks. As shown in \citet{harvey2024epochs} and Table~\ref{tab: fields parameters} the depth of the JADES data within the JWST bands ranges from 29.58 to 30.21, with the deepest band being F277W. This field also includes deep F606W data from ACS.

Across all filter bands, JADES ensures a high level of pixel diversity, thereby significantly reducing the impact of flat-field inaccuracies, cosmic ray interference, and other issues at the pixel level. 

\subsection{PEARLS and NEP-TDF}
\label{sec: NEP NIRCam}
The North Ecliptic Pole Time-Domain Fields (NEP-TDF) is part of the JWST PEARLS observational program. The Prime Extragalactic Areas for Reionization and Lensing Science project \citep[PEARLS; PIDs 1176, 2738, PI: Rogier Windhorst;][]{2023AJ....165...13W,2023A&A...672A...3D, 2023ApJ...952...81F}, is a JWST Guaranteed Time Observation (GTO) program. As a Cycle 1 GTO project, the PEARLS team was allocated an exposure time of 110 hours. The project's primary objective is to capture medium-deep NIRCam imaging of blank and cluster fields with an approximate depth of 28–29 AB magnitudes. 

Four of these targets are located in and around gravitationally lensing galaxy clusters, while one is within a blank field - the NEP. The four clusters include MACS 0416, Clio, and El Gordo.   The data acquired from PEARLS includes imaging in seven wide filter bands: F090W, F115W, F150W, F200W, F277W, F356W, and F444W, along with one medium band: F410M. In addition, we also include bluer F606W imaging from the HST Advanced Camera for Surveys (ACS) Wide Field Channel (WFC) in the NEP-TDF from the GO-15278 (PI: R.~Jansen) and GO-16252/16793 \citep[PIs: R.~Jansen \&
N.~Grogin, see][]{2024ApJS..272...19O} HST programs.

For the lensing fields of SMACS 0723, MACS 0416, and El Gordo, the observations include pointings with one NIRCam module centered on the lensing cluster, while the second module is offset by around 3 arcminutes in a 'blank' region. Although we process both modules in these fields, we do not include sources found in the module containing the lensing cluster in this study. For this paper we do not include sources found in the cluster module to avoid sometimes uncertain corrections for potential mergers where there is more contamination from the cluster members and uncertainties due to the effects of lensing such as magnification \citep[e.g.,][]{Griffiths2018, Bhatawdekar2021}.   For these clusters it is only the NIRCam module which is not centred on the cluster is used.



\subsection{NGDEEP}
The Next Generation Deep Extragalactic Exploratory Public Survey (NGDEEP; ID 2079, PIs: S. Finkelstein, Papovich, and Pirzkal) \citep{2023ApJ...952L...7A, 2023arXiv230205466B, 2023ApJ...954L..46L} is the deepest JWST NIRCam data during the first year of operations. Initially scheduled for late January to early February 2023, NGDEEP's primary focus involved NIRISS Wide Field Slitless Spectroscopy of galaxies within the Hubble UltraDeep Field. However, due to an unforeseen suspension of NIRISS operations during the survey's observation window, only half of the planned observations were executed, with the remainder taken in early 2024. In this work we use exposures from the first epoch of observations. The parallel NIRCam observations from NGDEEP encompass six wide-band photometric filters (F115W, F150W, F200W, F277W, F356W, F444W).  Some of our early work in this field was presented in \citet{austin2023large}.

\subsection{GLASS}
The Grism Lens Amplified Survey from Space (GLASS) observational program \citep[GLASS; PID 1324, PI: T. Treu;][]{2022ApJ...935..110T} is centered on the Abell 2744 Hubble Frontier Field lensing cluster. GLASS primarily employs NIRISS and NIRSpec spectroscopy to study galaxies lensed by the Abell 2744 cluster. In parallel, NIRCam is used to observe two fields offset from the cluster center. Due to the minimal lensing magnification in these offset areas \citep{2023A&A...670A..60B}, they effectively serve as blank fields. The observations include seven wide filter bands: F090W, F115W, F150W, F200W, F277W, F356W, and F444W.

\subsection{SMACS J0723-7327 ERO}
The SMACS-0723 lensing cluster is part of the very first ERO program (PID 2736, PI: Klaus Pontoppidan, \cite{2022ApJ...936L..14P}). The program includes NIRCam, MIRI, NIRISS, and NIRSpec observations, and we use the NIRCam observations which contains six bands: F090W, F150W, F200W, F277W, F356W, and F444W. The exposure time of this observation is set to roughly reach the depth of the Hubble Frontier Fields in F814W (which was reached in 100~ks rather than 4~ks), 1.5 times the F160W depth, and almost 10 times the Spitzer \citep{Werner2004} IRAC 1+2 depth. Due to the inaccuracies in photometric redshifts for galaxies around $z \sim 10$ in this field, most of the identified pairs tend to be located either at the lower ($z \sim 4.5$) or higher ($z \sim 11.5$) ends of our redshift range.

\begin{table*}
\caption{$5\sigma$ depths, unmasked areas, and the number of robustly detected galaxies of the eight JWST Cycle-1 fields used in this study. For CEERS, JADES GOODS-S, NEP-TDF, and NGDEEP, we also incorporate existing HST ACS/WFC observations. CEERSP9 is listed separately from the other 9 pointings in CEERS as it is significantly deeper than the other pointings.}
\label{tab: fields parameters}
\centering
\begin{tabular}{
                l
                S[table-format=2.2] 
                S[table-format=2.2] 
                S[table-format=2.2] 
                S[table-format=2.2] 
                S[table-format=2.2] 
                S[table-format=2.2] 
                S[table-format=2.2] 
                S[table-format=2.2] 
                S[table-format=2.2]} 
\toprule \toprule
{HST ACS/WFC} & {CEERS P1-8 + 10} & {CEERSP9} & {NEP-TDF} & {JADES GOODS-S} & {MACS-0416} & {NGDEEP} & {GLASS} & {El Gordo} & {SMACS-0723} \\
\midrule
F606W & 28.60 & 28.31 & 28.74 & 29.07 & {-} & 29.20/30.30 & {-}& {-} &{-} \\
F814W & 28.30 & 28.32 & {-} & {-} & {-} & 28.80/30.95 & {-} & {-} & {-} \\
\midrule \midrule
{JWST NIRCam} & {CEERS P1-8 + 10} & {CEERSP9} & {NEP-TDF} & {JADES GOODS-S} & {MACS-0416} & {NGDEEP} & {GLASS} & {El Gordo} & {SMACS-0723} \\
\midrule
F090W & {-} & {-} & 28.50 & 29.58 & 28.67 & {-}   & 29.14 & 28.23 & 28.75 \\
F115W & 28.70 & 29.02 & 28.50 & 29.78 & 28.62 & 29.78 & 29.11 & 28.25 & {-} \\
F150W & 28.60 & 28.55 & 28.50 & 29.68 & 28.49 & 29.52 & 28.86 & 28.18 & 28.81 \\
F200W & 28.89 & 28.78 & 28.65 & 29.72 & 28.64 & 29.48 & 29.03 & 28.43 & 28.95 \\
F277W & 29.20 & 29.20 & 29.15 & 30.21 & 29.16 & 30.28 & 29.55 & 28.96 & 29.45 \\
F335M & {-}   & {-}   & {-}   & 29.58 & {-}   & {-}   & {-}   & {-}   & {-}   \\
F356W & 29.30 & 29.22 & 29.30 & 30.17 & 29.33 & 30.22 & 29.61 & 29.02 & 29.55 \\
F410M & 28.50 & 28.50 & 28.55 & 29.64 & 28.74 & {-}   & {-} & 28.45 & {-}   \\
F444W & 28.85 & 29.12 & 28.95 & 29.99 & 29.07 & 30.22 & 29.84 & 28.83 & 29.28\\
\midrule
{(arcmin\(^2\))} & 66.40 & 6.08 & 57.32 & 22.98 & 12.3 & 6.31 & 9.76 & 3.90 & 4.31\\
\midrule
{N$_{\mathrm{galaxy}}$} & 
\multicolumn{1}{c}{1116} & 
\multicolumn{1}{c}{140} & 
\multicolumn{1}{c}{1090} & 
\multicolumn{1}{c}{932} & 
\multicolumn{1}{c}{49} & 
\multicolumn{1}{c}{46} & 
\multicolumn{1}{c}{28} & 
\multicolumn{1}{c}{21} & 
\multicolumn{1}{c}{30} \\
\bottomrule \bottomrule
\end{tabular}
\end{table*}

\section{Photometric Redshifts and Stellar Masses}

In the following two sections, we describe the tools and methods used to measure galaxy redshifts and stellar masses.  We discuss the methods for measuring the properties of our galaxy sample, including stellar masses and magnitudes and how we selected the mergers from the probability distribution function for each galaxy's photometric redshift.  We then select galaxies within redshift and stellar mass ranges to measure how the merger history changes within these bins.
\label{sec: Photometric Redshift and Sample Selection}

\subsection{Photometric Redshift Measures}
To ensure robust galaxy close-pair selection and reliable derived fractions, it is crucial to derive accurate photometric redshift measurements and assess their quality. In this section, we describe the methods used to measure the photometric redshifts of our sample.

We use \texttt{EAZY-PY} (hereafter \texttt{EAZY} \citep{brammer2008eazy}) as our primary photometric SED-fitting code. We employ the "$\mathrm{tweak\_fsps\_QSF\_12\_v3}$" templates alongside Sets 1 and 4 of the SED templates developed by \citet{2023ApJ...958..141L}. These additional templates are optimized for high-redshift galaxies, which exhibit bluer rest-frame UV colors \citep{2022ApJ...941..153T, 2023MNRAS.520...14C, 2023ApJ...947L..26N, 2024MNRAS.531..997C, austin2024epochs} and stronger emission lines \citep[e.g.][]{2023ApJ...958L..14W}, characteristics frequently observed in such young and distant galaxies.

The "$\mathrm{tweak\_fsps\_QSF\_12\_v3}$" templates are generated using the flexible stellar population synthesis (FSPS) package \citep{2010ascl.soft10043C}. Set 1 includes three pure stellar templates from \citet{2023ApJ...958..141L}, which are based on the BPASS v2.2.1 stellar population synthesis (SPS) model \citep{2017PASA...34...58E, 2018MNRAS.479...75S, 2022MNRAS.512.5329B} for ages of $10^6$, $10^{6.5}$, and $10^7$ years at a fixed metallicity $Z_\star = 0.05 \, \mathrm{Z}_\odot$, assuming a standard Chabrier \citep{2003PASP..115..763C} IMF. Set 4 incorporates nebular continuum and line emission calculated using CLOUDY v17 \citep{2017RMxAA..53..385F}, with an ionization parameter $\log U = -2$, gas metallicity $Z_{gas} = Z_\star$, Lyman continuum escape fraction $f_\mathrm{esc,LyC} = 0$, hydrogen density $n_\mathrm{H} = 300$ cm$^{-3}$, and a spherical geometry, excluding Ly$\alpha$ emission.

We execute the \texttt{EAZY-py} fitting three times: first with a uniform redshift prior of $0 \leq z \leq 25$, and subsequently with a constrained upper redshift limit of $z \leq 4$ and $z \leq 6$. This triple approach allows us to compare the goodness of fit between high and low-redshift solutions, as indicated by the $\chi^2$ values, ensuring robust identification of high-redshift galaxies.

To account for uncertainties in flux calibrations and aperture corrections, as well as potential mismatches between synthetic templates and observed galaxies, we impose a minimum flux uncertainty of 10\% \citep{2023PASP..135b8001R}. Our selection criteria rely predominantly on specific cuts in computed quantities to minimize biases and incompleteness. We perform a visual review of cutouts and SED-fitting  solutions, this step results in less than 5\% rejection of our original sample, a significantly lower rate compared to other studies \citep[e.g.,][]{2024ApJ...964...66H}. We refer readers to \textsection 3 of \cite{2024ApJ...965..169A}, \textsection 2.5 of \cite{austin2024epochs}, \textsection 3.2 of \cite{harvey2024epochs}, and \textsection 2 of \cite{2024arXiv240714973C} for details of our selection criteria for robust high-redshift galaxies.

The photometric redshift quality of the EPOCHS sample has been investigated previously \citep{2024ApJ...965..169A, 2024MNRAS.529.4728D} by comparison with spectroscopically confirmed sources from JWSTs Near-InfraRed Spectrograph \citep[NIRSpec;][]{ferruit2022near, Jakobsen2022, Boker2023}. We give a brief summary of the quality of the data here. In \cite{2024ApJ...965..169A}, we matched our photometric redshifts with spectroscopic redshifts for five fields—CEERS, JADES GOODS-S, SMACS-0723, MACS-0416, and El-Gordo—finding a $\sim10$\% outlier fraction (above $15\%$ of the spectroscopic solutions), with the largest discrepancies occurring around $2 < z < 3$. In \cite{2024MNRAS.529.4728D}, we investigated the properties of 43 galaxies at $z > 7$ with both NIRCam and NIRSpec PRISM observations in the CEERS and JADES GOODS-S fields. This investigation revealed an outlier fraction of 4.7\%, defined as $ \mathrm{z}_\mathrm{phot} > 1.15(\mathrm{z}_\mathrm{spec}+1)$ or $ \mathrm{z}_\mathrm{phot} < 0.85 * \mathrm{z}_\mathrm{spec}+ 0.15$ \citep[][]{2024MNRAS.529.4728D}, and a normalized median absolute deviation (NMAD) of 0.035. These results provide strong evidence of the high quality of our photometric redshifts, thereby ensuring the accuracy of close-pair selections in this study.

\subsection{Stellar Mass Measures}
\label{sec: Stellar Mass Estimation}

To infer galaxy stellar masses from the photometry we use the \texttt{Bagpipes} \citep{carnall2018inferring} bayesian SED fitting code, adopting redshift priors from the \texttt{EAZY} redshift probability distribution function for each galaxy. We use Log10 priors for dust, metallicity, and age. The reason for selecting Log10 priors is because we expect high redshift galaxies to be young, with lower metallicity and to contain little dust. We set prior limits for metallicity in the range of $[5\text{e-03}, 5] \, \text{Z}_{\odot}$, and we use a dust prior in the range of $[0.0001, 10.0]$ in $A_\text{V}$, ionization parameter $\mathrm{Log}_{10}(\mathrm{U})$ in range of $[-4, -1]$.  The time assumed in our fitting models for star formation to start is at 0.001 Gyr, while the time assumed for star formation to stop is at $\tau_\text{U}$, with $\tau_\text{U}$ denoting the age of the Universe at that redshift. In addition, \cite{kroupa2001MNRAS.322..231K} IMF, \cite{2003MNRAS.344.1000B} SPS model, and the  \cite{calzetti2000dust} dust attenuation model is implemented.

We fit the galaxy data with six parametric star formation history (SFH) models, the integral of which corresponds to the stellar mass—log-normal, delayed, constant, exponential, double delayed, and delayed burst—along with a non-parametric Continuity model \citep{2019ApJ...876....3L}. The integration of the SFH is the stellar mass. These models have been used in various works \citep[e.g.][]{2023Natur.619..716C,2023MNRAS.524.2312E,2023MNRAS.522.6236T, 2023MNRAS.519.5859W}. Different SFHs can have a large impact on inferred stellar mass and star formation rate (SFR) \citep{furtak2021robustly, 2022ApJ...927..170T,  harvey2024epochs, wang2024quantifying}. We have chosen to present results inferred from our fiducial model, the log-normal SFH, because we expect high-redshift galaxies to exhibit rising or bursting star formation. Please refer to \citet{harvey2024epochs} for a detailed discussion on SFH, the effects of different priors in SED fitting, and their impact on galaxy stellar mass and the stellar mass function.

\begin{figure}
    \centering
    \includegraphics[width=\columnwidth]{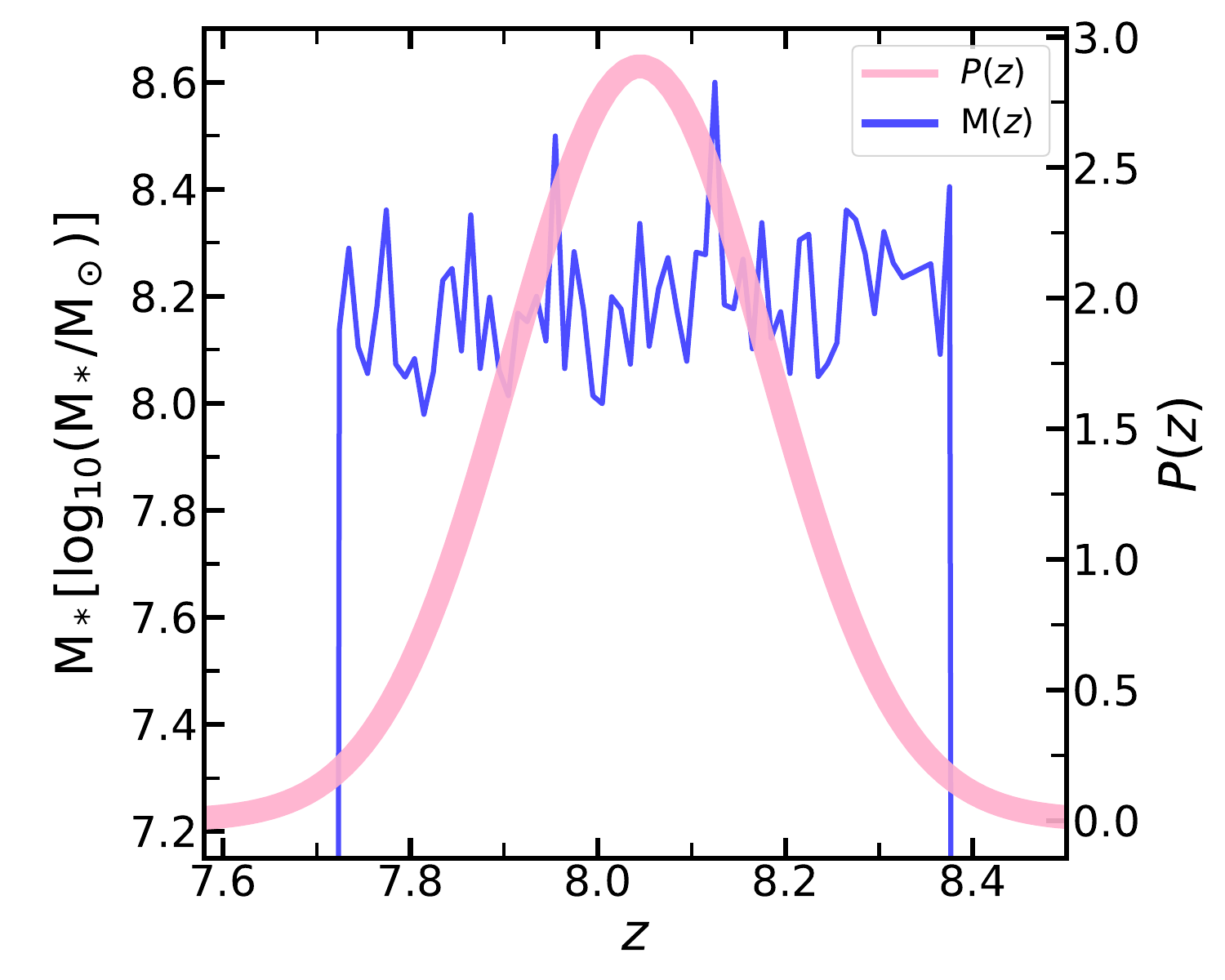}    
    \caption{Example of stellar mass $\mathrm{M}_*(z)$ and redshifts $P(z)$ distributions for an example galaxy from the NEP-TDF field, generated using \texttt{Bagpipes} and \texttt{EAZY}. The blue line represents the $\mathrm{M}_*(z)$ distribution, while the pink line depicts the $P(z)$ distribution. We compute the values of $\mathrm{M}_*(z)$ only at redshifts where the value of $P(z)$ is non-zero.}

    \label{fig:JWST pz mz}
\end{figure}

As we employ an integrated probability method that incorporates the entire photometric redshift probability distribution to compute the evolution of the merger history, we compute the stellar mass for each galaxy not only at the best-fit redshift but across all redshifts within the redshift posterior. For each galaxy, we take 400 independent draws from the best-fit posterior from the SED, each of which has a synthetic spectra for the galaxy, and taken from this the associated galaxy properties such as redshift and stellar mass. This lets us parameterise $\mathrm{M}_*(z)$, the best-fit galaxy stellar mass as a function of redshift. It is important to note that we give \texttt{Bagpipes} an informative redshift prior, taken from the highest likelihood draw from the \texttt{EAZY} posterior, which limits the possible range of redshifts \texttt{Bagpipes} will fit. An example of $\text{M}_*(z)$ together with its normalised redshift probability distribution for one of the NEP-TDF galaxies is shown in \autoref{fig:JWST pz mz}.

\section{Galaxy Pair Fraction Methodology}
\label{sec: Galaxy Pair Fraction Methodology}

The primary objective of this study is to accurately determine the fraction of galaxy close-pairs at high redshift, denoted as \( \text{f}_\text{p} \). Subsequently, we derive merger rates and mass accretion rates from this fraction. Accurately computing \fpp requires a statistically robust approach, particularly in light of the challenges presented by the photometric uncertainty, inherent incompleteness of  galaxy samples, and the prevalent bias towards the brightest high-z galaxies. To address these issues, we have developed a galaxy pair selection methodology, building upon the foundational work of \cite{2015A&A...576A..53L}, \cite{2017MNRAS.470.3507M}, \cite{2019ApJ...876..110D}, and \cite{2022ApJ...940..168C}. This method employs the integrated probability approach. Instead of using the best-fit redshifts of galaxies to search for mergers, we utilize their entire redshift probability distributions.  The steps are described in Sections \ref{sec: ppf} - \ref{sec: Sample Completeness Correction}. Each step is carried out at every redshift, and the results are integrated to derive the final pair fractions as detailed in Section \ref{sec: Final Integrated Pair Fractions}.

\subsection{Pair Probability Function}
\label{sec: ppf}
In this section, we illustrate the three close-pair selection criteria and describe how they combine to establish the final Pair Probability Function, $\mathcal{PPF}(z)$, at each redshift. 

\subsubsection{Projection Separation Selection}
\label{sec: Projection Separation Selection}
The initial selection of close pairs in our study is based on the projection separation. We consider the physical projection separation, \( r_p \), between two galaxies, which is a function of their angular separation \( \theta \) and the angular diameter distance \( d_A(z) \) at redshift \( z \). The projected separation \( r_p \) is thus mathematically expressed as:
\begin{equation}
    r_p(z) = \theta \times d_A(z).
\end{equation}

\noindent Subsequently, the projection separation mask is defined as follows: 
\begin{equation}
    \mathcal{M}^{r_p}(z) = 
    \begin{cases}
    1 & \text{if } 20 \, \text{kpc} < r_p(z) < 50 \, \text{kpc} \\
    0 & \text{otherwise},
    \end{cases}
\end{equation}
where a value of 1 signifies that the pair falls within the desired range for projected separation, thus meeting our first close pair selection criteria.

The selected projected separation range $20 < r_p < 50$ kpc differs from those used in previous studies by \cite{2017MNRAS.470.3507M}, \cite{2019ApJ...876..110D}, and \cite{2022ApJ...940..168C}. This range is strategically and physically motivated chosen. The lower limit of 20 kpc is chosen because, at the redshift range of this study ($4.5 < z < 11.5$), surface brightness dimming can be significant, making morphological indicators such as CAS unreliable. As a result, clumpy galaxies may appear as pairs with small projected separation less than 10 kpc. Additionally, the NIRCam pixel resolution of 0.031" is insufficient to resolve close pairs with very small projected separations. The upper limit of 50 kpc is chosen to exclude galaxy pairs whose separations are too large for merging within short timescales. Numerical simulations suggest that major bound companions separated by distances within this threshold typically merge within approximately 1 Gyr \citep{2008MNRAS.391.1489K, 2008ApJ...685..235P}. In Section \ref{sec: pair fractions with varying projection separation criteria}, we explore the effect of different $r_p$ ranges on the derived close pair fractions. We find that the pair fractions remain largely consistent even when using a lower limit of 5 kpc (see Figure~\ref{fig: Pair fractions separations}).

\subsubsection{Line of Sight (Redshift) Selection}
\label{sec: Line of Sight (Redshift) Selection}
For each close pair that meets the projection separation criteria, we designate the more massive galaxy as the primary galaxy and the less massive one as the secondary galaxy. The combined redshift probability function, \( \mathcal{Z}(z) \), is calculated to assess the likelihood that the pair is at the same redshift and, therefore, represents a true merger pair. This function is defined as:
\begin{equation}
\mathcal{Z}(z) = \frac{2 P_{1}(z) P_{2}(z)}{P_{1}(z) + P_{2}(z)} = \frac{P_{1}(z) P_{2}(z)}{N(z)},
\end{equation}
where \( P_1(z) \) and \( P_2(z) \) are the normalised redshift probability distribution functions for the primary and secondary galaxies, respectively, and \( N(z) \) serves as a normalization factor such that \( \mathcal{Z}(z) \) is properly scaled.

The combined redshift probability function \( \mathcal{Z}(z) \) incorporates the individual probability distributions of both galaxies, taking into account the uncertainty in their redshift determinations. This approach allows us to statistically account for the probability that both galaxies in a pair are at a similar redshift, which is a fundamental requirement for identifying potential mergers. However, we are careful to not double count pairs when examining the pair fraction.

An interesting quantity that can be derived from \( \mathcal{Z}(z) \) is \( \mathcal{N}_z \), which is calculated by integrating \( \mathcal{Z}(z) \) across the entire redshift range:
\begin{equation}
\mathcal{N}_z = \int_{0}^{\infty} \mathcal{Z}(z) \, dz,
\end{equation}
where \( \mathcal{N}_z \) values range between 0 and 1. This parameter quantifies the probability that a pair of galaxies in a potential merger system is a real pair. \(\mathcal{N}_z\) can be very useful for assessing the proximity of two galaxies along the line of sight direction, even though it is not directly utilized in the derivation of \(\mathcal{PPF}(z)\).

As each target galaxy can have more than one close companion, each potential galaxy pair is analysed separately and included in the total pair count. This ensures that our analysis accounts for potential multiple mergers for each galaxy. 

\begin{figure}
    \centering
    \begin{subfigure}{\columnwidth}
        \centering
        \includegraphics[width=\textwidth]{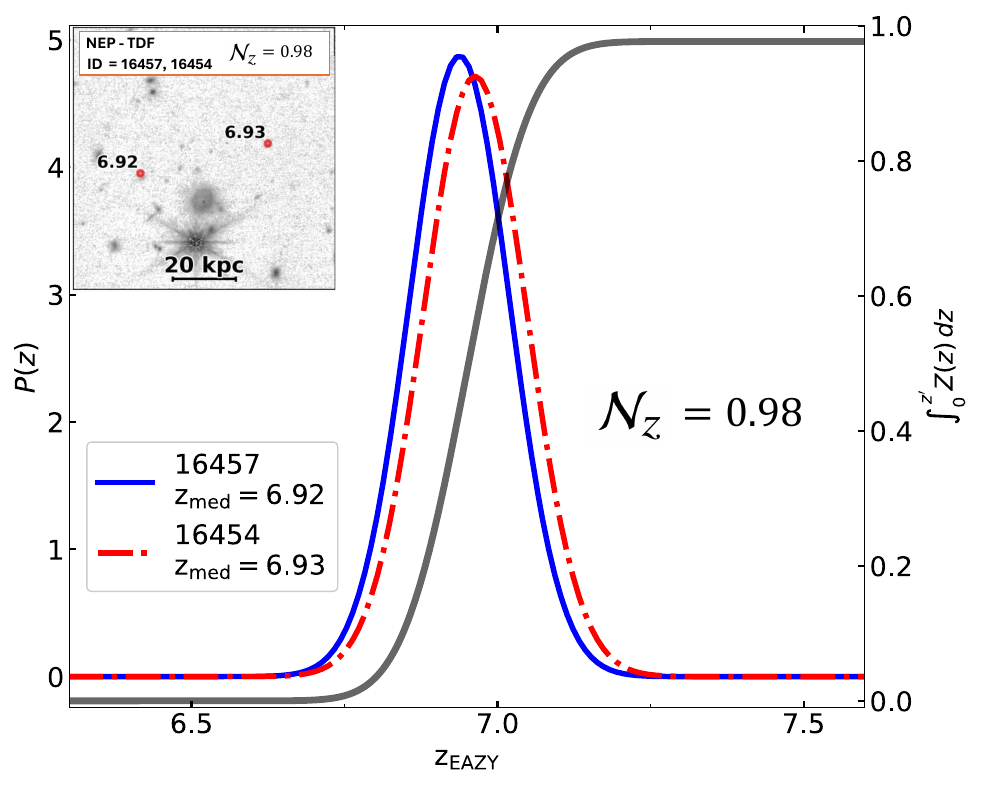}
        \caption{}
        \label{fig: nep Nz}
    \end{subfigure}
    
    \begin{subfigure}{\columnwidth}
        \centering
        \includegraphics[width=\textwidth]{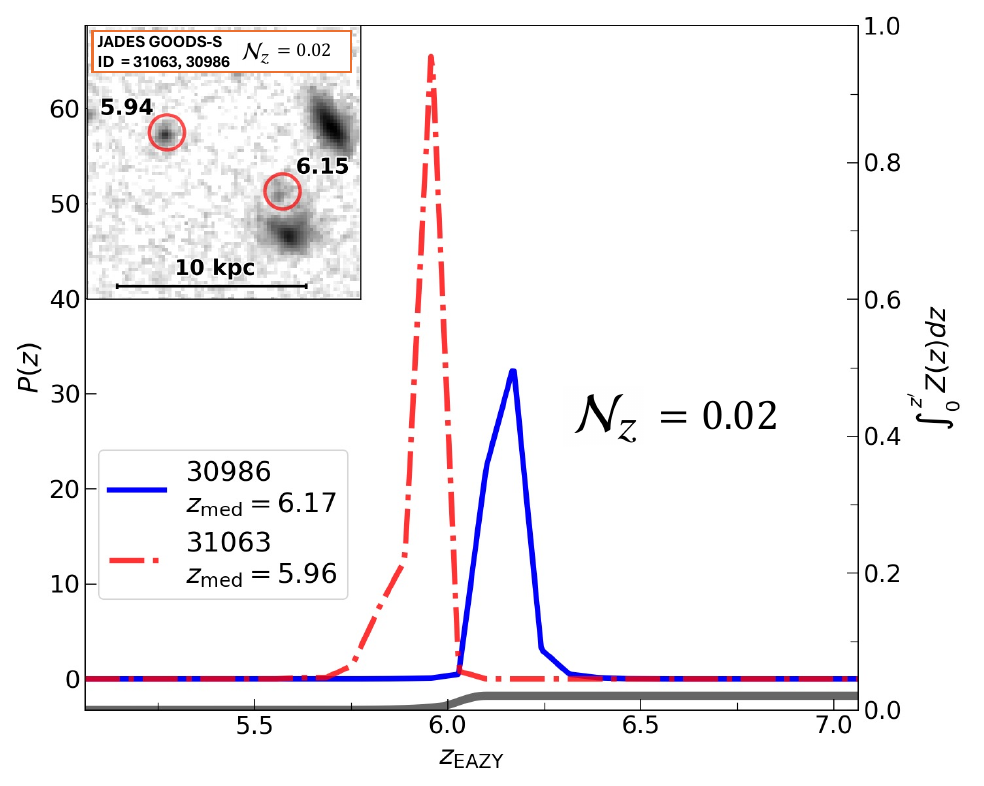}
        \caption{}
        \label{fig: JADES Nz}
    \end{subfigure}
    
    \caption{Example of \( \mathcal{N}(z) \) for two close pair systems from the NEP-TDF and JADES GOODS-S fields. The blue and red lines represent the normalized probability distributions of the galaxy redshifts within each pair. \( z_\mathrm{med} \) is the best fit photometric redshift. The black line denotes the integrated \( \mathcal{Z}(z) \), with \( \mathcal{N}(z) \) displayed in the bottom right corner of each panel. (a): Two galaxies with an angular separation of 7.32 arcsec and a physical projection separation of 41.3 kpc, with \(\mathcal{N}(z) = 0.98\), satisfy the projection separation criteria (20 kpc - 50 kpc) in this work and have a very high probability of being a real pair. (b): Two galaxies with an angular separation of 1.18 arcsec and a physical projection separation of 6.8 kpc, with \(\mathcal{N}(z) = 0.02\).  This system does not meet the projection separation criteria and has a very low probability of being a real pair.}
    
    \label{fig: ceers_jades_nz}
\end{figure}

We demonstrate the impact of \( \mathcal{Z}(z) \) and \( \mathcal{N}(z) \) on the selection of close pairs in \autoref{fig: nep Nz} and \autoref{fig: JADES Nz}. These figures present two close-pair systems from the NEP-TDF and JADES GOODS-S fields. In the example from NEP-TDF, two galaxies have a physical projection separation of 41.3 kpc and \(\mathcal{N}(z)\) of 0.98, satisfying the projection separation criteria (20 kpc - 50 kpc) in this work, and have a very high probability of being a real pair. Conversely, in the JADES GOODS-S example, two galaxies have a physical projection separation of 6.8 kpc and \(\mathcal{N}(z)\) of 0.02, which do not meet the projection separation criteria, and have a very low probability of being a real pair.

\subsubsection{Stellar Mass Selection}
\label{sec: stellar mass selection}
We have developed a modified version of the stellar mass selection criteria, building upon \cite{2015A&A...576A..53L, 2017MNRAS.470.3507M, 2019ApJ...876..110D, 2022ApJ...940..168C}. The primary distinction lies in the aspect of sample completeness. The aforementioned papers employ a redshift-dependent mass-completeness limit influenced by the flux limit set by the survey to accommodate varying survey depths across different fields. However, as can be seen from Table \ref{tab: fields parameters}, there are no big differences in survey depths across the eight JWST fields we are using. What is crucial to consider is that we are likely missing faint high-redshift galaxies, which could reduce the observed number of close-pairs. Thus, we have developed our novel completeness correction methodology, which is outlined in Section \ref{sec: Sample Completeness Correction}. In this section, we define our stellar mass selection to ensure the \textit{general} completeness of the sample, before proceeding to the completeness correction step.

We plot the stellar mass distributions for all galaxies in the redshift ranges $4.5 < z < 11.5$ and $6.5 < z < 11.5$ in \autoref{fig:jwst mass histogram}. Both distributions resemble Gaussian-like shapes with a noticeable right-skew. The median stellar masses are $\log_{10}(M_*/\mathrm{M}_\odot) = 7.92$ and $7.89$ for the two redshift bins, respectively. In a complete sample, we would expect a larger number of low-mass galaxies compared to high-mass ones. However, due to the depth limitations of our survey, many low-mass, faint galaxies are missed. As a result, the number of galaxies below $\log_{10}(M_*/\mathrm{M}_\odot) \sim 8$ is significantly lower. Furthermore, Beyond $\log_{10}(M_*/\mathrm{M}_\odot) = 10$, the number of detected galaxies becomes too small for meaningful statistical analysis, and so we adopt this value as the upper stellar mass limit. To ensure a robust and consistent sample, we then select galaxies with stellar masses in the range $8 < \log_{10}(M_*/\mathrm{M}_\odot) < 10$, and define our stellar mass selection mask in subsequent sections.

The stellar mass selection mask is defined as:
\begin{equation}
    \mathcal{M}^{\mathrm{M}}(z) = 
    \begin{cases}
    1 & \text{if } \mathrm{M}_{*}^{\text{lim},1}(z) \leq \mathrm{M}_{*,1}(z) \leq \mathrm{M}_{*}^{\text{max}} \\
      & \text{and } \mathrm{M}_{*}^{\text{lim},2}(z) \leq \mathrm{M}_{*,2}(z) \\
    0 & \text{otherwise},
    \end{cases}
    \label{eq: pair selection mask}
\end{equation}
where \( \mathrm{M}_{*,1}(z) \) and \( \mathrm{M}_{*,2}(z) \) represent the stellar masses of the primary and secondary galaxies at a given redshift, respectively. The primary galaxy is the higher mass one in a potential pair. The definitions of \( \mathrm{M}_{*}^{\text{lim},1}(z) \) and \( \mathrm{M}_{*}^{\text{lim},2}(z) \) are given by
\begin{equation}
    \mathrm{M}_{*}^{\text{lim},1}(z) = \mathrm{M}_{*}^{\text{min}},
\end{equation}
and
\begin{equation}
    \mathrm{M}_{*}^{\text{lim},2}(z) = \mu \mathrm{M}_{*,1}(z).
\end{equation}
The parameter $\mu$ represents the mass ratio, where values greater than $\mu$ characterize major mergers. In this study, we have set $\mu$ to a value of $1/4$ and defined major mergers as those with $\mu > 1/4$ \citep{2009ApJ...697.1369B, 2011ApJ...742..103L}. The stellar mass range of interest is defined by \( \mathrm{M}_{*}^{\text{max}} \) and \( \mathrm{M}_{*}^{\text{min}} \). In this work, we use the range $\log_{10}(\mathrm{M}_*/\mathrm{M}_\odot) = 8.0 - 10.0$.
The last mask is the primary galaxy selection mask, which is the first equation of \autoref{eq: pair selection mask}, given by:
\begin{equation}
    \mathcal{S}(z) = 
    \begin{cases}
    1 & \text{if } \mathrm{M}_{*}^{\text{lim},1}(z) \leq \mathrm{M}_{*,1}(z) \leq \mathrm{M}_{*}^{\text{max}} \\ 
    0 & \text{otherwise}
    \end{cases}.
\end{equation}

\subsubsection{Pair Probability Function}
The final pair probability function at a given redshift, \( \mathcal{PPF}(z) \), is derived by combining three selection criterias: projection separation, line of sight, and stellar mass. It is defined as:
\begin{equation}
    \mathcal{PPF}(z) = \mathcal{M}^{r_p}(z) \times \mathcal{Z}(z) \times \mathcal{M}^{M}(z).
    \label{eq: ppf}
\end{equation}
It is important to reiterate that the above processes are performed for each redshift within the entire photometric redshift fitting range. That is, at every redshift, we first generate an initial merger catalog based on projection separation criteria, then apply a mass selection mask, and compute $\mathcal{Z}(z)$ to produce the $\mathcal{PPF}(z)$ of each pair at that redshift. This is why we compute masses at all redshifts where the probability is non-zero for each galaxy (Section \ref{sec: Stellar Mass Estimation}). In the following sections, we will discuss two additional parameters that, when combined with $\mathcal{PPF}(z)$, will be integrated over the redshift bin of interest to compute pair fractions.

\subsection{Correction for Selection Effects}
\label{sec: selection effect}
The \( \mathcal{PPF}(z) \) function defined in \autoref{eq: ppf} is affected by two selection effects:
\begin{itemize}
    \item[(I)] The incompleteness in the search volume near the boundaries of the images.
    \item[(II)] Photometric redshift quality differences between survey regions.
\end{itemize}
We detail the corrections applied to address these issues in this section and discuss their implementation in the context of JWST data.

\subsubsection{Incompleteness Due to Image Borders and Masked Areas}
\label{sec: Incompleteness Due to Image Borders and Masked Areas}
Primary galaxies located close to image boundaries or masked regions (e.g., bright stars) possess a reduced spatial search area for potential companions, consequently reducing the observed number of pairs. Since the search area is contingent upon a fixed physical search radius, \( r_p \), this correction factor varies with redshift and necessitates computation at each redshift of interest.

For every primary galaxy, at each redshift, we delineate an annular region defined by an inner circle with a physical radius of 20 kpc and an outer circle with a radius of 50 kpc (projection separation criteria defined in \ref{sec: Projection Separation Selection}), centered on the galaxy. We then calculate the fraction of the annular's area, \( f_{\mathrm{area}}(z) \), that does not lie outside the image boundaries or within masked regions. Subsequently, each secondary galaxy is assigned a weight inversely proportional to the fraction of the area unmasked or inside the image. This weight is defined as:
\begin{equation}
    w^{\text{area}}(z) = \frac{1}{f_{\mathrm{area}}(z)}.
\end{equation}
This approach inherently compensates for the irregular geometries of survey areas and minor computational inaccuracies arising from finite pixel sizes. The impact of this correction to pair fractions is $\sim 7\% - 15\%$. This is analogous to the volume correction in luminosity functions. 

Another parameter we consider is the odds parameter, \( \mathcal{O} \), which is conceptualized by \citet{2000ApJ...536..571B} and \citet{2014MNRAS.441.2891M}, quantifies the confidence of a galaxy's photometric redshift falling within a certain interval around the best-fit value. This parameter has been effectively utilized in studies by \citet{2015A&A...576A..53L} and \citet{2017MNRAS.470.3507M}. As analyzed in Section~\ref{sec: Photometric Redshift and Sample Selection}, our photometric redshifts exhibit exceptional quality, with only 4.7\% outliers, and an NMAD value of 0.035 at \(z > 7\). Consequently, we do not use the \( \mathcal{O} \) parameter in our analysis. Similarly, \citet{2019ApJ...876..110D} did not employ the \( \mathcal{O} \) parameter, attributing this to the comprehensive magnitude-dependent calibration of photometric redshift posteriors, which ensures that the redshift distributions are accurately calibrated across all magnitudes.

\subsection{Completeness Correction for the JWST High-z Sample}
\label{sec: Sample Completeness Correction}
The incompleteness of the JWST high-redshift sample is evident from the mass distribution presented in \autoref{fig:jwst mass histogram}. A comprehensive sample would feature an increasing number of galaxies at lower masses, not the Gaussian-like distribution observed. More importantly, surface brightness dimming at $z = 4.5 - 11$ becomes significantly high. This results in a higher likelihood of missing faint companions at high redshifts. Consequently, the only objects for which pairs can be counted at $z > 4.5$ are those with companions that are point sources with very high surface brightness. We aim to correct this effect and retrieve those missing faint companions. The details of our completeness correction methodology are outlined in this section.

After applying a stellar mass selection of $\log_{10}(\mathrm{M}_*/\mathrm{M}_\odot) = 8.0 - 10.0$ to ensure general completeness, as described in Section \ref{sec: stellar mass selection}, we proceed to correct for the incompleteness of each individual galaxy. The method we developed for correction involves simulating different incompleteness scenarios with low-redshift data ($z < 6$) and investigating its impact on the derived pair fractions. We employ the same pair fractions methodology and use the same HST CANDELS catalogs from \cite{2019ApJ...876..110D} to calculate pair fractions under various scenarios of incompleteness. The CANDELS dataset encompasses five fields with a total number of 181,086 galaxies: GOODS-S \citep{2013ApJS..207...24G}, GOODS-N \citep{2023yCat..22430022B}, COSMOS \citep{2017ApJS..228....7N}, UDS \citep{2013ApJS..206...10G}, and Extended Groth Strip (EGS) \citep{2017ApJS..229...32S}. This dataset is complete for $\log_{10}(\mathrm{M}_*/\mathrm{M}_\odot) > 10.3$ and has undergone multiple quality checks and flag applications to ensure its cleanliness and reliability, thus enabling us to trace pair fraction evolution up to $z = 6.0$.

We simulate incompleteness scenarios by systematically removing portions of the complete lower redshift sample, ranging from \(10\%\) to \(90\%\), with a step size of \(10\%\), thus generating catalogs with varying degrees of completeness from 0.9 to 0.1. For each level of completeness, we implement the pair fraction methodology defined above. This allows us to evaluate the impact of sample completeness on the pair fractions. To minimize random errors due to the random exclusion of galaxies, we repeat the entire procedure five times. The data from all fields are utilized for \( z < 3.5 \), while for \( 3.5 < z < 6.5 \), we exclusively use data from the HST GOODS-S field, noted for its geatest depth among the five.

Upon calculating the pair fraction, \( \text{f}_\text{p} \), across different redshift bins under nine scenarios of incompleteness, we plot the \( \text{f}_\text{p}^{\, \text{ratio}} \)  against completeness for each redshift bin, as illustrated in \autoref{fig: Candels completeness}. Here, the \(\text{f}_\text{p}^{\, \text{ratio}}\) is defined as the pair fraction at a given completeness level divide by the complete sample case (completeness = 1). The resulting trend, indicative of a direct linear correlation between increasing completeness and \( \text{f}_\text{p}^{\, \text{ratio}} \), is modeled with a linear fit (\autoref{eq: fp_ratio v.s completeness}) using the Bayesian Markov Chain Monte Carlo (MCMC) method and \texttt{emcee} package. Specifically, we employ 100,000 steps and 50 walkers to generate candidate gradients and y-intercept values. For both sets of values, we adopt the median as the representative value and use the 1$\sigma$ deviation as the associated uncertainty, as the distributions follow a perfect Gaussian.
\begin{equation}
    \text{f}_\text{p}^{\, \text{ratio}}= \text{Gradient}(z) \times \text{Completeness} (z, \mathrm{M}_*) + \text{Y-Intercept}(z),
    \label{eq: fp_ratio v.s completeness}
\end{equation}
Subsequently, we chart the evolution of the best-fit lines' gradients and y-intercepts over different redshift bins in \autoref{fig: Candels comp gradient y-intercept}. This reveals an increasing trend in the gradient at higher redshifts and a decreasing trend in the y-intercept. These trends are further fitted linearly, with the best-fit parameters presented in the caption of \autoref{fig: Candels comp gradient y-intercept}, using the MCMC method described above. 

\subsubsection{The $\mathcal{C} (z, \mathrm{M}_*)$ correction factor}
The completeness correction factor, $\mathcal{C}(z, \mathrm{M}_*)$, depends on both mass and redshift. We compute the completeness level of JWST galaxies using the \texttt{JAGUAR} simulation \citep{2018ApJS..236...33W}, with details outlined in Appendix \ref{sec: Completeness Level} and in \autoref{fig: completeness all}. Additionally, we obtain the gradient and y-intercept information for the best-fit line (\autoref{eq: fp_ratio v.s completeness}) at different redshifts based on \autoref{fig: Candels comp gradient y-intercept}. By substituting these three parameters: gradient, completeness, and y-intercept to \autoref{eq: fp_ratio v.s completeness}, we derive the $\mathrm{f}_\mathrm{p}^{\mathrm{ratio}}$, which ranges between 0 and 1. The completeness correction factor is then defined as:
\begin{equation}
    \mathcal{C}(z, \mathrm{M}_*) = \frac{1}{\text{f}_\text{p}^{\, \text{ratio}}(\text{Completeness}(z, \mathrm{M}_*))}.
\end{equation}

\noindent In \autoref{fig: completeness all correction factor} we present $\mathcal{C} (z, \mathrm{M}_*)$ values across all eight fields.  We then implement this correction into our pair fraction measurement as described in the next section. 

\subsection{Final Integrated Pair Fractions}
\label{sec: Final Integrated Pair Fractions}
Integrating the $\mathcal{PPF}(z)$,  weights and $ \mathcal{C}(z, \mathrm{M}_*)$ discussed in Sections \ref{sec: ppf}, \ref{sec: selection effect}, and \ref{sec: Sample Completeness Correction}, we compute the final integrated pair fractions within each redshift interval of interest, denoted by $[z_{\text{min}}, z_{\text{max}}]$. Recall the more massive galaxy in each pair is designated as the primary, with the less massive serving as the secondary.

Letting \( i \) index each primary galaxy and \( j \) index every potential secondary galaxy surrounding the primaries, we express the number of pairs for each primary galaxy within each redshift interval as
\begin{equation}
    N_{\text{pair},i} = 
    \displaystyle\sum_j \int_{z_{\text{min}}}^{z_{\text{max}}} \mathcal{C}(z,M_*) \times \mathcal{PPF}_j(z) \times w^{\mathrm{area}}(z) \, dz
\end{equation}
where \( \mathcal{PPF}(z) \) is the pair probability function detailed in Section \ref{sec: ppf}, and \( \mathcal{C}(z,M_*) \) is the completeness correction factor introduced in Section~\ref{sec: Sample Completeness Correction}. Note that usually the summation over $j$ is just for one single pair, with multiple pairs and higher $j$ values being relatively rare. 

Each primary galaxy's contribution to the overall distribution of primary galaxies within each redshift range is given by:
\begin{equation}
    N_{\text{1},i} = 
    \displaystyle \int_{z_{\text{min}}}^{z_{\text{max}}}  \mathcal{P}_{1,i}(z) \times \mathcal{S}_{1,i}(z) \, dz
\end{equation}
where \( \mathcal{P}_{1,i}(z) \) represents the normalized redshift distribution function for each primary galaxy, \( \mathcal{S}_{1,i}(z) \) denotes the selection mask for primary galaxies as defined in Section \ref{sec: stellar mass selection}.

Finally, we define the pair fraction for each redshift interval as:
\begin{equation}
    \text{f}_{\text{p}} = \frac{\sum_i N_{\text{pair},i}}{\sum_i N_{\text{1},i}}
\end{equation}

\noindent This is then the pair fraction which we use through the remainder of the paper and which we also use to calculate the merger rate and the merger mass accretion rate.
 
\subsubsection{Uncertainty in Pair Fractions}
To compute the statistical uncertainties associated with the pair fraction, we employ the bootstrap resampling method \citep{e89fac9c-03d7-3e22-aa30-08f5596f8fce,dca15e5b-b3f7-3417-8555-955fe36eb045}. The standard deviation of the pair fraction, denoted as $\sigma_{\text{f}_{\text{p}}}$, is defined by the equation:
\begin{equation}
\sigma_{\text{f}_{\text{p}}} = \sqrt{\frac{\sum_{i=1}^{N} (f_{m,i} - \langle \text{f}_{\text{p}} \rangle)^2}{N - 1}},
\end{equation}
where $f_{m,i}$ represents the pair fractions estimated from a randomly selected sample of primary galaxies, $N$ denotes the number of realizations, and $\langle \text{f}_{\text{p}} \rangle$ is the average pair fraction, calculated as $\langle \text{f}_{\text{p}} \rangle = \frac{1}{N}\sum_{i=1}^{N} \mathrm{f}_{\text{p, i}}$. During the bootstrap process, we perform resampling with replacement data to generate new realizations of the data. We use 100,000 iterations to ensure statistical stability. In fact, we find the pair fractions to be stable after 10,000 iterations.

Note during the bootstrapping process, we do not propagage the uncertainties of $\mathcal{C}(z,M_*)$ into account. The 50th percentile of the $\mathcal{C}(z,M_*)$ values is used in the bootstrapping method. As shown in \autoref{fig: Candels comp gradient y-intercept}, the uncertainties in the gradient and y-intercept of the completeness correction are small, indicating that the 16th and 84th percentiles of $\mathcal{C}(z,M_*)$ are close to the 50th percentile (small $\sigma$). Therefore, while incorporating the full uncertainty in $\mathcal{C}(z,M_*)$ would slightly increase the uncertainty in the derived pair fractions, this effect is minimal and not noticeable on the figure given that the pair fractions results are presented on a log10 scale.

\begin{figure}
    \centering
    \includegraphics[width=\columnwidth]{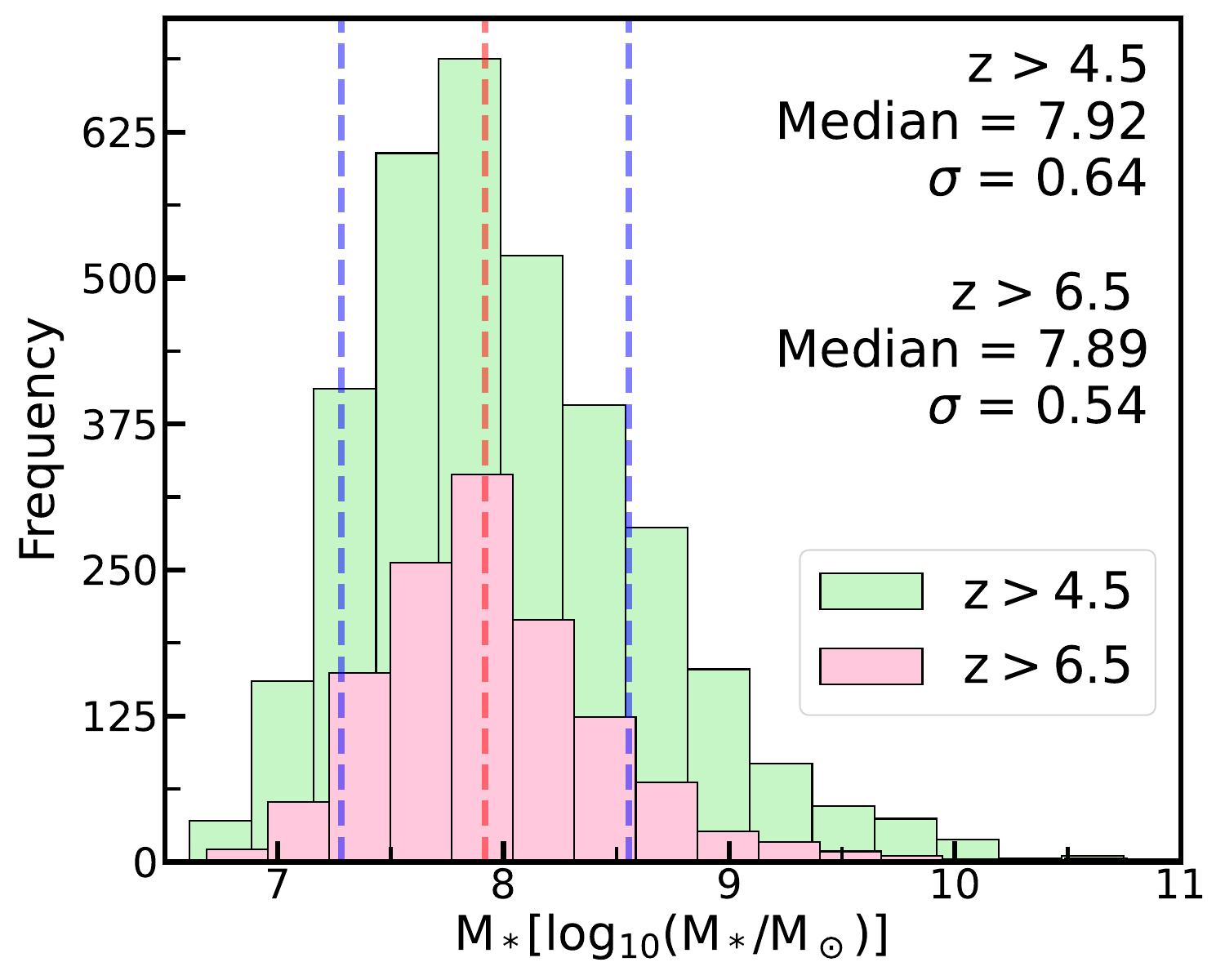}    \caption{Histogram displaying the distribution of best-fit stellar masses for all 1276 galaxies at $z > 6.5$ and for 3452 galaxies at $z > 4.5$ (an additional 2376 galaxies from the CEERS, JADES GOODS-S, and NEP-TDF fields). These samples are from eight  JWST fields: CEERS, JADES GOODS-S, NEP-TDF, NGDEEP, GLASS, El-Gordo, SMACS-0723, and MACS-0416. The median stellar masses are found to be 7.89 [$\mathrm{log}_{10}(\mathrm{M}_* / \mathrm{M}_\odot)$] and 7.92 [$\mathrm{log}_{10}(\mathrm{M}_* / \mathrm{M}_\odot)$] for each group, respectively.  At the same time, it is clear that there is a drop in completeness at $\mathrm{log}_{10} (\mathrm{M}_* / \mathrm{M}_\odot) < 8$ which is also seen in image simulations of these fields \citep[][]{harvey2024epochs}.}
    \label{fig:jwst mass histogram}
\end{figure}

\subsection{Simulation Data Pair Counts}
\label{sec: Simulation Data Pair Counts}
Comparing observational results with simulations is essential for bridging the gap between theoretical models of galaxy formation and evolution and what is observed in the real universe. We derive pair fractions using the \textit{Semi-analytic Forecasts for JWST} simulation catalog (hereafter SC-SAM). The "Semi-analytic Forecasts for the Universe" project is an ongoing project that offers physically grounded predictions regarding the characteristics and distributions of galaxies and quasars throughout cosmic history. These forecasts are generated using the well-established Santa Cruz semi-analytic model (SAM) for galaxy formation. This model stands out for its balance of physical rigor and computational efficiency, making it a valuable tool for understanding galactic evolution. Notably, the model has demonstrated its capability to reliably reproduce observational constraints up to a redshift of approximately $z\sim10$ \citep{2019MNRAS.483.2983Y}.

The project encompasses two primary data series: \textit{Semi-analytic Forecasts for JWST} and \textit{Semi-analytic Forecasts for Roman}. For our research, we utilize the \textit{Semi-analytic Forecasts for JWST} data product. The methodology of the simulation, along with its comprehensive details, is thoroughly documented in a series of studies \citep{2019MNRAS.483.2983Y, 2019MNRAS.490.2855Y, 2020MNRAS.494.1002Y, 2020MNRAS.496.4574Y, 2021MNRAS.508.2706Y, 2022MNRAS.515.5416Y}. This simulation comprises three products—\textit{wide}, \textit{deep}, and \textit{ultra-deep}—spanning areas of approximately $1000 \, \mathrm{arcmin}^2$, $132 \, \mathrm{arcmin}^2$, and $2 \, \mathrm{deg}^2$, respectively. We use the \textit{deep} dataset, as it overlaps with the HUDF, has a comparable area to our observations ($190 \,\mathrm{arcmin}^2$), and includes the JADES GOODS-S field. 

The methodology for calculating pair fractions from the SC-SAM catalogs differs from that used with actual observational data. In the simulation, precise redshift values, $z$, are available, in contrast to the redshift probability distribution function, $P(z)$, used in observational analyses, thereby simplifying the pair counting process.

The first step involves filtering the catalog. The simulation includes galaxies fainter than JWST's observational limit, which could potentially contaminate the pair fraction calculation. The median $\mathrm{m}_{\mathrm{F444W}}$ of our JWST sample is $27.8$, with a $1\sigma$ spread of $1.05$. Consequently, we impose a magnitude cutoff of $m_{\mathrm{F444W}} < 30$ on the SC-SAM catalog, utilizing the catalog's \textit{NIRCam F444W dust} column for this purpose. Additionally, we apply a mass selection criterion of $\log_{10}(\mathrm{M}_*/\mathrm{M}_\odot) = 8.0 - 10.0$ to the SC-SAM catalog, ensuring consistency with the completeness selections of our high-redshift JWST galaxies. The \textit{deep} dataset contains eight different realizations. After applying magnitude and mass selections, a total of 522,249 galaxies remain across all eight realizations.

Next, we select galaxy pairs from the SC-SAM catalog based on the projection separation criteria: $20 \, \text{kpc} < r_p(z) < 50 \, \text{kpc}$. For redshift (line-of-sight) selection, we employ the velocity difference upper limit method. The rest-frame velocity difference between two galaxies is given by
\begin{equation}
    \Delta v = \frac{c \times |z_1 - z_2|}{(1 + z_{\mathrm{mean}})},
\end{equation}
where $z_1$ and $z_2$ are the redshifts of the two galaxies, respectively, and $z_\mathrm{mean}$ is their average redshift. While \cite{2017A&A...608A...9V} use an upper limit of 500 km/s, higher limits have also been used, such as 1000 km/s \citep{2020MNRAS.493.1178E} and 10,000 km/s \citep{2023ApJ...957L..35G}. We calculate pair fractions at various velocity upper limits scenarios, with the highest value set at 2000 km/s, to better match what we are likely obtaining with our photo-zs. We find that there are not big differences in the derived pair fractions using different velocity upper limits, consistent with the findings of \cite{2019ApJ...876..110D}.

For each identified pair, we continue to apply the same definitions as in the observational data: the primary galaxy is the one with the greater mass in the pair, and the secondary galaxy is the less massive one. Only pairs associated with primary galaxies are counted to avoid over-counting identical pairs. We then calculate the pair fraction for each redshift bin using the formula:
\begin{equation}
    \mathrm{f}_{\text{p}} = \frac{N_{\text{pair}}}{N_{\text{total}}},
\end{equation}
which reflects the ratio of the number of pairs to the total number of galaxies in each redshift bin. The calculation of the errors in the pair fraction is performed using the binomial statistics \citep{1986ApJ...303..336G}.

As SC-SAM simulation \textit{deep} dataset comprises eight different realizations, we have conducted analyses on all these realizations and averaged the pair fractions derived from these multiple datasets to compute the final pair fractions for every redshift bin.

\section{high-redshift Merger Evolution at z > 6 }
\label{sec: high redshift Merger Evolution}
In this section, we present the first close-pair analysis of the pair fraction, merger rate, and mass accretion rate evolutions at $z = 4.5 - 11.5$. Along with our high-redshift JWST data points, we also include lower redshift results from the literature to trace the evolution from the Early Universe to the Local Universe. We present catalogs and cutout images of high-redshift close-pair merger systems in Appendix \ref{sec: merger catalog}.

\subsection{Pair Fraction}
\label{sec: pair fraction}
We compute pair fractions across eight JWST fields and the SC-SAM simulation for \(z = 4.5 - 11.5\), using the method detailed in Section \ref{sec: Galaxy Pair Fraction Methodology}. Considering the limited field area of each JWST field and the possibility of overdense regions to potentially elevate pair fractions \citep{2024arXiv240311615P}, we have combined the results from all eight fields to mitigate these effects.

In certain redshift bins within some fields, the samples are almost all incomplete to some extent (Completeness < 0.2).  If we blindly follow the completeness correction methodology outlined in Section \ref{sec: Sample Completeness Correction}, it will produce an extremely large completeness correction factor, \(\mathcal{C}(z, \mathrm{M}_*)\), which will greatly contaminate the derived pair fractions at very low completeness levels (please see \autoref{fig: completeness all correction factor} for the values of \(\mathcal{C}(z, \mathrm{M}_*)\) across different stellar mass and redshift ranges for all eight fields). As a result, these redshift bins are discarded and not included in the final combined $\mathrm{f}_\mathrm{p}$ derivation due to their being very incomplete and the corrections for these pair fractions is very high.  These discarded redshift bins are shown in \autoref{tab: discarded redshift bins}. Furthermore, we find a notably limited number of close pairs in SMCAS-0723, GLASS, MACS-0416, and El-Gordo fields due to small field area as shown in \autoref{tab: fields parameters}. Consequently, our calculation of \fpp is dominated by data from CEERS, JADES GOODS-S, NEP-TDF, and NGDEEP fields, where the more substantial number of pairs were detected.  The completeness values are discussed in more detail in the appendix and are shown graphically in Figure~\ref{fig: completeness all}.  Most of our corrections for this completeness in the pair fraction are relatively modest.   We only plot and analyze these relatively complete bins in redshift and stellar mass for our analysis. 

The evolution of the major merger pair fraction ($\mathrm{f}_\mathrm{p}$) with redshift is shown as yellow-orange stars in \autoref{fig: JWST Pair Fractions}. The corresponding pair fraction values for each redshift bin are listed in \autoref{tab: pf, mr, mr, smar values}. In addition to our high-redshift JWST measurements, we include lower-redshift results from \cite{2014MNRAS.445.1157C, 2015A&A...576A..53L, 2019ApJ...876..110D, 2022ApJ...940..168C}, spanning redshift ranges of $z \sim 0$, $0.4 < z < 1.0$, $0 < z < 6$, and $0 < z < 3$, respectively. The result from \cite{2014MNRAS.445.1157C} is plotted as blue triangles, while \cite{2015A&A...576A..53L} is shown in green pentagons. The work by \cite{2019ApJ...876..110D} presents pair fractions across different CANDELS sub-fields, each shown as colored rings, with the combined result depicted as a black ring. Finally, \cite{2022ApJ...940..168C} includes measurements from both VIDEO and UltraVISTA, represented by dark-blue and red diamonds, respectively.

It is also worthwhile to include many great lower redshift works  \citep[e.g.,][]{2013A&A...553A..78L, 2016ApJ...830...89M, 2017MNRAS.468..207S, 2017A&A...608A...9V,2017MNRAS.470.3507M, 2018MNRAS.475.1549M, 2021MNRAS.507.3113K, 2024arXiv240700594I}. However, considering the JWST data and methodology (integrated pair probability method) used in this work, it would be more appropriate to integrate with lower-redshift results that utilize similar datasets (e.g., HST-based) and methodologies to produce pair fraction evolution from $z \sim 11.5$ to the local Universes. 


Looking more closely at the evolution of the pair fraction in \autoref{fig: JWST Pair Fractions}, we observe a gradual increase in the major merger pair fraction up to $z \sim 8$, reaching a value of $0.211 \pm 0.065$. Although the last two points at \(z = 9.0\) and \(z = 10.5\) are lower than the point at \(z = 8.0\), these two points have high uncertainty. Taking this uncertainty into account, we conclude that the overall evolution of pair fractions shows a flat trend for \(z > 6\). We treat the point at \(z = 6.0\) as an outlier, as it is noticeably high. The reason for this is that among the three largest fields that dominate the \fpp values - CEERS, JADES GOODS-S, and NEP-TDF - the JADES GOODS-S and NEP-TDF galaxies exhibit a very high degree of incompleteness at $z = 5.5 - 6.5$ (\autoref{fig: completeness all},\autoref{tab: discarded redshift bins}), with the majority of galaxies having a completeness of less than 0.1, resulting in their exclusion. Thus, the \fpp at \(z = 6.0\) is dominated by the CEERS point, which is inherently high and could not be averaged down due to the exclusion of the two fields that could have balanced that value. The observed decline in $f_\mathrm{p}$ at $z > 8$ could be attributed to increased statistical uncertainty due to the smaller surveyed volume. In smaller volumes, pair fraction measurements become more sensitive to sampling variance, potentially leading to fluctuations that are consistent with the larger error bars expected from limited-volume statistics. Overall, the average \(\mathrm{f}_\mathrm{p}\) between \(z = [4.5 - 11.5]\) is \(0.17 \pm 0.05\).

This indicates that \(\mathrm{f}_\mathrm{p}\) at higher redshifts does not change as dramatically as observed at lower redshifts. Comparatively, the morphological selection method used by \cite{2024arXiv240311428D} finds no clear evolution in \(\mathrm{f}_\mathrm{p}\) between \(z = 4.0 - 9.0\), reporting an average pair fraction of \(\mathrm{f}_\mathrm{p} = 0.11 \pm 0.04\). This value is approximately 0.06 lower than our results based on the close-pair method. This difference arises primarily from variations in the timescales associated with each method. Morphological methods depend critically on the observability of merger-induced signatures, which can be shorter than the actual dynamical merging timescales (e.g., \citealt{2019MNRAS.486.3702S}). Additionally, not all galaxies identified as close pairs will necessarily merge into a single galaxy, as merger outcomes depend on various dynamical conditions and parameters (e.g., \citealt{2010MNRAS.404..575L}). Thus, the discrepancy between the two methods reflects both differences in merger observability windows and intrinsic variability in merger outcomes.

To model and assess the pair fraction evolution as a function of redshift, we take inspiration from various previous studies that fitted a Power-Law or a Power-Law + Exponential model to the evolution of pair fractions with redshift is appropriate \citep[e.g.,][]{2002ApJ...565..208P,2003ApJS..147....1C, 2007ApJ...659..931B,conselice2008structures, 2016ApJ...830...89M, 2017A&A...608A...9V, 2018MNRAS.475.1549M, 2019ApJ...876..110D, 2022ApJ...940..168C}, expressed as:
\begin{equation}
    \mathrm{f}_\mathrm{p} = f_0 \times (1+z)^m, 
    \label{eq: power law}
\end{equation}
and
\begin{equation}
    \mathrm{f}_\mathrm{p} = f_0 \times (1+z)^m \times e^{\tau(1+z)}.
    \label{eq: power exponential}
\end{equation}

\noindent We apply both a power-law and a power-law + exponential model to fit our data. The fitting is performed using only our JWST data, supplemented by a zero-point data from \citet{2014MNRAS.445.1157C}, represented as a blue triangle. The fitted lines are shown as black and red lines in \autoref{fig: JWST Pair Fractions}, and the fitted parameters are detailed in \autoref{tab: pair fractions fitting parameters}. The observation of a gradual increase in \fpp up to \(z = 8\), followed by a statistically flat trend towards \(z = 11.5\), indicates that a power-law + exponential model is better suited to describe the evolution of \fpp with redshift. This can be directly observed from the figure, as the power-law fitting results in a continuous increase in pair fractions even at very high redshifts (\(z > 8\)), where a flat trend is expected. Furthermore, we find that the fitted power-law + exponential line aligns very well with established literature, even though they are not included in the fitting. The higher \(\mathrm{f}_\mathrm{p}\) values from VIDEO, GAMA, and UltraVISTA by \citet{2022ApJ...940..168C} are because this work presents total pair fractions, which are, by definition, higher than major pair fractions, as pair fractions include both major and minor pairs (\(1/10 < \mu < 1/4\)).

It is also arguable that the stellar mass range of the literature results we used to compare and fit with our results are not consistent. We use lower redshift results with a generally higher stellar mass selection than our stellar mass range. However, this is simply because galaxies at high redshifts are less massive, and we are integrating our corrected mass-completed high-redshift sample with other mass-completed results at lower redshifts. We have done our best to normalize out the incompleteness with this, and as long as the pair fraction does not change significantly with mass at a given redshift without our range, then these comparisons are fine for a first examination of this merger evolution. 

Minor mergers (\(1/10 < \mu < 1/4\)) are observed to be more dominant than major mergers at low redshift \citep[e.g.,][]{2019ApJ...876..110D, 2019A&A...631A..51P, 2022ApJ...940..168C}. However, minor mergers are harder to detect because they are less luminous and more compact in size, especially at our redshift range \(z > 4.5\) where galaxies are so distant. Thus, the mass ranges of minor companions are incomplete and fall below our mass completeness limit. When using our method to compute minor merger pair fractions, we obtain values 3.5 times lower than those for major mergers, which is inaccurate due to the omission of many faint close-pair systems. Therefore, in this work, we do not present minor merger pair fractions and focus solely on major mergers.

The pair fractions presented so far are calculated using a physical projected separation range of $20 < r_p(z) < 50$ kpc, which we justified in detail in Section \ref{sec: Projection Separation Selection}. Additionally, these fractions were determined within the stellar mass range of $8.0 < \log_{10}(\mathrm{M}_* / \mathrm{M}_\odot) < 10.0$ (Sections \ref{sec: Stellar Mass Estimation} and \ref{sec: stellar mass selection}). Beyond our fiducial criteria, it is insightful to explore how varying the projected separation criteria and selecting narrower stellar mass sub-ranges affect the derived pair fractions. We perform these comparisons in the following two sections.

\begin{table}
    \centering
    \caption{Redshift bins discarded due to the presence of a large number of incomplete samples, to prevent contamination of pair fractions.}

    \label{tab: discarded redshift bins}
    \begin{tabular}{c|c} \toprule \toprule
        Field & Discarded Redshift-Bin \\\hline
        JADES GOODS-S & 5.5 - 6.5 \\ 
        NEP & 5.5 - 6.5 \\  
        El-Gordo & 9.5 - 11.5 \\ 
        MACS-0416 & 6.5 - 7.5, 9.5 - 11.5 \\
        SMACS-0723 & 9.5 - 11.5 \\
        \toprule \toprule
    \end{tabular}
\end{table}

\begin{table*}
\centering
\caption{Pair Fractions, Merger Rates, and Mass accretion rates and Specific Mass Accretion rates derived in this work across six different redshift bins.}
\label{tab: pf, mr, mr, smar values}
\begin{tabular}{c|cccccc}
\toprule \toprule
\textbf{Redshift} & {$[4.5,5.5]$} & $[5.5,6.5]$ & $[6.5,7.5]$ & $[7.5,8.5]$ & $[8.5,9.5]$ & $[9.5,11.5]$\\
\hline
\textbf{Pair Fractions} & \multirow{2}{*}{$0.125 \pm 0.020$} & \multirow{2}{*}{$0.252 \pm 0.035$} & \multirow{2}{*}{$0.162 \pm 0.025$} & \multirow{2}{*}{$0.211 \pm 0.065$} & \multirow{2}{*}{$0.140 \pm 0.070$} & \multirow{2}{*}{$0.117 \pm 0.084$}\\
($\mathrm{f}_\mathrm{p}$) &&&&&&\\

\textbf{Merger Rates} & \multirow{2}{*}{$1.87 \pm 0.30$} & \multirow{2}{*}{$5.15 \pm 0.71$} & \multirow{2}{*}{$4.32 \pm 0.66$} & \multirow{2}{*}{$7.12 \pm 2.21$} & \multirow{2}{*}{$5.85 \pm 2.90$} & \multirow{2}{*}{$6.45 \pm 4.62$} \\
($\mathcal{R}_\mathrm{M}$) [$\mathrm{Gyr}^{-1}$] &&&&&&\\
\textbf{Mass Accretion Rates} & \multirow{2}{*}{$0.86 \pm 0.14$} & \multirow{2}{*}{$2.40 \pm 0.33$} & \multirow{2}{*}{$2.04 \pm 0.31$} & \multirow{2}{*}{$3.66 \pm 1.16$} & \multirow{2}{*}{$2.88 \pm 1.76$} & \multirow{2}{*}{$3.19 \pm 2.31$} \\
($\rho$) [$\mathrm{M}_\odot \mathrm{Gyr}^{-1}$] /$10^{9}$ &&&&&&\\
\textbf{Specific Mass Accretion Rates} & \multirow{2}{*}{$0.93 \pm 0.15$} & \multirow{2}{*}{$2.55 \pm 0.35$} & \multirow{2}{*}{$2.14 \pm 0.33$} & \multirow{2}{*}{$3.52 \pm 1.07$} & \multirow{2}{*}{$2.92 \pm 1.42$} & \multirow{2}{*}{$3.23 \pm 2.32$} \\
($\dot{\mathrm{M}}/{\mathrm{M}_*}$) [$\mathrm{Gyr}^{-1}$] &&&&&&\\

\toprule \toprule
\end{tabular}
\end{table*}

\begin{table*}
\centering
\caption{Fitted Parameters for Power Law and Power Exponential Law Models applied to Pair Fractions, Merger Rates, and Specific Mass Accretion Rates, as described in \autoref{eq: power law} and \autoref{eq: power exponential}. These fittings incorporate results from our high-redshift JWST data, supplemented by zero-point data from \citet{2014MNRAS.445.1157C}.}
\label{tab: pair fractions fitting parameters}
\begin{tabular}{c|cccc}
\toprule \toprule
\textbf{Quantity}                             & \textbf{Function Type}    & \textbf{f$_0$} & \textbf{m}    & \textbf{tau}   \\ \hline
\multirow{2}{*}{Pair Fraction}                & Power-Law               & 0.031 ± 0.014 & 0.818 ± 0.233 & -              \\
                                              & Power-Law + Exponential & 0.030 ± 0.016 & 1.359 ± 0.725 & -0.138 ± 0.175 \\ \hline
\multirow{2}{*}{Merger Rate}                  & Power-Law               & 0.013 ± 0.006 & 2.818 ± 0.233 & -              \\
                                              & Power-Law + Exponential & 0.013 ± 0.007 & 3.359 ± 0.725 & -0.138 ± 0.175 \\ \hline
\multirow{2}{*}{Specific Mass Accretion Rate} & Power-Law               & 0.003 ± 0.001 & 3.169 ± 0.240 & -              \\
                                              & Power-Law + Exponential & 0.003 ± 0.002 & 4.051 ± 0.713 & -0.227 ± 0.173 \\ \toprule \toprule
\end{tabular}
\end{table*}

\subsubsection{Pair Fractions with Varying Projection Separation Criteria}
\label{sec: pair fractions with varying projection separation criteria}
In Section  \ref{sec: Projection Separation Selection}, we discuss the physical projection separation criteria used to select initial merger samples, with the range being $r_p = [20,50]$ kpc. However, the optimal range for this criterion is debatable and still uncertain; some studies set an upper limit of $\sim 30$ kpc \citep{2016ApJ...830...89M, 2017MNRAS.470.3507M, 2017A&A...608A...9V, 2019ApJ...876..110D, 2022ApJ...940..168C}, while others opt for $50$ kpc  \citep{2015A&A...576A..53L, 2017MNRAS.468..207S, 2023ApJ...957L..35G}. Therefore, investigating the impact of different projection separation criteria on pair fractions at $z = 4.5 - 11.5$ is crucial.

In \autoref{fig: Pair fractions separations}, we present pair fractions derived from five different projection separation criteria— with inner, outer radii in kpc: $[5,10]$, $[5,20]$, $[5,30]$, $[5,40]$, $[5,50]$ kpc—for \(z = 4.5 - 11.5\), while maintaining consistency in all other methodologies. The result from our fiducial separation criteria, $[20,50]$ kpc, is shown as the black dotted line. We observe that the \(\mathrm{f}_\mathrm{p}\) derived from all projection separations exhibits a similar trend—a gentle increase in the first few data points followed by a statistically flat trend or gradual decline at higher redshifts. Additionally, as expected, smaller projection separation measures result in lower pair fractions. For every 10 kpc reduction in the projection separation search, \(\mathrm{f}_\mathrm{p}\) decreases by a factor of $1.72 \pm 0.47$.

\subsubsection{Pair Fractions with Varying Stellar Mass Criteria}
In this section, we investigate the evolution of pair fractions across different stellar mass bins, while maintaining consistency in all other methodologies. As outlined in Section \ref{sec: stellar mass selection}, we initially apply a stellar mass selection of $\log_{10}(\mathrm{M}_*/\mathrm{M}_\odot) = 8.0 - 10.0$ to create a general mass completed sample, before proceeding to correct for completeness using our novel methodology discussed in Section \ref{sec: Sample Completeness Correction}. The range $\log_{10}(\mathrm{M}_*/\mathrm{M}_\odot) = 8.0 - 10.0$ is sufficiently broad to encompass all mass-completed samples, but it is also intriguing to investigate how pair fractions vary within this range. Therefore, in addition to computing the pair fractions at this mass range with a fiducial projection separation of $20 - 50$ kpc (shown as the black dotted line), we also calculate \(\mathrm{f}_{\mathrm{p}}\) for narrower mass ranges: $\log_{10}(\mathrm{M}_*/\mathrm{M}_\odot) = 8.0 - 9.0$ and $\log_{10}(\mathrm{M}_*/\mathrm{M}_\odot) = 9.0 - 10.0$.

\begin{figure*}
    \centering
    \begin{subfigure}{\linewidth}
        \centering
        \includegraphics[width= 0.8 \linewidth]{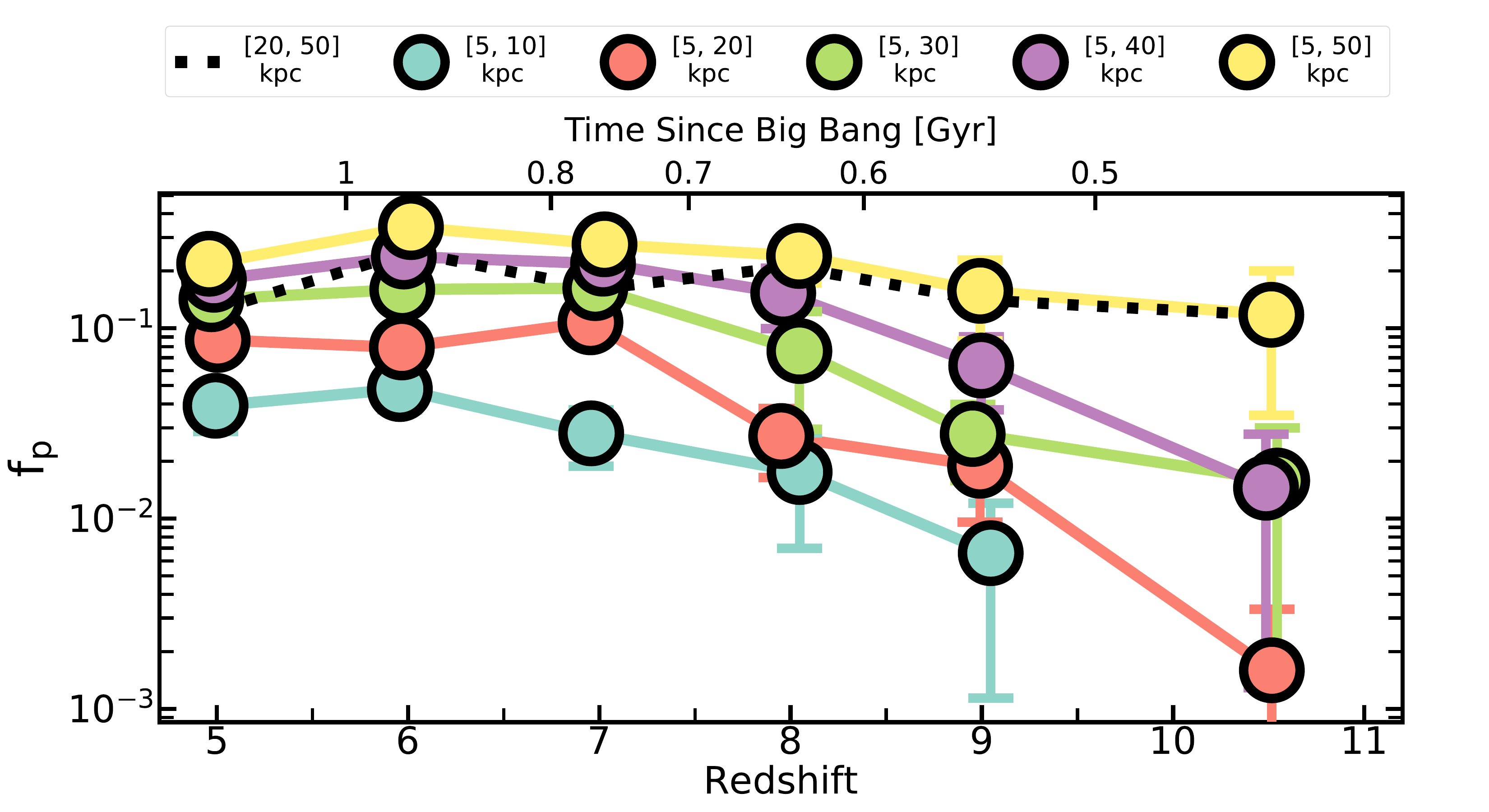}
        \caption{The pair fractions from $z = 4.5$ to $z = 11.5$, calculated using different projection separation criteria while maintaining consistent methodologies and a stellar mass range of $\log_{10}(\mathrm{M}_*/\mathrm{M}_\odot) = 8.0 - 10.0$. Results from our fiducial setting, [20,50] kpc, is shown by the black dotted line, while \fpp computed at [5,10], [5,20], [5,30], [5,40], and [5,50] kpc are shown by different colored circles. For every 10 kpc reduction in the projection separation search, \(\mathrm{f}_\mathrm{p}\) decreases by a factor of $1.72 \pm 0.47$.  Note that if we use a 5kpc inner limit we obtain very similar results as our default 20 kpc inner limit.  }
        \label{fig: Pair fractions separations}d
    \end{subfigure}
    
    \begin{subfigure}{\linewidth}
        \centering
        \includegraphics[width = 0.8 \linewidth]{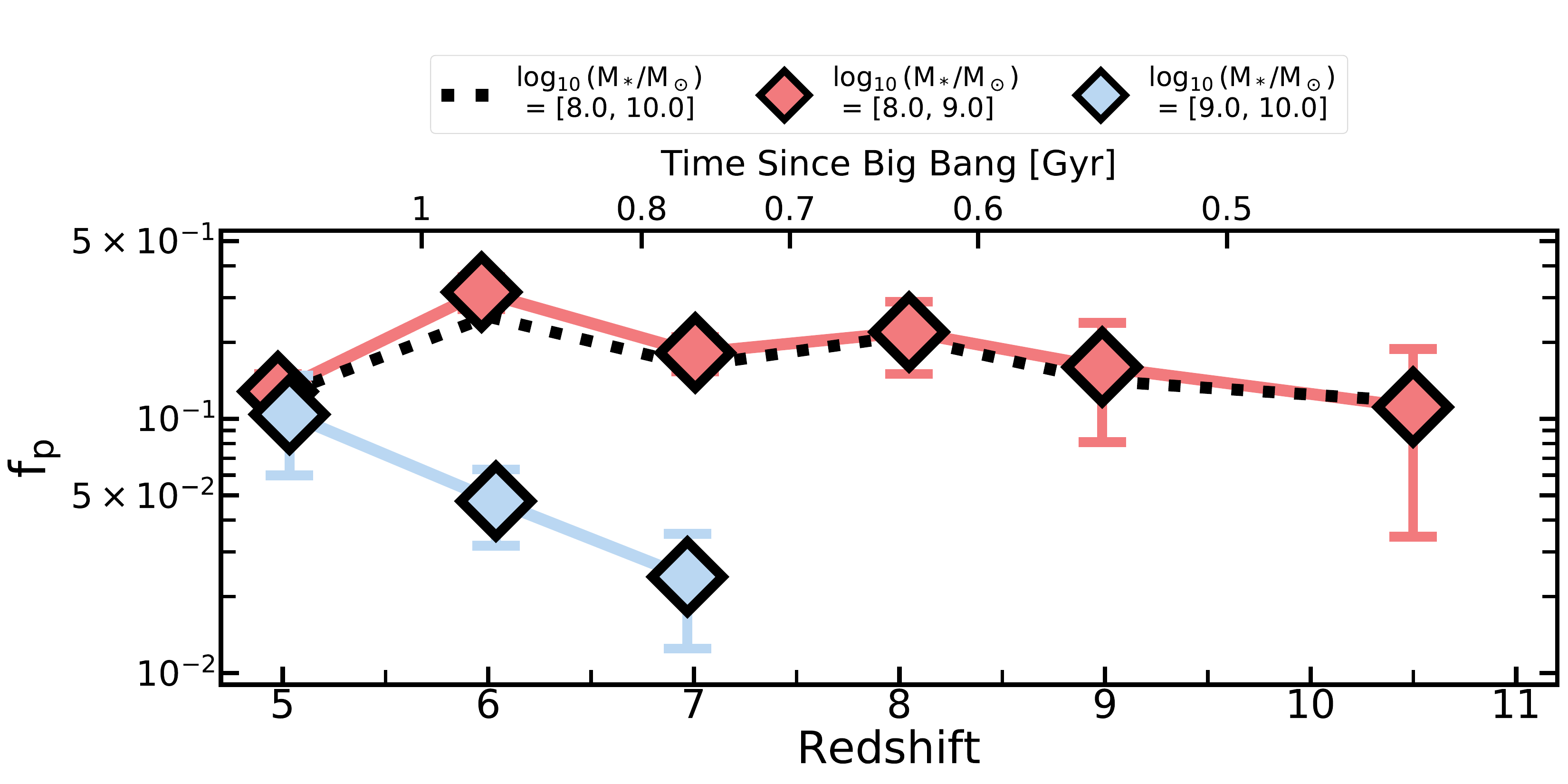}
        \caption{Pair fractions from $z = 4.5$ to $z = 11.5$, calculated using different stellar mass selection criteria while maintaining consistent methodologies and a projection separation range of $20 - 50$kpc. Results from our fiducial setting, $\log_{10}(\mathrm{M}_*/\mathrm{M}_\odot) = 8.0 - 10.0$, are shown by the black dotted line, while $\mathrm{f}_\mathrm{p}$ computed at $\log_{10}(\mathrm{M}_*/\mathrm{M}_\odot) = 8.0 - 9.0$ and $\log_{10}(\mathrm{M}_*/\mathrm{M}_\odot) = 9.0 - 10.0$ are shown as red and blue diamonds. Pair fractions computed at $\log_{10}(\mathrm{M}_*/\mathrm{M}_\odot) = 8.0 - 9.0$ is very similar to our fiducial model. Pair fractions from $\log_{10}(\mathrm{M}_*/\mathrm{M}_\odot) = 9.0 - 10.0$ shows a sharp decline compared to the other two, which is because there are very few massive galaxies in this range at high redshift.
}
        \label{fig: Pair fractions varying mass}

    \end{subfigure}
    \caption{In these two figures, we present the pair fractions computed with varying (a) Projection Separation Criteria, and (b) Stellar Mass Selection Criteria, with all other methodologies remaining consistent.}
\end{figure*}

We present this comparison in \autoref{fig: Pair fractions varying mass}, with \(\mathrm{f}_\mathrm{p}\) computed for stellar mass ranges of \(\log_{10}(\mathrm{M}_*/\mathrm{M}_\odot) = 8.0 - 9.0\) and \(\log_{10}(\mathrm{M}_*/\mathrm{M}_\odot) = 9.0 - 10.0\) represented in red and blue diamonds. We observe that \fpp values for the \(\log_{10}(\mathrm{M}_*/\mathrm{M}_\odot) = 8.0 - 9.0\) and \(\log_{10}(\mathrm{M}_*/\mathrm{M}_\odot) = 8.0 - 10.0\) ranges are quite similar. However, the \(\mathrm{f}_\mathrm{p}\) from the \(\log_{10}(\mathrm{M}_*/\mathrm{M}_\odot) = 9.0 - 10.0\) range shows a markedly sharp declining trend with redshift. This distinction is mainly attributable to the stellar mass distribution of high-redshift galaxies, as outlined in \autoref{fig:jwst mass histogram} and other recent literature \citep[e.g.,][]{2024MNRAS.529.4728D, harvey2024epochs}. High-redshift galaxies are generally less massive than low-redshift ones, and those within the \(\log_{10}(\mathrm{M}_*/\mathrm{M}_\odot) = 9.0 - 10.0\) range are particularly rare, resulting in statistically insufficient sample sizes, thereby leading to distinct trends in \(\mathrm{f}_\mathrm{p}\).

\subsubsection{Simulation Pair fractions}
\label{sec: sc-sam pair fractions}
\begin{figure*}
    \centering
    \begin{subfigure}{\textwidth}
        \centering
        \includegraphics[width=\linewidth]{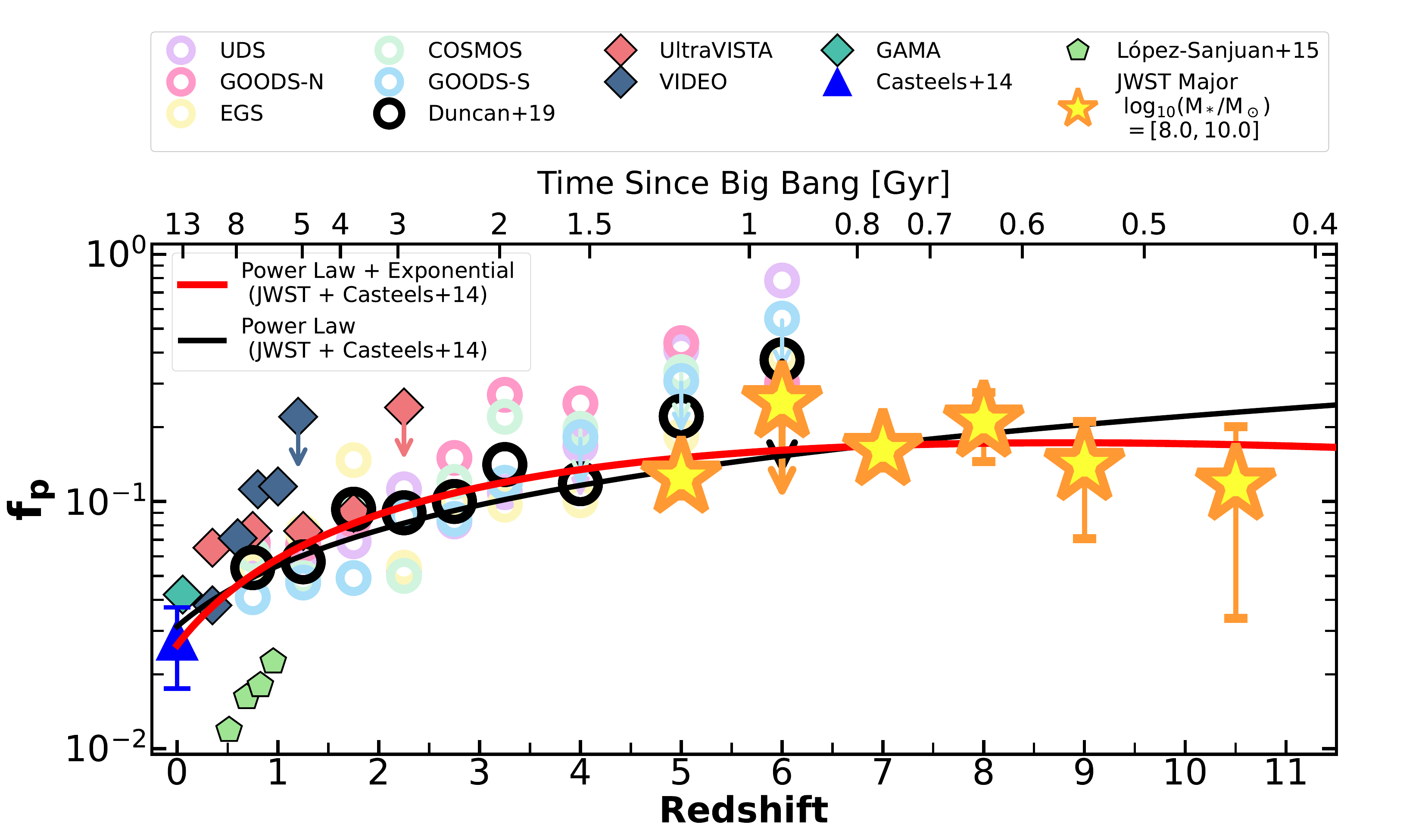}
        \caption{The evolution of our measured pair fractions with redshift. The new high-redshift major merger fraction ($\mathrm{f}_\mathrm{p}$) from our JWST analysis is depicted by yellow-orange stars, with $1\sigma$ error bars shown. The mass range used for our JWST data is $\log_{10}(\mathrm{M}_*/\mathrm{M}_\odot) = 8.0 - 10.0$. Additionally, we present lower redshifts results from four studies: green pentagons denote data from ALHAMBRA as reported by \protect\cite{2015A&A...576A..53L} for $M_B < -20.0$; rings represent CANDELS data by \protect\cite{2019ApJ...876..110D}, with colored rings indicating each sub-field and black rings representing the combined results for masses $\log_{10}(\mathrm{M}_*/\mathrm{M}_\odot) > 10.3$; diamonds symbolize data from the REFINE survey by \protect\cite{2022ApJ...940..168C} for masses $\log_{10}(\mathrm{M}_*/\mathrm{M}_\odot) > 11.0$; and the blue triangle denotes GAMA data computed by \protect\cite{2014MNRAS.445.1157C}. All literature except \protect\cite{2022ApJ...940..168C} present major merger $\mathrm{f}_\mathrm{p}$; \protect\cite{2022ApJ...940..168C} presents total $\mathrm{f}_\mathrm{p}$. Power-law and power-law + exponential fits (\autoref{eq: power law}, \autoref{eq: power exponential}) are applied to our JWST data, with \protect\citep{2014MNRAS.445.1157C} data used as a zero point. The fitted parameters are detailed in \autoref{tab: pair fractions fitting parameters}. We find an increase in pair fractions up to \(z \sim 8\), reaching \(0.211 \pm 0.065\), followed by a statistically flat evolution to \(z = 11.5\). The average \(\mathrm{f}_\mathrm{p}\) between \(z = [4.5 - 11.5]\) is \(0.17 \pm 0.05\).}
        \label{fig: JWST Pair Fractions}
    \end{subfigure}
    
    \begin{subfigure}{\textwidth}
        \centering
        \includegraphics[width=\linewidth]{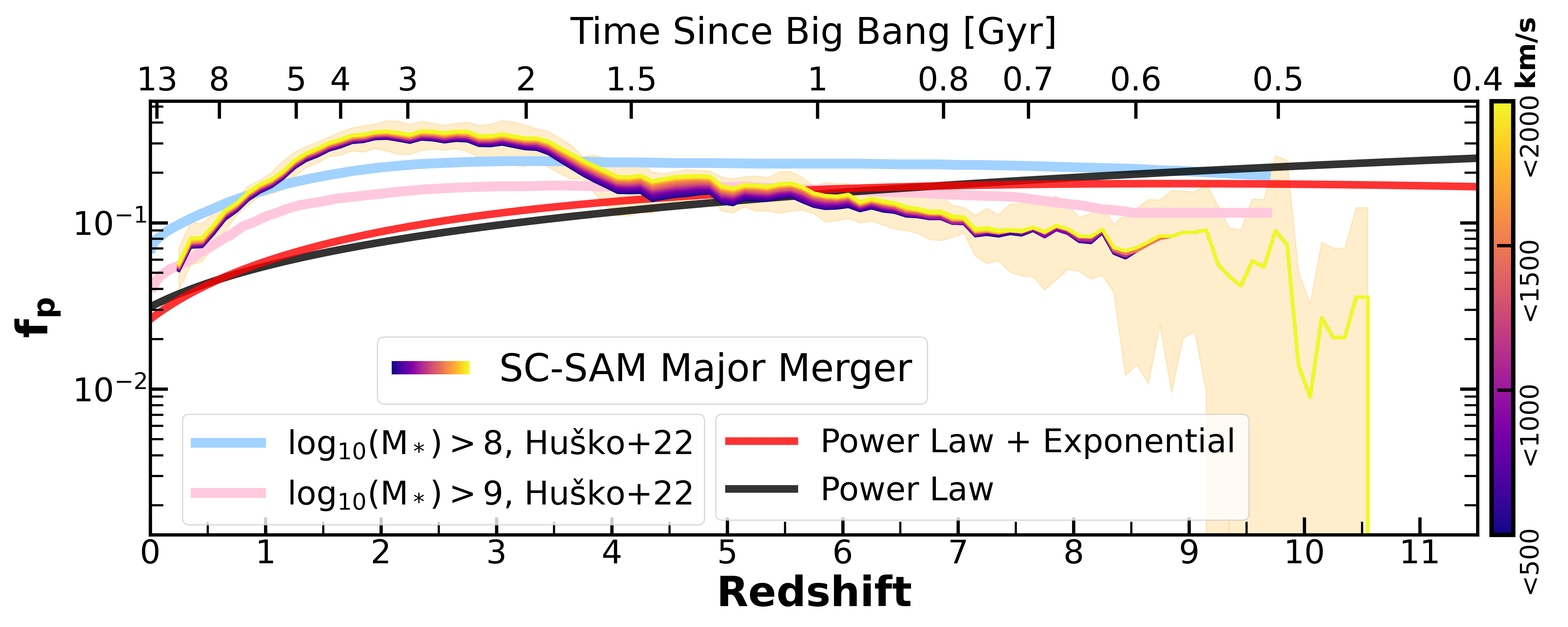}
        \caption{Evolution of pair fractions with redshift derived from the SC-SAM simulations and \protect\cite{2022MNRAS.509.5918H}. The fitted power and power exponential lines from observational data in \autoref{fig: JWST Pair Fractions}  are shown in black and red lines. Results from \cite{2022MNRAS.509.5918H} are displayed in blue and pink lines, where blue represents \(\mathrm{f}_\mathrm{p}\) for galaxies with \(\log_{10}(\mathrm{M}_*/\mathrm{M}_\odot) > 8.0\), and pink for \(\log_{10}(\mathrm{M}_*/\mathrm{M}_\odot) > 9.0\). SC-SAM \fpp for the major merger (\(\mu > 1/4\)) is depicted by purple-orange lines, respectively. The color represents different upper rest-frame velocity limits between two galaxies in km/s, with each velocity maximum corresponding to a unique pair fraction curve. The orange shaded regions indicate the combined binomial uncertainties and the variation across the eight different realizations for the pair fractions with $\Delta v < 2000$ km/s. Major merger pair fractions from SC-SAM peak at \(z = 2 - 3\) before declining. Observational data shows a gradual increase in the major merger pair fraction from \(z = 4.5\) to \(z = 8\) with a \fpp value of \(0.211 \pm 0.065\), followed by a moderate decline to \(0.117 \pm 0.084\) at \(z = 10.5\).}
        \label{fig: Pair Fractions Simulations}
    \end{subfigure}
    \caption{Two figures illustrate the evolution of pair fractions, \(\mathrm{f}_{\mathrm{p}}\), derived from observations (top panel) and the Santa Cruz-SAM simulations (bottom panel). A notable discrepancy in pair fractions between observations and simulations is observed at most redshifts.}
\end{figure*}

We calculate major merger pair fractions ($\mu > 1/4$) from the SC-SAM simulations' \textit{deep} dataset, employing the method described in Section \ref{sec: Simulation Data Pair Counts}. We consider various upper limits for rest-frame velocity differences between galaxies, with the maximum set at $\Delta v = 2000$ km/s. The evolution of major merger pair fractions with redshifts is illustrated in \autoref{fig: Pair Fractions Simulations}, represented by purple-yellow lines. Fitted power-law and power-law + exponential curves based on observational data are depicted in black and red lines. Pair fractions from the simulation by \cite{2022MNRAS.509.5918H} are displayed in blue and pink lines, where blue represents \(\log_{10}(\mathrm{M}_*/\mathrm{M}_\odot) > 9.0\) and pink for \(\log_{10}(\mathrm{M}_*/\mathrm{M}_\odot) > 8.0\).

From SC-SAM simulation, we observe that major merger pair fractions reach a maximum value of approximately $0.3$ at $z = 1.5 - 3.5$, before starting to decline at higher redshifts. Additionally, we find that the SC-SAM simulation inadequately describes the pair fractions at $z > 6$, or at any redshift except near $z \sim 4$, due to the limited number of pairs seen in the simulation. A moderate agreement between the SC-SAM simulations and observational results is observed at \(z < 4\), where pair fractions can be adequately described by a power-law. However, at higher redshifts (\(z > 4\)), the observational results significantly deviate from the simulations. Our analysis reveals an increase in pair fractions up to \(z = 8\), reaching \(0.211 \pm 0.065\), followed by a statistically flat evolution to \(z = 11.5\). This finding sharply contrasts with the fast decreasing trend predicted by the simulations after $z = 3$, where the pair fractions decrease to around 0.01 at $z = [9-11.5]$. 

These are major differences between simulations and observations in the merger history, and this difference is now the most important conflict between simulations and observations for distant galaxies. There are several possibilities for explaining the difference we find, including observational as well as simulation input. For example, over-quenching of satellite galaxies may explain these distinctions \citep[e.g.,][]{2008MNRAS.387...79V, 2020MNRAS.499..230B, 2020A&A...638A.112V}. The Santa Cruz SAM, like many early models, excessively quenches satellite galaxies. In the current configuration, cosmological accretion halts once a galaxy becomes a satellite, rapidly depleting its gas supply and stifling star formation. This process is likely more extreme than what occurs in the real Universe, leading to simulated satellite galaxies that are \textit{less luminous} and \textit{less massive} than their observed counterparts.  This would then imply that when these systems merge with most massive galaxies the merger is not seen as major, but minor. Updates to semi-analytic and empirical satellite models are necessary, and many researchers, such as Behroozi et al. and Pandya et al., are actively developing these enhancements.

In addition to the SC-SAM simulation, we compare our results with major merger pair fractions derived by \cite{2022MNRAS.509.5918H}, who employed the Planck Millennium cosmological dark matter simulation and the \texttt{GALFORM} semi-analytical model of galaxy formation. They computed pair fractions using a projection separation of \(r_p < 20\) kpc and a velocity difference of \(\Delta v < 500\) km/s. Their results are illustrated in \autoref{fig: Pair Fractions Simulations} with blue and pink lines, where blue lines represent \(\mathrm{f}_\mathrm{p}\) for galaxies with \(\log_{10}(\mathrm{M}_*/\mathrm{M}_\odot) > 8.0\), and pink lines represent \(\mathrm{f}_\mathrm{p}\) for galaxies with \(\log_{10}(\mathrm{M}_*/\mathrm{M}_\odot) > 9.0\). When comparing with the SC-SAM \(\mathrm{f}_\mathrm{p}(z)\), the results from \cite{2022MNRAS.509.5918H} seems to show better agreement with our observational data points. However, it is important to note that the selection criteria used in \cite{2022MNRAS.509.5918H} are significantly more restrictive compared to the broader selection adopted in our study. Since more restrictive criteria tend to yield systematically lower pair fractions, the apparent agreement at high redshift may in fact mask a more substantial discrepancy. This highlights that even with conservative selections, some simulations may still overpredict pair fractions at high redshift, further reinforcing the observed tension between simulations and observations.

Overall, regardless of which simulation we compare against, we do not find strong agreement between our observations and theoretical predictions—consistent with findings from other studies at lower redshifts \citep[e.g.,][]{2009MNRAS.396.2345B, 2009ApJ...697.1971J, 2018MNRAS.475.1549M, 2019ApJ...876..110D}. \cite{2018MNRAS.475.1549M} also compared galaxy merger rates at $0 < z < 3$ with halo major merger rate predictions for halos of $\mathrm{M_{halo}} \sim 10^{12} \, \mathrm{M}_\odot$, finding that galaxy merger rates broadly agree with halo merger rates at $z < 1$, but deviate significantly at $z > 1$. This discrepancy—up to a factor of five—greatly exceeds the typical factor-of-two uncertainties arising from different halo occupation models \citep{2010ApJ...715..202H}.

To explain this difference between models and observations, \cite{2016ApJ...830...89M} proposes that the discrepancy arises from the different ways in which galaxies are selected. Observational studies typically rely on stellar mass, while simulations often use baryonic or dark matter mass. Since high-redshift galaxies tend to have higher gas fractions, some mergers classified as minor ($\mu < 1/4$) based on stellar mass may actually be major mergers when selected by baryonic mass. As this is the first observational work using the close-pair method to study pair fractions at $z > 6.0$, we currently lack information on the gas content of these systems to test this directly. However, incorporating gas content into merger classifications would be a valuable step in future studies.

In addition to mass selection effects, one could argue that our use of a completeness correction factor—applied to recover the underlying pair fractions—differs from the approach used in simulations, which are intrinsically complete by construction. However, studies such as \cite{2019ApJ...876..110D, 2025arXiv250201721P} adopt redshift-dependent stellar mass completeness limits \citep[e.g.,][]{2010A&A...523A..13P} and restrict their galaxy samples to those above these limits when measuring pair fractions. Despite this, they also report poor agreement with simulation predictions.

It is also important to emphasize that the comparison between observations and theoretical simulations in this work is based on pair fractions, which represent the primary observable derived from our method. When extending the comparison to merger rates (Section \ref{sec: merger rates})—computed from pair fractions using assumed values for $C_\mathrm{merg}$ (the fraction of close pairs that eventually merge) and $\tau_{\mathrm{m}}(z)$ (the merger timescale)—the discrepancy with simulations becomes even larger. This is because both $C_\mathrm{merg}$ and $\tau_{\mathrm{m}}(z)$ may depend on redshift and stellar mass, yet most merger studies (including this one) treat one as a constant and the other evolving with redshifts. This can introduce additional uncertainties when comparing observations with theoretical predictions.

\subsection{Merger Rates}
\label{sec: merger rates}
The merger rate is defined as the number of merger events per galaxy per unit time, denoted as $\mathcal{R}_\text{M}(z, \mathrm{M}_*)$. Compared to pair fractions, merger rates are more fundamental in nature because the pair fraction is an observational metric, subject to variations due to differing methodologies across redshift bins. Such variations affect the sensitivity to the timescales being observed, thereby complicating direct comparisons across different studies. In contrast, comparing merger rates between literature becomes meaningful when the typical timescale for observing a galaxy merger is well-defined. The methodology for calculating $\mathcal{R}_\text{M}(z, \mathrm{M}_*)$ in this paper follows \cite{ 2019ApJ...876..110D, 2020ApJ...895..115F,2022ApJ...940..168C}, with specific details provided in following paragraphs.

The merger rate per galaxy is expressed as
\begin{equation}
\label{eq: merger rate}
    \mathcal{R}_\mathrm{M}(z, \mathrm{M}_*) = \frac{\mathrm{f}_\mathrm{p}(z, \mathrm{M}_*) \times C_\mathrm{merg}}{\tau_{\mathrm{m}}(z)},
\end{equation}
where $\mathrm{f}_\mathrm{p}(z, \mathrm{M}_*)$ is the pair fraction at each redshift bin, $C_\mathrm{merg}$ is the fraction of pairs that will eventually result in a merger event, and $\tau_{\mathrm{m}}(z)$ is the merger timescale at a certain redshift. 

In the literature, a value of $0.6$ is commonly assigned to $C_\mathrm{merg}$ \citep{2011ApJ...742..103L, 2014ARA&A..52..291C, 2017MNRAS.470.3507M}. However, this value is subject to significant uncertainty, as it depends on various factors including stellar mass, mass ratio, and redshift. Considering the high redshift scenario of our study, which further adds to the complexity of accurately defining $C_\mathrm{merg}$, we opt to set this value to 1. This implies, under our working assumptions, that each pair we observe is anticipated to eventually merge.

\begin{figure}
    \centering
    \includegraphics[width=\linewidth]{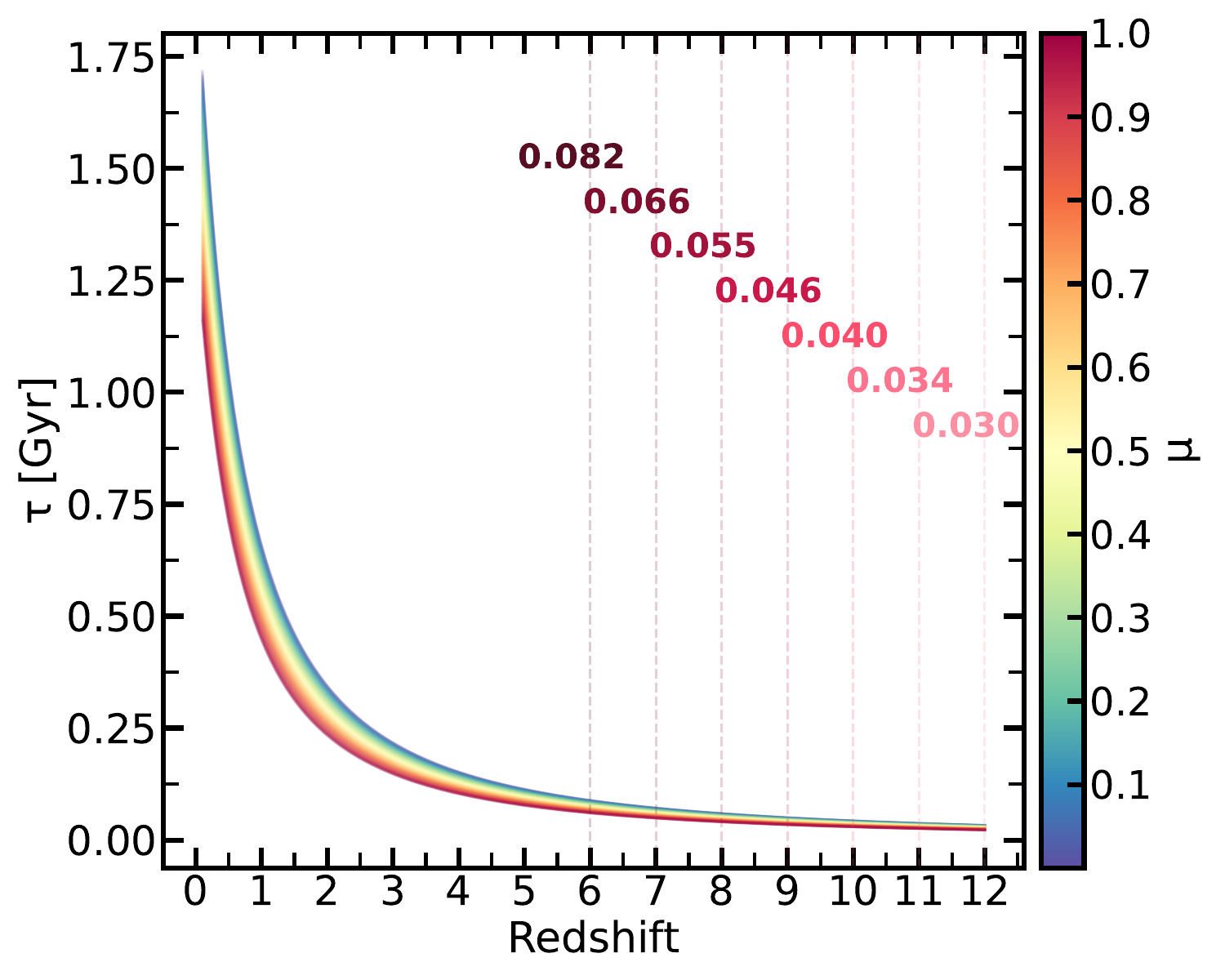}
    \caption{Evolution of Merger Timescales as a Function of Redshift as well as mass ratio $\mu$. The depicted data are derived using parameters reported by \protect\cite{2022ApJ...940..168C}, based on analyses conducted with the {\tt IllustrisTNG 300-1} cosmological simulation. The numbers on the vertical lines correspond to the timescales (y-value) at corresponding redshifts, expressed in units of Gyr.}

    \label{fig: Merger Timescales}
\end{figure}

The merger timescale, $\tau_{\mathrm{m}}(z)$, refers to the duration over which a merging system remains in a specific observed physical state. Various methods have been developed to derive merger timescales, ranging from empirical measurements \citep{2009MNRAS.399L..16C} to simulations \citep{2006ApJ...638..686C, kitzbichler2008calibration, 2008ApJ...672..177L, 2014MNRAS.445..175G,2014MNRAS.444.1518V, 2025arXiv250402930X}. 
Typical values have been found to be around $\tau_{\mathrm{m}}(z) = 0.3 - 0.7$ Gyr for pairs at $z < 3$. Using {\tt IllustrisTNG} simulation \citep{genel2014introducing, vogelsberger2014introducing}, \cite{2019ApJ...876..110D, 2022ApJ...940..168C} found this timescale evolve with redshifts and mass ratio $\mu$. They determined that merger timescales scale as follows:
\begin{equation}
    \tau_{\mathrm{m}}(z) = a \times (1+z)^{b} \, \text{[Gyr]},
\end{equation}
where 
\begin{equation}
    \begin{cases}
    a = -0.65 \pm 0.08 \times \mu + 2.06 \pm 0.01\\[10pt]
    b = -1.60 \pm 0.01.
    \end{cases}
\end{equation}
In \autoref{fig: Merger Timescales}, we plot the evolution of merger timescales with redshifts and $\mu$.

\begin{figure*}
    \centering
    \includegraphics[width=\linewidth]{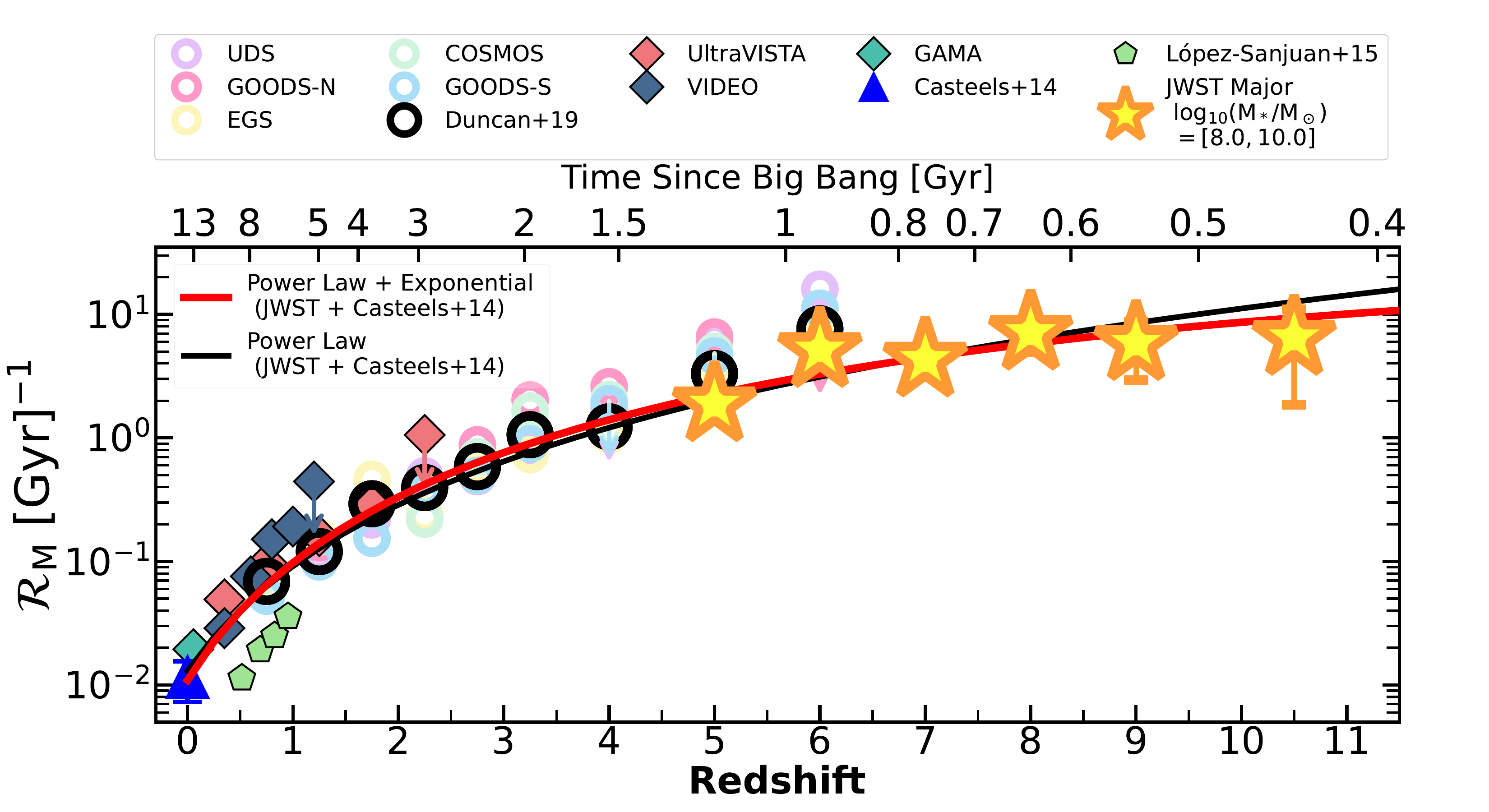}
    \caption{Evolution of the  galaxy merger rate with redshift. High-redshift major merger rates from our JWST results are represented by yellow-orange stars. The mass range used for our JWST data is $\log_{10}(\mathrm{M}_*/\mathrm{M}_\odot) = 8.0 - 10.0$. Additionally, we present lower redshifts results from four studies: green pentagons denote data from ALHAMBRA as reported by \protect\cite{2015A&A...576A..53L} for $M_B < -20.0$; rings represent CANDELS data by \protect\cite{2019ApJ...876..110D}, with colored rings indicating each sub-field and black rings representing the combined results for masses $\log_{10}(\mathrm{M}_*/\mathrm{M}_\odot) > 10.3$; diamonds symbolize data from the REFINE survey by \protect\cite{2022ApJ...940..168C} for masses $\log_{10}(\mathrm{M}_*/\mathrm{M}_\odot) > 11.0$; and the blue triangle denotes GAMA data computed by \protect\cite{2014MNRAS.445.1157C}. All literature except \protect\cite{2022ApJ...940..168C} present major merger rate; \protect\cite{2022ApJ...940..168C} presents total merger rate. A power law and power-law + exponential models are fitted to the combined data, with parameters shown in \autoref{tab: pair fractions fitting parameters}. We observe a saturation in merger rates at $z > 6.0$. Beyond this point, the merger rate becomes stable with a mean value of $5.78 \pm 0.98$ Gyr$^{-1}$.}
    \label{fig: JWST Merger Rates}
\end{figure*}

We present major-merger rates in \autoref{fig: JWST Merger Rates} using this measure of the timescales. Merger rates derived in this work are represented by yellow-orange stars, with values for each redshift bin shown in \autoref{tab: pf, mr, mr, smar values}. We include lower redshift results from \cite{2019ApJ...876..110D, 2022ApJ...940..168C} in the plot and compute the merger rates from the pair fractions of \cite{2014MNRAS.445.1157C} and \cite{2015A&A...576A..53L} using our method. Combined with these previously published values from the literature, we observe that \(\mathcal{R}_\text{M}\) increases between $z = 0.0 - 6.0$ and stabilizes after $z = 6.0$, with a mean number of merger events per galaxy of $5.78 \pm 0.98$ Gyr$^{-1}$. From $z = 8 - 11.5$, although the pair fraction decreases, the merger timescale also decreases by approximately the same ratio. According to \autoref{eq: merger rate}, this results in a stable \(\mathcal{R}_\text{M}\) within this redshift range.

Previous studies have demonstrated a continuous power-law increase in \(\mathcal{R}_\text{M}\) from \(z = 0.35\) to \(z = 6.0\). In this study, we model the \(\mathcal{R}_\text{M}\) evolution using both a power-law and a power-law + exponential (as presented in \autoref{eq: power law} and \autoref{eq: power exponential}). The fitting is performed using only our JWST data, supplemented by a zero-point data from \citet{2014MNRAS.445.1157C}, represented as a blue triangle. The fitted lines are shown as black and red lines in \autoref{fig: JWST Merger Rates}, and the fitted parameters are detailed in \autoref{tab: pair fractions fitting parameters}. The discovery of saturation in merger rates at higher redshift suggests that the pattern is more accurately described by a power-law + exponential rather than a power law. The rapid merger rates observed during the Universe's first 1 Gyr indicate a rapid transformation of galaxy properties, which we will explore further in the remainder of this paper.


\subsection{Merger-Driven Stellar Mass Growth}
\label{sec: Merger-Driven Stellar Mass Growth}
In this section we elucidate the impact of merger activities on the stellar mass of galaxies, focusing specifically on the mass increase attributable to major mergers. This is important as it for the first time allows us to determine the role of mergers in galaxy formation - that is, the establishment of stellar mass within galaxies at $z > 6$.  We quantify this by the mass accretion rate (MAR), total mass accretion rate density (tMARD), and specific mass accretion rate (sMAR). These three quantities are examined and analyzed in this section.

\subsubsection{Mass Accretion Rates}
The mass accretion rate,  $\rho_{1/4}(z)$, is the average amount of mass added through mergers per unit time per galaxy. The subscript $1/4$ signifies major mergers characterized by a mass ratio $\mu > 1/4$. Recall that the primary galaxy is the one with the higher mass in a pair, while the secondary galaxy is the one with the lower mass. For each primary galaxy within a given redshift bin, the additional stellar masses added through mergers is the average stellar mass of secondary galaxies, symbolized as $\mathcal{M}_{*,2}(z)$. Consequently, the mass accretion rate is expressed as:
\begin{equation}
    \label{eq: mass accretion rate}
    \rho_{1/4}(z) = \mathcal{R}_\mathrm{M}(z) \times \mathcal{M}_{*,2} (z),
\end{equation}
where $\mathcal{M}_{*,2}(z)$ is calculated using the following integral:
\begin{equation}
    \mathcal{M}_{*,2}(z) = \frac{\displaystyle \int_{\mu \mathcal{M}_{*,1}}^{\mathcal{M}_{*,1}} \phi(z, \mathrm{M}_{*}) \,  \mathrm{M}_{*}  \, d\mathrm{M}_{*}}{\displaystyle \int_{\mu \mathcal{M}_{*,1}}^{\mathcal{M}_{*,1}} \phi(z, \mathrm{M}_{*}) \,  d\mathrm{M}_{*}}.
\end{equation}
Here, $\phi(z, \mathrm{M}_{*})$ is the Galaxy Stellar Mass Function (GSMF) at each redshift bin. We use stellar mass functions from \citet{2014MNRAS.444.2960D} for $z = 4.5 - 6.5$, and \citet{harvey2024epochs} for $z = 6.5 - 11.5$.   
 The value $\mathcal{M}_{*,1}$ represents the average stellar mass of the primary galaxy in each redshift bin, computed as:
\begin{equation}
    \mathcal{M}_{*,1}(z) = \frac{\displaystyle \int_{\mathrm{M}_{*,1}^{\mathrm{min}}}^{\mathrm{M}_{*,1}^{\mathrm{max}}} \phi(z, \mathrm{M}_{*}) \,  \mathrm{M}_{*}  \, d\mathrm{M}_{*}}{\displaystyle \int_{\mathrm{M}_{*,1}^{\mathrm{min}}}^{\mathrm{M}_{*,1}^{\mathrm{max}}} \phi(z, \mathrm{M}_{*}) \,  d\mathrm{M}_{*}},
\end{equation}
where ${\mathrm{M}_{*,1}^{\mathrm{min}}}$ and ${\mathrm{M}_{*,1}^{\mathrm{max}}}$ represent the stellar mass range we used, being $10^8$ and $10^{10}$ $\mathrm{M}_\odot$, respectively. 

By integrating the mass accretion rate over a specific time interval, we can retrieve the average amount of mass added to the primary galaxy's stellar mass during that time interval, denoted as
\begin{equation}
    \delta \displaystyle \mathcal{M}(t) = \int_{t_\mathrm{min}}^{t_\mathrm{max}} \rho_{1/4}(t)\, dt.
\end{equation}
We use the point at $z = 10.5$ as the $t_{\mathrm{min}}$ value, corresponding to $441$ Myr after the Big Bang. The redshift of each subsequent data point is used as $t_\mathrm{max}$.

\begin{figure}
    \centering

    \begin{subfigure}{\linewidth}
        \centering
        \includegraphics[width=\linewidth, height=0.23\paperheight, keepaspectratio]{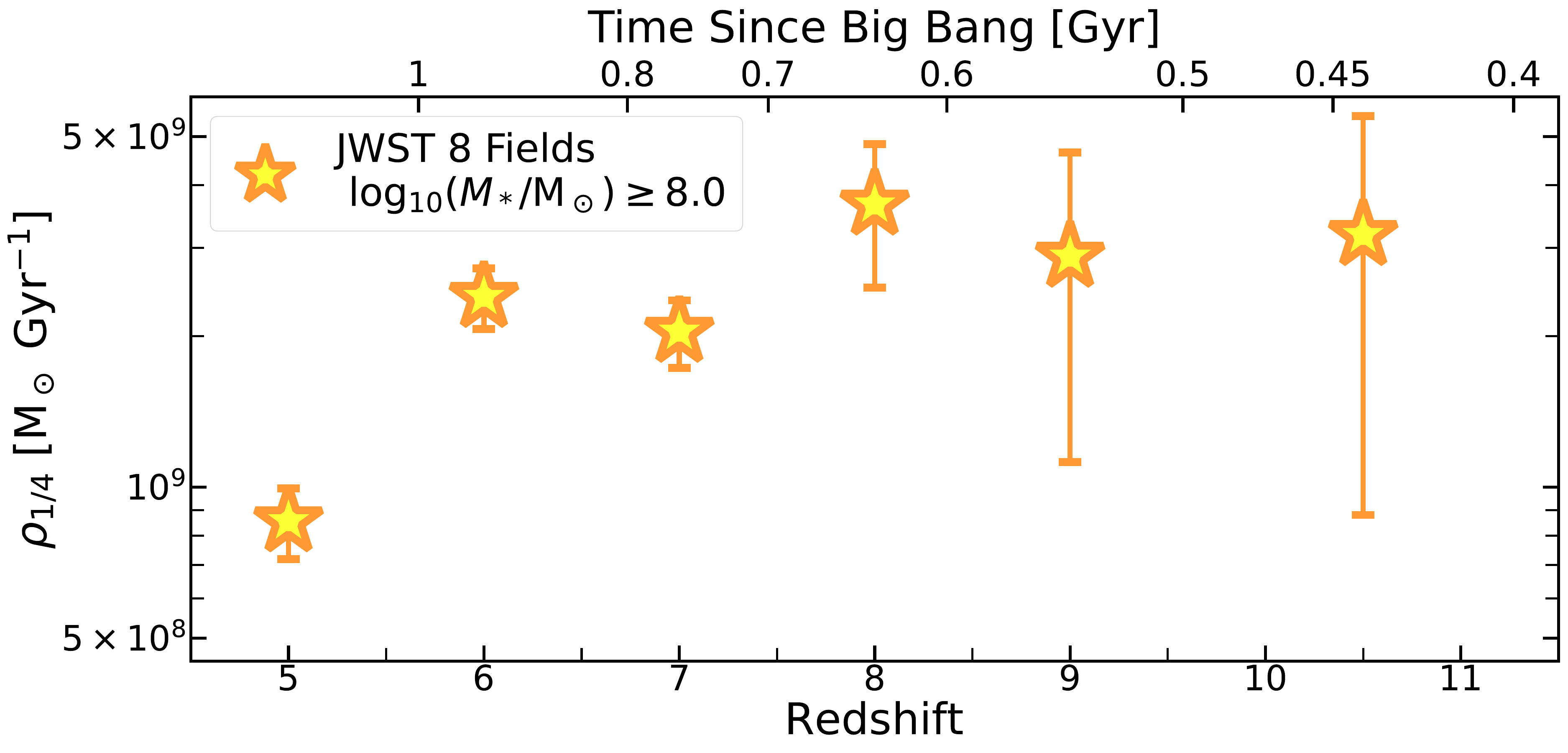} 
        \caption{Average Mass Accretion Rate per Galaxy vs. Redshift.   The value on the y-axis is the amount of mass accretion onto an `average' primary galaxy within each redshift bin due to major mergers.  We find that this value increases up to \(z = 6\), and then becomes stable. Values at each redshift are listedin \autoref{tab: pf, mr, mr, smar values}.}
        \label{fig: Mass Accretion Rate}
    \end{subfigure}

    \begin{subfigure}{\linewidth}
        \centering
        \includegraphics[width=\linewidth, height=0.23\paperheight, keepaspectratio] {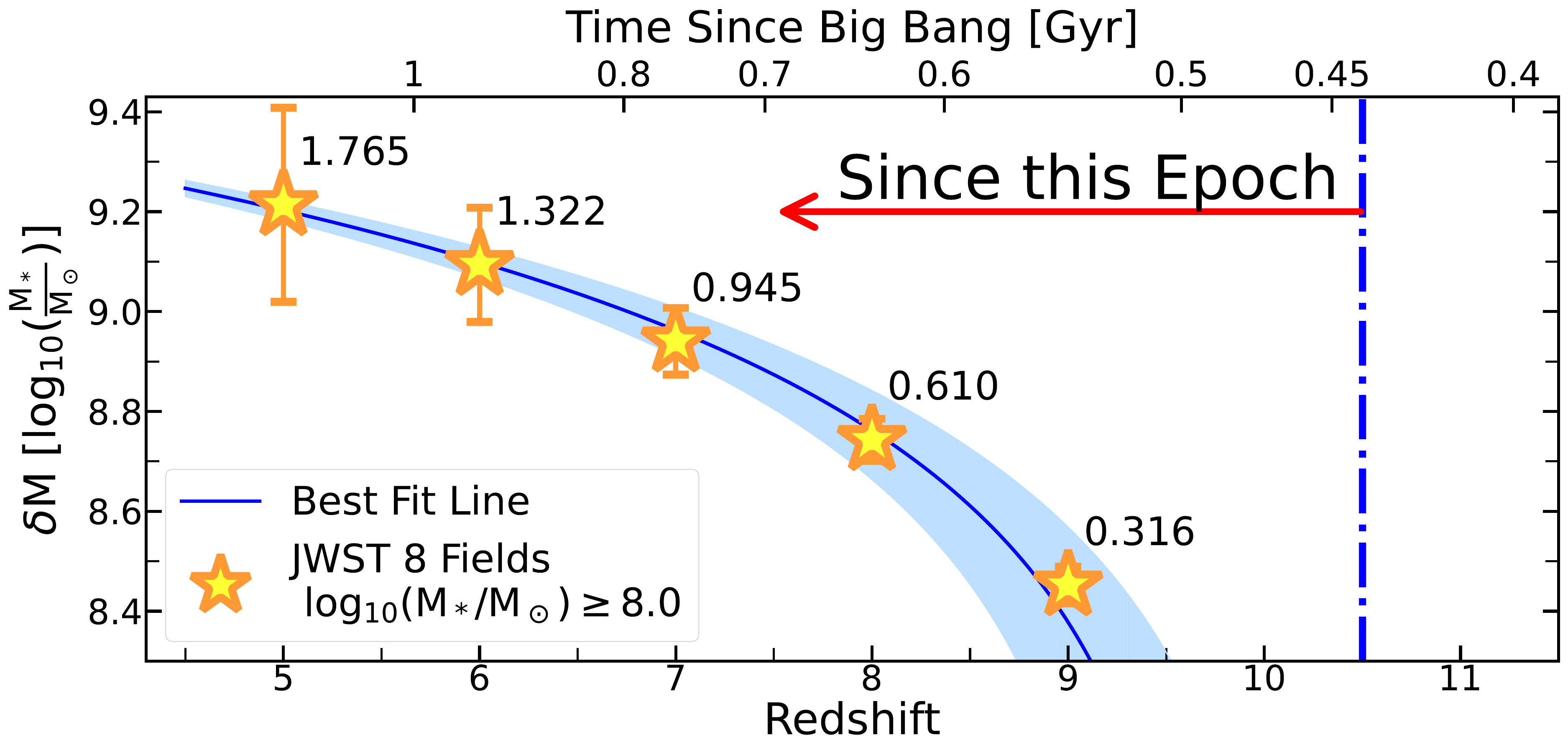}
        \caption{The integrated average stellar mass accreted per galaxy, attributed exclusively to major mergers.  This figure is the integral of the top panel, such that the average total amount of accumulated stellar mass from mergers starting at $z = 10.5$ is shown in the bottom panel. The text above each data point indicates the ratio of the total mass accreted due to mergers compared to the average stellar mass of the primary galaxies at the start of the integration. The fit shown with the blue line, applied to the unlogged mass scale,  effectively captures this trend.}
        \label{fig: Mass Accretion Rate Delta M}
    \end{subfigure}

    \caption{Two figures demonstrating: (a) Stellar mass accretion rates from major mergers, (b) Integrated stellar mass accreted from major mergers.}
    \label{fig:combined all three}
\end{figure}

Evolution of $\rho_{1/4}(z)$ and $\delta \mathcal{M}(t)$ with redshifts are shown in \autoref{fig: Mass Accretion Rate} and \autoref{fig: Mass Accretion Rate Delta M}, respectively. Values of $\rho_{1/4}(z)$ at each redshift bin are presented in \autoref{tab: pf, mr, mr, smar values}. From \autoref{fig: Mass Accretion Rate}, we observe that the galaxy mass accretion rates are generally higher at higher redshifts, but becomes stable after $z > 6.0$. From $z = 4.5$ to $z = 11.5$, the average mass of secondary galaxies ($\mathcal{M}_{*,2}$) decreases, but only by 2.5\%. Thus, we conclude that the higher mass accretion rates at $z > 6.0$ are primarily due to the observed higher merger rates.

In \autoref{fig: Mass Accretion Rate Delta M}, we present the increase in stellar mass resulting from major merger events. The value of each point indicates the integration of $\rho_{1/4}(z)$ from $z = 10.5$ to the redshift at that point. Consequently, this value is higher at lower redshifts, owing to a longer time interval allowing more mass to accumulate. On top of each point is the value of $\delta \mathcal{M}(t)$ divided by the average primary galaxy stellar mass, $\mathcal{M}_{*,1}$, in the corresponding redshift bin. We then fit the unlogged \(\delta \mathcal{M}(z)\) versus redshift using a linear line, depicted as the blue line in the plot. This linear relationship describes the trend very well, with uncertainties in the gradient and y-intercept being 4.24\% and 3.12\%, respectively.

Examining this figure in detail, we find that, on average, galaxy stellar masses have nearly tripled ($2.77 \pm 0.99$) due to merger events from \(z = 10.5\) to \(z = 5.0\), over a period of $\sim 730$ Myr. This trend is similar to findings at redshifts (\(z < 3\)), which suggest a doubling of stellar mass from mergers, according to \cite{2022ApJ...940..168C}. Therefore, we conclude merger events in the early Universe play a significant role in growing galaxy stellar masses.  One major question however is how important this process of stellar mass assembly is compared to the process of star formation, which we will investigate in next two sections.

\subsubsection{Total Mass Accretion Rate Density}
\label{sec: mass accretion rate density}
In this section we investigate the contribution of mergers to galaxy stellar mass comparing with galaxy intrinsic star formation rates. To achieve this, we compare the Total Mass Accretion Rate Density (tMARD) with Total Star formation Rate Density (tSFRD), and Specific Mass Accretion Rate (sMAR) with Specific Star Formation Rate (sSFR). The later will be discussed in the next section.

We compute the tMARD(z) by dividing the mass accretion rate per galaxy, \(\rho_{1/4}(z)\), by the comoving volume, \(V_{\text{comoving}}(z)\), and then multiplying by the total number of galaxies in the corresponding redshift bin, before applying the area normalization to account for the entire sky. The mathematical way to express this is
\begin{equation}
\mathrm{tMARD} (z) = \frac{\rho_{1/4}(z) \times N(z)_{\mathrm{galaxies}}}{V_{\mathrm{comoving}}(z)} * \frac{A_\mathrm{Total}}{A_\mathrm{Survey}},
\end{equation}
with unit of $\mathrm{M}_\odot \, \mathrm{yr}^{-1} \, \mathrm{Mpc}^{-3}$.   Note that this is a very narrow mass range for which the mass accretion is being measured, which we later compare with the star formation at the same mass range.  We are not attempting here to obtain a 'total' tMARD value over all stellar masses, which is often done to the star formation rate density measured as a function of redshift \citep[e.g.,][]{2023ApJS..265....5H, 2024ApJ...965..169A}.

The tSFRD for each redshift bin is calculated by summing the SFRD of all galaxies within that bin, and then multiplying by the same area normalization factor. Note that we multiply the SFR by \((1 - R)\), where \(R\) is the return fraction that we calculated as \(R = 0.423\) from \cite{2014ARA&A..52..415M}, to account for the fraction of stellar mass that is returned to the interstellar medium through stellar processes such as supernovae and stellar winds. This adjustment ensures that we reflect only the mass that remains (the observed mass). Thus, the tSFRD at each redshift bin in this work can be written as 
\begin{equation}
    \mathrm{tSFRD (z)} = \frac{\sum_{i}^{N} \mathrm{SFR}_i \times (1 - R)}{V_{\mathrm{comoving}}(z)} \times \frac{A_{\mathrm{Total}}}{A_{\mathrm{Survey}}}.
\end{equation}
\noindent We derive SFRs from three sources: \texttt{Bagpipes}, using both 10 Myr and 100 Myr averaged timescales based on a log-normal star formation history, and the SFRD inferred from the UV luminosity. As we directly utilize all galaxies from these eight JWST fields without any completeness corrections for galaxies fainter than the observational limit, the tSFRD reported here is generally significantly lower than tSFRD derived from UV luminosity functions \citep[e.g.,][]{2023ApJS..265....5H, 2024ApJ...965..169A}.

We present the tMARD and tSFRD at \(z = 4.5 - 11.5\) in \autoref{fig: Mass Accretion Rate Density}. tMARD is represented by yellow-orange stars, while tSFRD is depicted with lines: red for \texttt{Bagpipes} 10 Myr averaged timescale, green for 100 Myr averaged timescale, and purple for the UV luminosity based measurements. By performing this cross-comparison and computing the ratio of tMARD to tSFRD, we find that the contribution of major mergers to galaxy stellar mass assembly during this epoch is, on average, $45 \pm 16\%$, $69 \pm 23\%$, and $108 \pm 44\%$, relative to in-situ star formation rates derived from \texttt{Bagpipes} (10 Myr average), UV-based estimates, and \texttt{Bagpipes} (100 Myr average), respectively. Thus, a typical value is that major mergers are adding $74\, ( \pm 28 )\%$ to the stellar masses of galaxies through this redshift range. 

\subsubsection{Specific Mass Accretion rate}
The specific mass accretion rate (sMAR) is defined as the mass accretion rate divided by the average stellar mass of primary galaxies:
\begin{equation}
\mathrm{sMAR} (z) = \frac{\dot{\mathrm{M}}}{\mathrm{M}_*} = \frac{\rho_{1/4}(z)}{\mathcal{M}_{*,1}(z)} = \frac{\mathcal{R}_\mathrm{M}(z) \times \mathcal{M}_{*,2}(z)}{\mathcal{M}_{*,1}(z)}.
\end{equation}
\noindent Similar to the previous section, we compare the specific mass accretion rate (sMAR) from mergers with the specific star formation rate (sSFR) from intrinsic star formation in this section, to evaluate their respective contributions to the total galaxy stellar mass. The sSFR is computed from three sources: \texttt{Bagpipes} using 10 Myr and 100 Myr averaged, and UV luminosity. Note that we also multiply the SFR by \((1 - R)\) to reflect the observed SFR, as we did in the previous section.

\begin{figure*}
    \begin{subfigure}{\linewidth}
        \centering
        \includegraphics[width=\linewidth, height=0.23\paperheight, keepaspectratio]{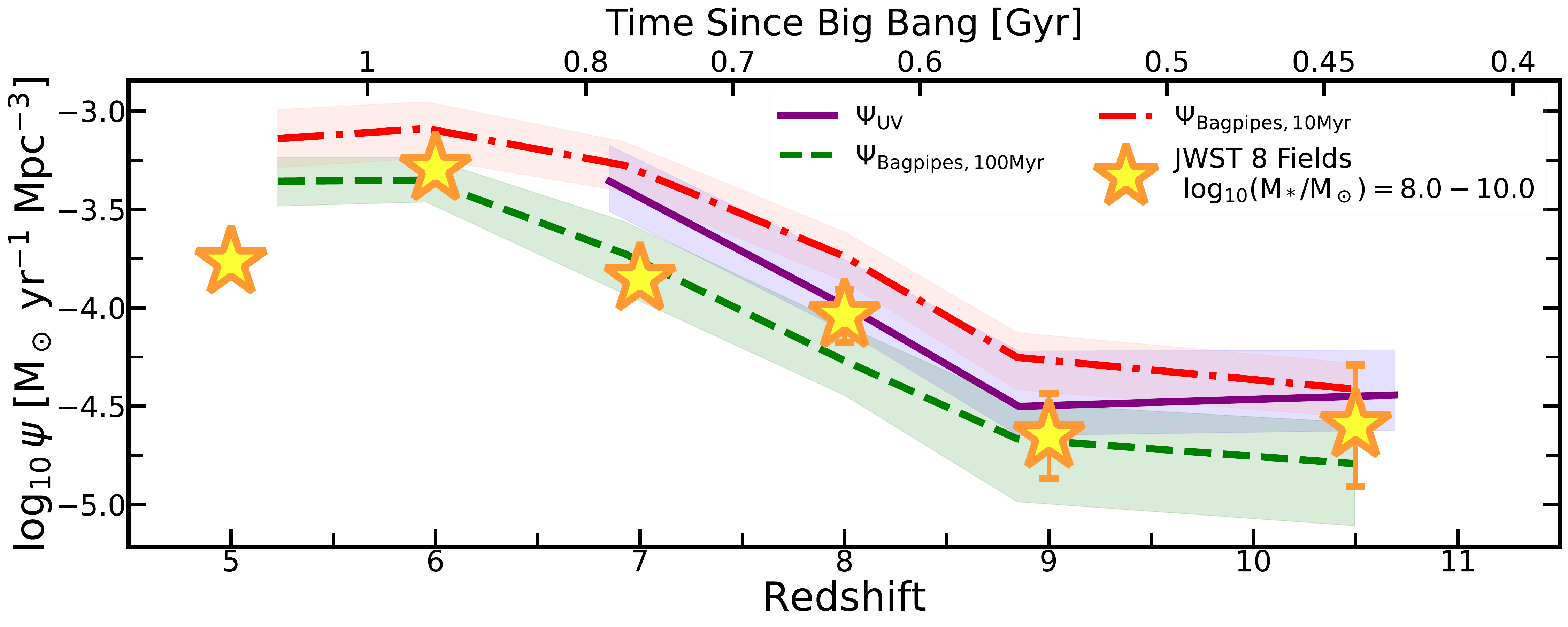}
        \caption{In this figure, we present the Total Mass Accretion Rate Density (tMARD) from major mergers, in units of $\mathrm{M}_\odot \, \mathrm{yr}^{-1} \, \mathrm{Mpc}^{-3}$, versus Redshift alongside the Total Star Formation Rate Density (tSFRD) versus Redshift. The values of tMARD are represented by yellow-orange stars, while we present three sources of tSFRD: red for \texttt{Bagpipes} 10 Myr averaged timescales (log-normal SFH), green for 100 Myr averaged timescales, and purple for the UV luminosity measurement. From this comparison, we find that the contribution of mergers to galaxy stellar mass assembly, is on average, is $45\, ( \pm 16 )\%$, $69\, ( \pm 23 )\%$, and $108\, ( \pm 44 )\%$, equivalent to intrinsic star formation computed from Bagpipes 10 Myr averaged, UV, and Bagpipes 100 Myr averaged.}
        \label{fig: Mass Accretion Rate Density}
    \end{subfigure}

    \begin{subfigure}{\linewidth}
        \centering
        \includegraphics[width=\linewidth]{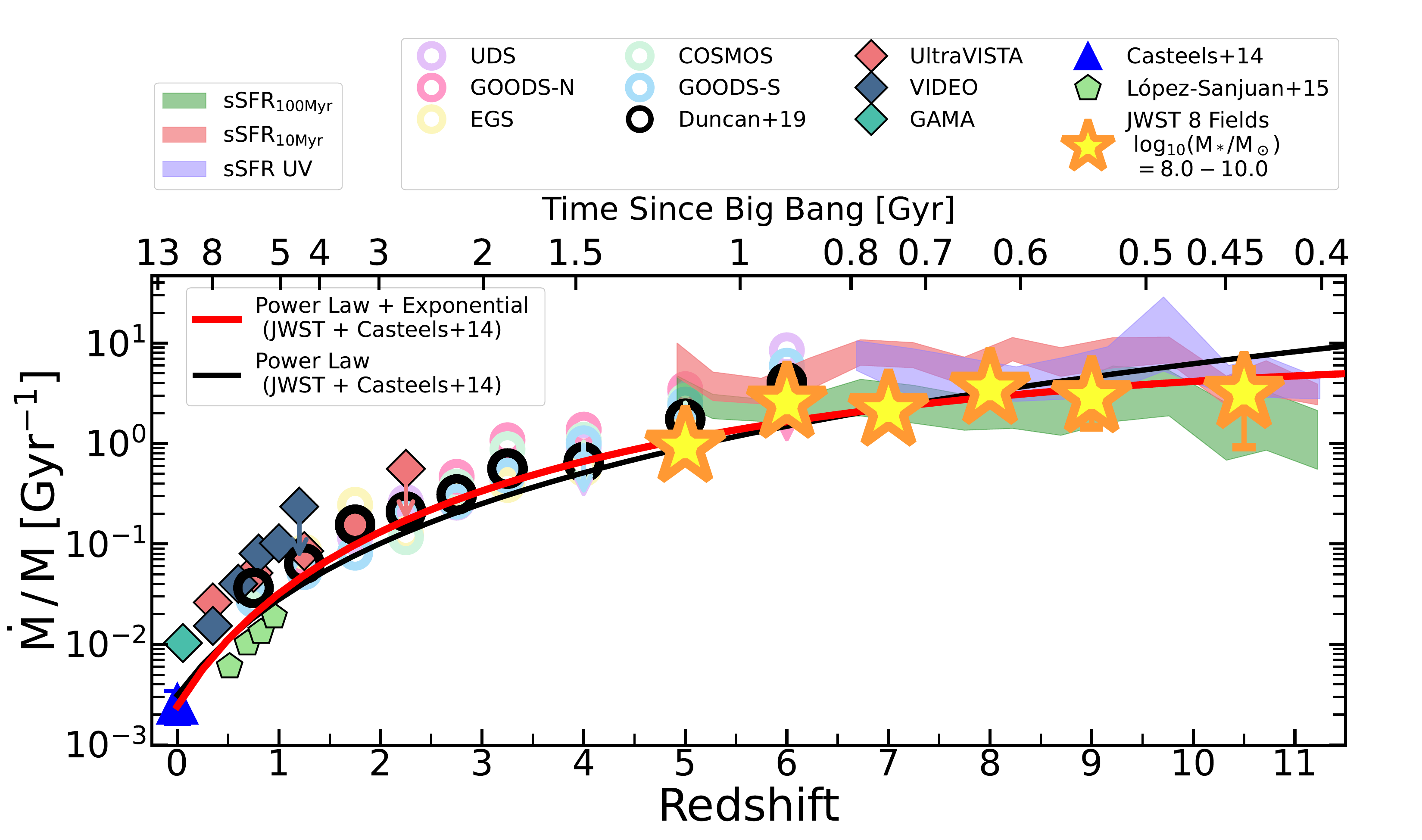}
        \caption{Evolution of specific mass accretion rates (sMAR) from mergers and specific star formation rates (sSFR) with redshift. We present three sources of sSFR: \texttt{Bagpipes} with a 100 Myr average and 10 Myr average timescale, and UV luminosity, which are depicted by green, red, and purple shaded regions. These sSFR values are computed from the entire EPOCHS V1 high-redshift sample. High-redshift specific mass accretion rates derived in this work are represented by yellow-orange stars. Additionally, we use our methodology to derive the values of the sMAR from pair fractions from four lower redshifts results: green pentagons denote data from ALHAMBRA as reported by \protect\cite{2015A&A...576A..53L} for $M_B < -20.0$; rings represent CANDELS data by \protect\cite{2019ApJ...876..110D}, with colored rings indicating each sub-field and black rings representing the combined results for masses $\log_{10}(\mathrm{M}_*/\mathrm{M}_\odot) > 10.3$; diamonds symbolize data from the REFINE survey by \protect\cite{2022ApJ...940..168C} for masses $\log_{10}(\mathrm{M}_*/\mathrm{M}_\odot) > 11.0$; and the blue triangle denotes GAMA data computed by \protect\cite{2014MNRAS.445.1157C}. All literature except \protect\cite{2022ApJ...940..168C} present major merger result; \protect\cite{2022ApJ...940..168C} presents total merger result. From this comparison, we find the contribution of mergers to galaxy stellar mass is, on average, $45\, ( \pm 14 )\%$, $51\, ( \pm 16 )\%$, and $110\, ( \pm 34)\%$ equivalent to the intrinsic star formation rates computed from \texttt{Bagpipes} averaged over 100 Myr, UV luminosity, and \texttt{Bagpipes} averaged over 10 Myr, respectively.}
        \label{fig: Specific Mass Accretion Rate}
    \end{subfigure}
    \caption{These two figures compare the contributions of stellar mass assembly from major mergers with the amount of mass added due to intrinsic star formation: (a) Shows the Total Mass Accretion Rate Density (tMARD) and the Total Star Formation Rate Density (tSFRD) across redshifts, and (b) Shows the Specific Mass Accretion Rates (sMAR) and Specific Star Formation Rates (sSFR) across redshifts.}
\end{figure*}

We display these two quantities in \autoref{fig: Specific Mass Accretion Rate}. The shaded red and green regions represent the sSFR$_{\mathrm{Bagpipes}}$ averaged over 10 and 100 Myr respectively,  and the purple region gives the sSFR$_{\mathrm{UV}}$, estimated directly from the rest UV photometry. The sMAR values from this study, alongside those derived using our methodology from literature-based pair fractions, are represented using distinct markers. Values of our JWST sMAR at each redshift bin are shown in \autoref{tab: pf, mr, mr, smar values}. We fit the evolution of sMAR with a power-law and a power-law + exponential form, with the best-fitted parameters presented in \autoref{tab: pair fractions fitting parameters}. Discrepancies in sSFR between \texttt{Bagpipes} and the UV method arise from differences in star formation history assumptions. For the sSFR calculated with \texttt{Bagpipes}, we utilize a log-normal SFH, whereas for the UV-derived sSFR, we adopt the constant SFH model by \cite{madau2014cosmic}, incorporating the dust correction factor from \cite{1999ApJ...521...64M}.

From this comparison, we find results similar to we discussed in Section \ref{sec: mass accretion rate density}. The contribution of mergers to galaxy stellar mass is, on average, $45\, ( \pm 14 )\%$, $51\, ( \pm 16 )\%$, and $110\, ( \pm 34)\%$ of the intrinsic star formation rates computed from \texttt{Bagpipes} averaged over 100 Myr, UV luminosity, and \texttt{Bagpipes} averaged over 10 Myr, respectively. In a broader context, combined with the findings from Section \ref{sec: mass accretion rate density}, we conclude that the contribution of mergers to galaxy stellar mass is $71 \, (\pm 25)\%$ equivalent to the contribution from intrinsic star formation from gas. This implies that at least $42\% \pm 24\%$  of the total stellar mass of galaxies arises from merger events. The reason for saying "at least" is that we are searching for close pairs, and some of the galaxies in these pairs may have already merged before we observe them (post-mergers). Identifying post-merger galaxies requires a more complex analysis, such as using deep learning convolutional neural networks \citep{2020ApJ...895..115F}, which is beyond the scope of this paper. Thus, the intrinsic star formation rates from gas that we computed using SED and UV luminosity inevitably include contributions from mergers as well.

In \textbf{\textit{Paper II}} of \textbf{\textit{Galaxy Mergers in the Epoch of Reionization}} by \cite{2024arXiv241104944D}, we specifically investigate the extent of star formation enhancement in close pairs as a function of projected separation. We find that significant SFR enhancement occurs only at $r_p < 20$ kpc, with increases of $0.25 \pm 0.10$ dex and $0.26 \pm 0.11$ dex above the non-merger samples' median for redshift ranges $z = [4.5, 6.5]$ and $z = [6.5, 8.5]$, respectively. We also examine AGN activity triggered by major mergers and find an AGN excess of $1.34^{+0.23}_{-0.11}$ in the close-pair sample relative to the non-merger sample. Additionally, we find that nearly all AGNs have a close companion within 100 kpc.

\newpage
\section{Discussion}

Utilizing the deepest and widest observational NIRCam imaging programs to date from JWST, we have conducted a new close-pair major merger analysis spanning redshifts from $z > 4.5$ to $z = 11.5$.  This is a redshift range where the merger history is largely unknown observationally.  Based on this imaging data and using new methods we have computed pair fractions, merger rates, and stellar mass accretion rates at $z > 6$, highlighting the significant role of major mergers in galaxy formation. 

Notably, we observed a roughly constant rapid merger rate at the highest redshifts of $5.78 \pm 0.98 \, \text{Gyr}^{-1}$ mergers per galaxy, which is a rapid merger rate several orders of magnitude higher than the merger rate locally \citep[e.g.,][]{2014MNRAS.445.1157C, 2015A&A...576A..53L}.  We estimate that an average galaxy from $z = 10.5$ to $z = 5.0$ would become approximately $2.77 \pm 0.99$ times more massive by $z = 5.0$.  This is roughly similar to the amount of stellar mass added to these galaxies due to the star formation process. This also shows that mergers are a critical factor in galaxy formation early in the Universe.  

As a result, mergers are most certainly a significant factor in galaxy evolution at early times and far more important than even at modest redshifts at $3 < z < 6$ \citep[e.g.,][]{2019ApJ...876..110D}. In this paper, we undertake the first assessment of the contribution of mergers to stellar mass assembly across the redshift range \(z = 4.5 - 11.0\) (see Section \ref{sec: Merger-Driven Stellar Mass Growth}, \autoref{fig: Mass Accretion Rate}, \autoref{fig: Mass Accretion Rate Density}, and \autoref{fig: Specific Mass Accretion Rate}). We find that the contribution of mergers to stellar mass is $71 \pm 25\%$ relative to in-situ star formation, implying that at least $42 \pm 24\%$ of an average galaxy's total stellar mass is assembled through mergers over this redshift range. Additionally, our calculated pair fractions and merger rates are critical for predicting detectable gravitational wave events at $z > 6$, providing key forecasts for LISA (Laser Interferometer Space Antenna; \cite{lisa2020}), which is scheduled for launch in 2035.

Despite using nearly all available public JWST NIRCam Cycle-1 programs, our combined masked area of 189.36 arcmin$^2$ covers only a minuscule portion of the entire sky, and our results could be skewed by cosmic variance. However, considering our field locations in both the Northern and Southern hemispheres and their random selection, the impact of cosmic variance is likely negligible, as discussed as well in the luminosity function based on these data \citep[][]{2024ApJ...965..169A}. Our study's accuracy is primarily challenged by three factors: merger timescales, sample completeness, and photometric redshift uncertainties. Merger timescales are challenging to measure and have historically been assumed to be constant. In this study, we utilize merger timescale estimates from the \texttt{IllustrisTNG} simulation, which suggests a redshift evolution characterized by shorter timescales at higher redshifts, although these estimates are associated with significant uncertainties.  In fact, it is the time-scale of these mergers which is the most uncertain part of our analysis or any analysis of galaxy merger histories, as there are few observational ways to determine this time-scale \citep[c.f.][]{Conselice2009time}.

Furthermore, while we utilize the \texttt{JAGUAR} simulation to compute sample completeness, and have developed an innovative method to correct for incompleteness, achieving an entirely complete model is unfeasible. Moreover, the absence of spectroscopic observations for high-redshift photometric samples necessitates reliance on photometric redshifts to conduct our merger analysis, which introduces potential uncertainties similar to those noted in lower redshift studies \citep{2015A&A...576A..53L, 2017MNRAS.470.3507M, 2022ApJ...940..168C}. One issue is that the $\mathrm{Ly}\alpha$ damping wing, which could potentially affect photometric redshifts due to $\mathrm{Ly}\alpha$ photons passing through varying amounts of neutral IGM, which is also not taken into account. This omission may lead to under-predicting the number of close pairs, an effect that worsens at higher redshifts when there is more neutral IGM to traverse and at greater separations when the sightlines become more independent. This effect cannot be averaged out, but it does not play a role when the sightlines are exactly the same and is negligible when the sightlines are similar (as in close-pair systems). However, we have minimized these three factors (merger timescales, sample completeness, and photometric redshift uncertainties) using a statistical probabilistic pair counting methodology as in the previous literature. 

Compared with a recent morphological merger study at \(z = 4 - 9\) by \cite{2024arXiv240311428D}, which reported pair fractions of \(\mathrm{f}_\mathrm{p} = 0.11 \pm 0.04\) with no observed evolution with redshift, our findings show a higher average \(\mathrm{f}_\mathrm{p}\) of \(0.17 \pm 0.05\) in this redshift range. We find an increase in pair fractions up to \(z = 8\), reaching \(0.211 \pm 0.065\). The fitted power-law + exponential model presents a flat trend at \(z > 6\). Our $\mathrm{f}_\mathrm{p}$ values at $z = 4 - 9$ consistently exceeds those reported in the morphological study, a discrepancy likely because either merger timescales are shorter, as we find, or that not all close-pair systems ultimately merge, resulting in lower morphology-based pair fractions.  There needs to be a careful study of merging galaxies based on pairs and structure to resolve this issue, which sohuld inovlve the use of simulations. 

As discussed, we also compare our observational pair fractions with those predicted by simulations, specifically the SC-SAM and the Millennium cosmological dark matter simulation integrated with the \texttt{GALFORM} semi-analytical model. We find a significant disagreement between our observations and the simulations. Several reasons can explain this discrepancy.  First it must be noted that these simulations are often calibrated on observed galaxies properties, such as the mass and luminosity function, but the merger history is not considered in this.  Thus the merger history is an excellent test of the predictive power of simulations and models, and the disagreements should be considered in some detail.

Some of the possible explanations for the observed difference include: overly quenched satellite galaxies in the semi-analytical model, the different ways galaxies are selected between observational works and simulations \citep[e.g.,][]{2009MNRAS.396.2345B}, and the use of stellar mass for observations versus baryonic mass for simulations, as discussed in Section \ref{sec: sc-sam pair fractions}, as well as in past comparison papers such as \citet{2019ApJ...876..110D} and \citet{2009MNRAS.396.2345B}.  More simulation work focused on the merger history is needed to reconcile observations with theory, as well as how to calibrate stellar and total mass as is done in simulations \citep[][]{Conselice2018}.

In light of our findings, it is also important to consider alternative physical processes—beyond major mergers—that may contribute significantly to stellar mass assembly at high redshifts. A number of theoretical works have emphasized the role of secular processes and cold gas-driven compaction as key drivers of galaxy growth in the early Universe. In particular, studies such as \citet{2009Natur.457..451D, 2009ApJ...703..785D} and \citet{2010MNRAS.404.2151C} propose that cold gas accretion along cosmic filaments can fuel high star formation rates without the need for frequent major mergers. These inflows can induce violent disk instabilities and the formation of massive clumps, which migrate inward to build up central bulges \citep{2014MNRAS.443.3675M, 2017MNRAS.464..635M}. Moreover, simulations show that gas compaction events—short-lived phases of central starburst activity—can be triggered by smooth gas inflow rather than mergers \citep[e.g.,][]{2015MNRAS.450.2327Z}, highlighting the complexity of mass assembly pathways.

Our measurements, which show that major mergers contribute between $\sim40$--$70\%$ of the in-situ star formation rate density at $z > 4.5$, imply that a \textit{substantial fraction of stellar mass growth must still arise from non-merger-driven processes}. This is consistent with the picture in which secular evolution, smooth gas accretion, and compaction episodes all play complementary roles in building galaxies during the Epoch of Reionization. While our study focuses on the role of close-pair mergers, the remaining stellar mass growth not accounted for by these events provides indirect evidence for these alternative mechanisms. Future observational studies incorporating gas content, kinematics, and resolved structural evolution will be essential to disentangle the relative importance of these processes and build a more complete view of early galaxy formation.

Our results are revealing as galaxy formation occurs through a variety of mechanisms, with star formation from gas being largely the predominant process for building stellar mass. However, the literature indicates that the rate of galaxy mass growth often surpasses the contributions possible from the in situ gas within the galaxies themselves \citep[e.g.,][]{2012arXiv1212.5641C, 2016MNRAS.461.1112O, 2020MNRAS.496.1124P, Walter_2020}.   From our observations we are now able to trace this process back to the earliest times in which galaxies can be detected with JWST, and thus determine how galaxy assemble occurs rather than just when.

\section{Conclusions}
\label{sec: Conclusions}

In this study, we conduct an early analysis of galaxy merger evolution up to $z = 12$, utilizing the deepest and widest data from eight JWST Cycle-1 field programs, which collectively cover an unmasked area of 189.36 arcmin$^2$. We have accurately measured major-merger pair fractions, merger rates, and stellar mass accretion rates for galaxies with stellar masses between $8.0 <$ log$_{10}(\mathrm{M_*} / \mathrm{M}_\odot) < 10.0$. Our methodology, while grounded in established literature, introduces a novel completeness correction approach to precisely measure these quantities in the epoch of reionization. Our key findings are summarized as follows:

I. We find an increase in the measured pair fractions up to \(z \sim 8\), reaching a value of \(0.211 \pm 0.065\), followed by a statistically flat evolution to \(z = 11.5\). We conclude that the power-law + exponential model, \fpp $ = (0.030 \pm 0.016) * (1 + z)^{1.359 \pm 0.725} * e^{(-0.138 \pm 0.175)(1 + z)},$ is more appropriate than a traditional power law to describe the evolution of pair fractions with redshifts.  This suggests that we have measured the peak merger fraction at $z \sim 6$ which stays constant at higher redshifts.

II. Using merger time-scales based on simulations we find that merger rates (number of mergers per unit time) increases from the local Universe to \(z = 6.0\), at which time they  stabilize at a value of \(5.78 \pm 0.98\) Gyr\(^{-1}\). The rapid merger rates observed during the Universe's first 1 Gyr indicate a rapid transformation of galaxy stellar masses and thus also galaxy properties, showing that major mergers in the early Universe are of major importance.

III. Based on our measured merger rates we calculate that an average galaxy during the epoch $z = 10.5$ to $z = 5.0$  grows by a factor of $2.77 \pm 0.99$ by $z = 5.0$.  It is important however to keep in mind that this is the average stellar mass addition, and that some galaxies will have more than this while others will have less.

IV. We analyze the stellar mass accretion into galaxies due to mergers and compare it with the intrinsic star formation process within  galaxies themselves by examining the total mass accretion rate density (tMARD) from mergers versus the total star formation rate density (tSFRD) as well as the specific mass accretion rates (sMAR) from mergers with its star formation equivalent - the specific star formation rates (sSFR). We find that at $z = 4.5 - 11.5$, the stellar mass accreted from major mergers is $71\pm 25\%$ of the stellar mass created from star formation.  

Based on this approximately $42\pm 24\%$ of a galaxy's total stellar mass accrues from major merger events at this time, while the remainder originates from star formation.  This shows that galaxy mergers are at least as important as star formation for building the stellar masses of galaxies at early times. This does not include the star formation induced the mergers themselves, or minor mergers, which would only increase the importance of the merger process for adding stellar mass to galaxies.

V. We compare pair fractions from observational data with those derived from SC-SAM and Planck Millennium cosmological dark matter simulations. We find significant disagreements between our observations and simulations. Several reasons can explain this discrepancy, such as the overly quenched satellite galaxies in the semi-analytical model, the different ways galaxies are selected between observational works and simulations, and the use of stellar mass for observations versus baryonic mass for simulations. More simulation work focused on the merger history is needed to reconcile observations with theory.

Overall, using an integrated probabilistic pair counting method, we have made the first attempt to extend the investigation of merger evolution beyond the \(z = 6\) limit to \(z = 11.5\). This study currently represents the highest redshift analysis of mergers and may continue to be one of the most advanced. A worthwhile future endeavor, when a substantial number of photometric and spectroscopic observations are available, would be to compare these results with more extensive datasets to trace the minor merger history as well as determine the properties of mergers and how they correlate with SF and AGN activity. 

\section*{Data Availability}
The data underlying this article is made available by \cite{2023MNRAS.518.4755A, 2024ApJ...965..169A, 2023ApJ...952L...7A, austin2024epochs, harvey2024epochs, 2024arXiv240714973C}. The catalogues of the sample discussed herein may be acquired by contacting the corresponding author. 

\section*{Acknowledgement}
We acknowledge support from the ERC Advanced Investigator Grant EPOCHS (788113), as well as a studentship from STFC. LF acknowledges financial support from Coordenação de Aperfeiçoamento de Pessoal de Nível Superior - Brazil (CAPES) in the form of a PhD studentship. CCL acknowledges support from the Royal Society under grant RGF/EA/181016. CT acknowledges funding from the Science and Technology Facilities Council (STFC). We also extend our gratitude to Dr. Steven Willner for his constructive comments and suggestions on the manuscript. This work is based on observations made with the NASA/ESA \textit{Hubble Space Telescope} (HST) and NASA/ESA/CSA \textit{James Webb Space Telescope} (JWST) obtained from the \texttt{Mikulski Archive for Space Telescopes} (\texttt{MAST}) at the \textit{Space Telescope Science Institute} (STScI), which is operated by the Association of Universities for Research in Astronomy, Inc., under NASA contract NAS 5-03127 for JWST, and NAS 5–26555 for HST. Some of the data products presented herein were retrieved from the Dawn JWST Archive (DJA). DJA is an initiative of the Cosmic Dawn Center, which is funded by the Danish National Research Foundation under grant No. 140. This research made use of the following Python libraries: \textsc{Numpy} \citep{harris2020array}; 
\textsc{Scipy} \citep{2020SciPy-NMeth}; 
\textsc{Matplotlib} \citep{Hunter:2007}; 
\textsc{Astropy} \citep{2013A&A...558A..33A, 2018AJ....156..123A, 2022ApJ...935..167A}; 
\texttt{EAZY-PY} \citep{brammer2008eazy};
\textsc{LePhare} \citep{1999MNRAS.310..540A, 2006A&A...457..841I};
\textsc{Bagpipes} \citep{carnall2018inferring}; 
\textsc{mpi4py} \citep{dalcin2021mpi4py}; 
\textsc{Pickle} \citep{van1995python}.



\bibliographystyle{mnras}
\bibliography{main} 




\appendix

\section{The JWST Completeness Levels}
\label{sec: Completeness Level}

There are various incompletenesses that we have to consider within this merger work. Some of these issues are unique and differ from say incompletenesses that arise in detection of galaxies at various magnitudes \citep[e.g.,][]{harvey2024epochs} that are needed to correct galaxy mass and luminosity functions \citep[][]{2024ApJ...965..169A}. Namely, because we are incomplete in detections, we have to additionally consider how to deal with this in terms of galaxies which would be detected as pairs if the data was complete, but due to the depth of the imaging and the brightness of these galaxies remain undetected. 

\begin{figure*}
    \centering
    \includegraphics[width=\textwidth]{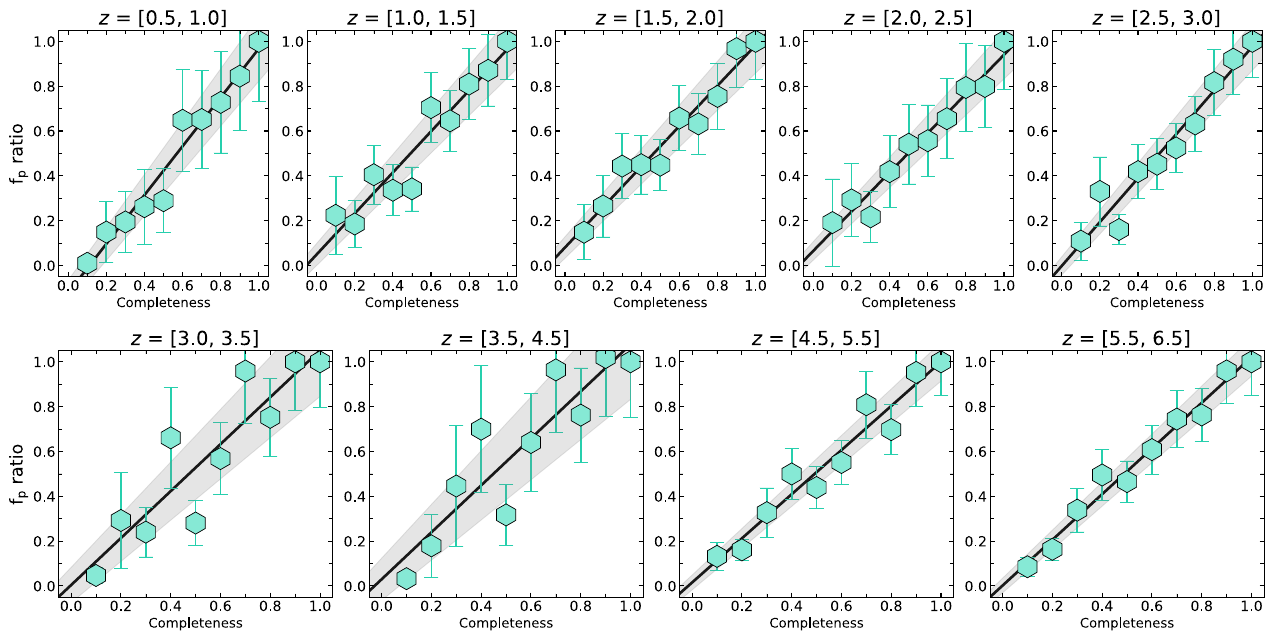}
    \caption{Evolution of the value $\text{f}_\text{p}^{\, \text{ratio}}$ (\autoref{eq: fp_ratio v.s completeness}) across various levels of incompleteness in different redshift bins. $\text{f}_\text{p}^{\, \text{ratio}}$ is the pair fraction at a given completeness level divide by the complete sample case (completeness = 1). To reduce random errors resulting from the arbitrary exclusion of samples, the procedure for introducing incompleteness into the pair fraction calculation is repeated five times, and the average $\text{f}_\text{p}^{\, \text{ratio}}$ values are shown for each redshift bin. An observable trend shows a lower $\text{f}_\text{p}^{\, \text{ratio}}$ corresponding to reduced completeness. Linear fits have been applied to each redshift bin, and the resulting gradients and y-intercepts of these best-fit lines are presented in \autoref{fig: Candels comp gradient y-intercept}.}
    \label{fig: Candels completeness}
\end{figure*}

\begin{figure*}
    \centering
    \includegraphics[width=\textwidth]{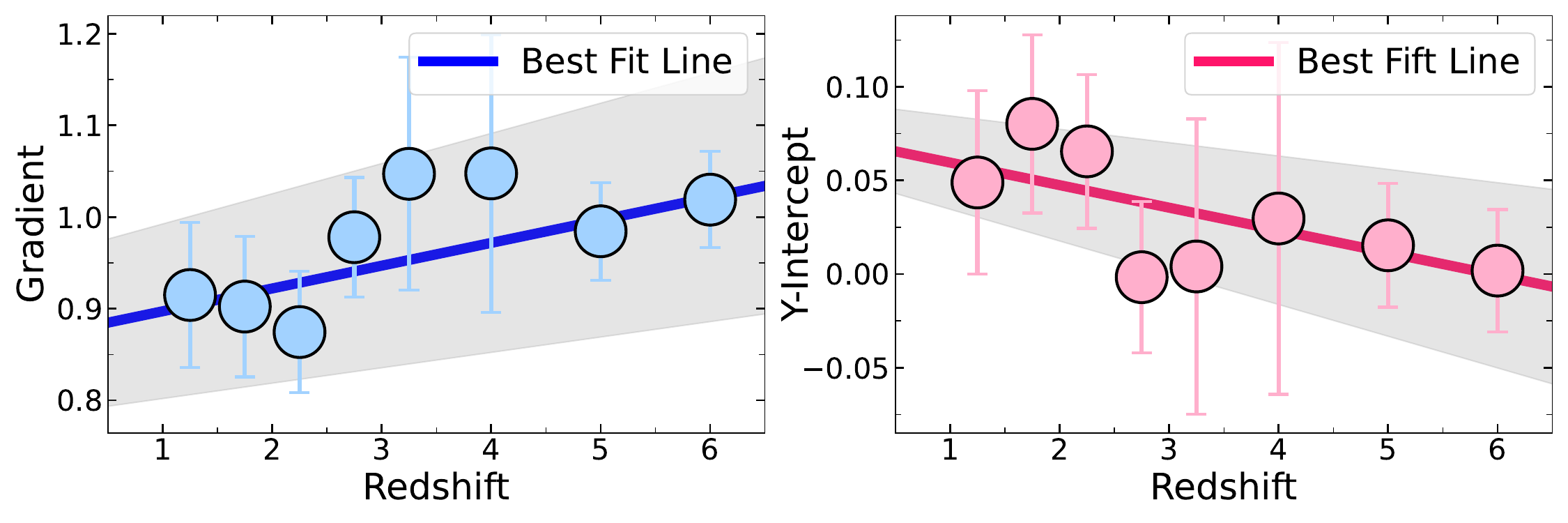}
    \caption{Evolution of the gradient and y-intercept for Best-Fit Lines Across Different Redshift Bins as shown in \autoref{fig: Candels completeness}. Straight lines are fitted to both the gradient and the y-intercept using the MCMC method, revealing an increasing trend in the gradient and a decreasing trend in the Y-intercept with redshift. The line of best fit for the gradient is $y = (0.028 \pm 0.012)x + (0.879 \pm 0.044)$, and for the Y-intercept, it is $y = (-0.013 \pm 0.006)x + (0.072 \pm 0.021)$.}
    \label{fig: Candels comp gradient y-intercept}
\end{figure*}

The completeness level for our JWST high redshift galaxy sample at each redshift bin and mass range is computed utilizing the \texttt{JAGUAR} simulation \citep{2018ApJS..236...33W}.  We consider this within each of our fields, which will have their own unique properties of incompleteness.  The \texttt{JAGUAR} simulation produces an output catalog of simulated galaxies, each characterized by attributes such as redshift, stellar mass, photometry, etc. For a given field, we consider the known average depths (i.e., the average uncertainty in flux) and apply a Gaussian scatter to the galaxy's photometry accordingly. The fluxes, once scattered, are processed through our comprehensive \texttt{EAZY} SED fitting and selection pipeline to determine whether a galaxy is selected and if its redshift is accurately estimated. We apply the same \texttt{Bagpipes} fitting to obtain masses and compare these masses to known catalog masses. Applying this process to the entire \texttt{JAGUAR} catalog allows us to determine the fraction of true high redshift galaxies in our final sample (completeness) and the number of low-redshift interlopers. This varies depending on the field, as both the filters and depths differ across fields. We can parameterize the completeness in terms of known variables (stellar mass, $\mathrm{M}_\mathrm{UV}$, apparent magnitude), by categorizing them into bins. A more detailed explanation is outlined in previous EPOCHS papers \cite{2024ApJ...965..169A, austin2024epochs, harvey2024epochs}. We present the completeness levels for all eight JWST fields used in this study in \autoref{fig: completeness all}. \autoref{fig: Candels completeness} and \autoref{fig: Candels comp gradient y-intercept} illustrate the impact of incompleteness on the derived pair fractions using the lower-redshift HST CANDELS catalog. Finally, the completeness correction factors across all fields are shown in \autoref{fig: completeness all correction factor}. This step is a crucial part of the completeness correction, which we have discussed in depth in Section \ref{sec: Sample Completeness Correction}.

\begin{figure*}
    \centering
    \includegraphics[width=\linewidth]{completeness_all.pdf}
    \caption{The level of completeness for our Eight JWST Fields computed using the \texttt{JAGUAR} Simulation. The black dashed bounding region indicates the redshift and stellar mass range used in this study. The grey areas indicate where no galaxies in those mass-redshift ranges within the simulation are located. As our analysis focuses on the mass range $\log_{10}(\mathrm{M}_*/\mathrm{M}_\odot) = 8.0 - 10.0$ and the redshift range $z = 4.5 - 11.5$, these gaps in the simulation data have a very minor impact on our results.}

    \label{fig: completeness all}
\end{figure*}

\begin{figure*}
    \centering
    \includegraphics[width=\linewidth]{completeness_all_correction_factor.pdf}
    \caption{Completeness correction factor $\mathcal{C} (z, \mathrm{M}_*)$ computed using our completeness correction method for our Eight JWST Fields. The black dashed bounding region indicates the redshift and stellar mass range used in this study. The grey areas indicate where no galaxies in those mass-redshift ranges within the simulation are located. As our analysis focuses on the mass range $\log_{10}(\mathrm{M}_*/\mathrm{M}_\odot) = 8.0 - 10.0$ and the redshift range $z = 4.5 - 11.5$, the gaps in the simulation data have a very minor impact on our results. }
    \label{fig: completeness all correction factor}
\end{figure*}


\section{Merger Catalog}
\label{sec: merger catalog}
\begin{figure*}
    \centering
    \includegraphics[width=\linewidth]{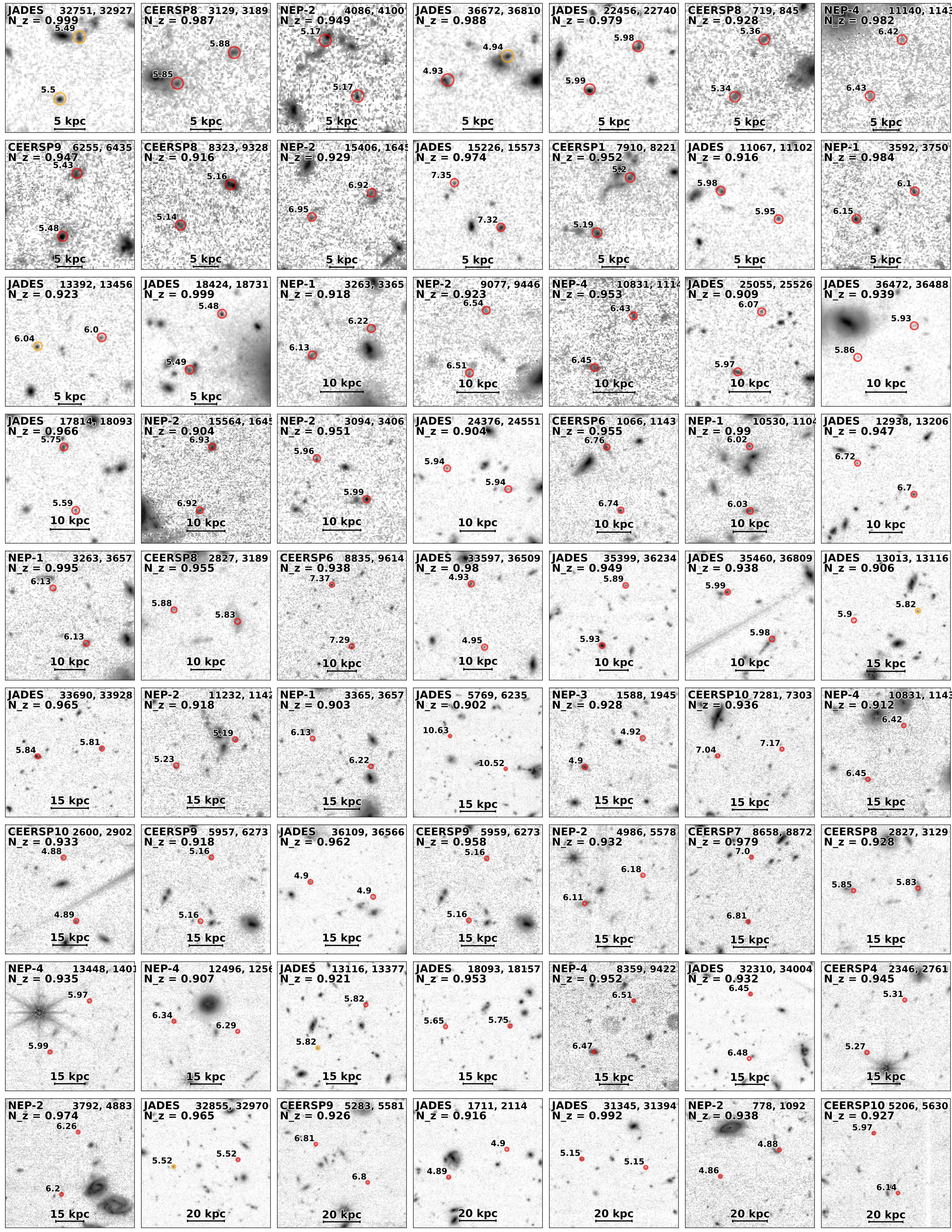}
    \caption{Cutout images for all close pair galaxies with $\mathcal{N}_z > 0.90$ and Projection Separations within $10 - 50$ kpc.}
    \label{fig: combined_first}
\end{figure*}

\begin{figure*}
    \centering
    \includegraphics[width=\linewidth]{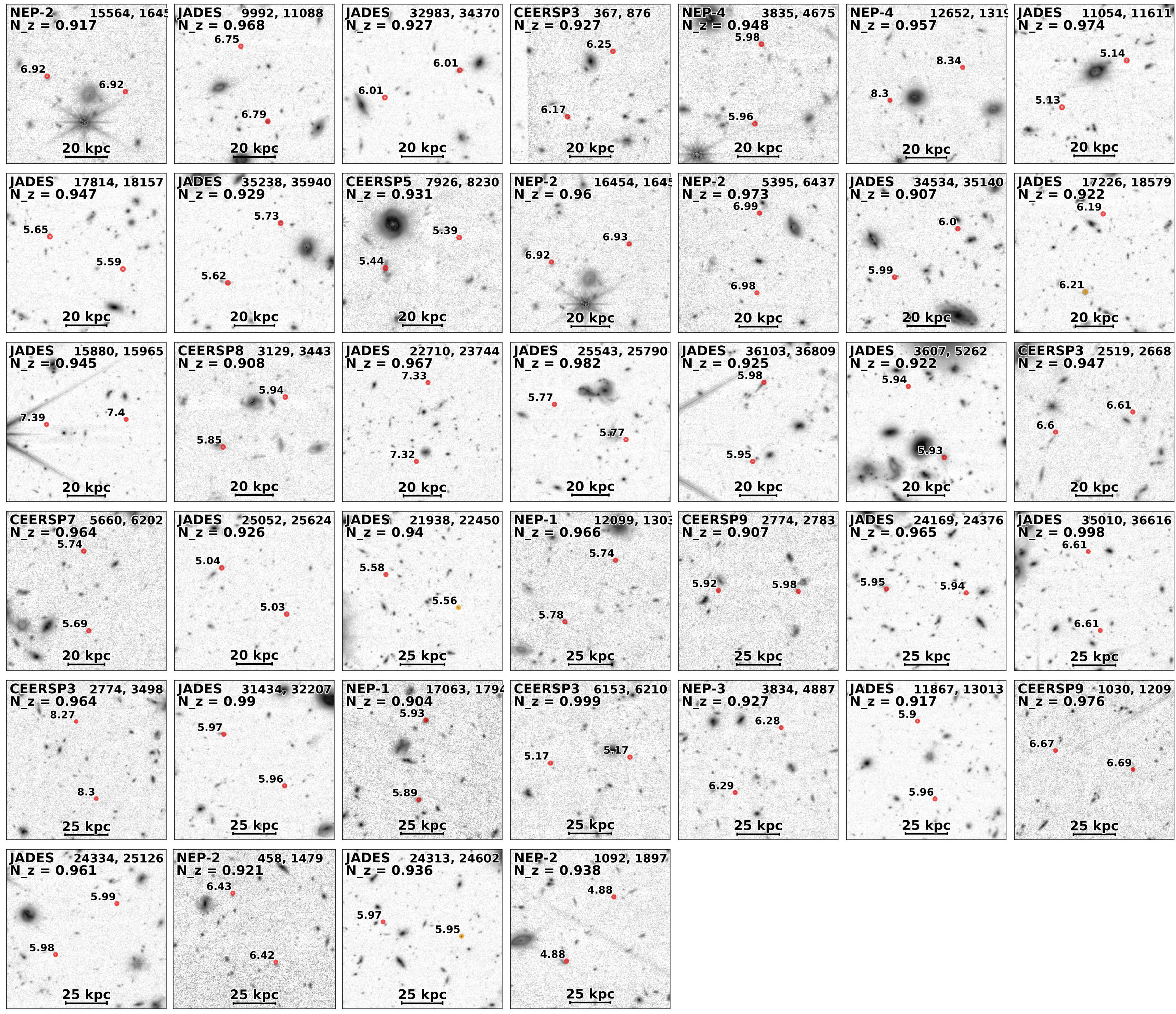}
   \caption{Stacked images of all close-pair galaxies with $\mathcal{N}_z > 0.90$ and projection separations ranging from $10$ to $50$ kpc. The images are generated by averaging data across multiple wavelength bands, with blue representing the ['F090W', 'F115W', 'F150W'] bands, green the ['F200W', 'F277W', 'F335M'], and red the ['F356W', 'F410M', 'F444W']. A 0.32 arcsecond aperture is drawn in red on each galaxy, with the redshift labeled at the top left of each aperture. Galaxies identified photometrically as AGN using the Nakajima template \protect\citep{Nakajima2022} are highlighted in orange.}

    \label{fig: comobined_second}
\end{figure*}

\begin{figure*}
    \centering
    \includegraphics[width=\linewidth]{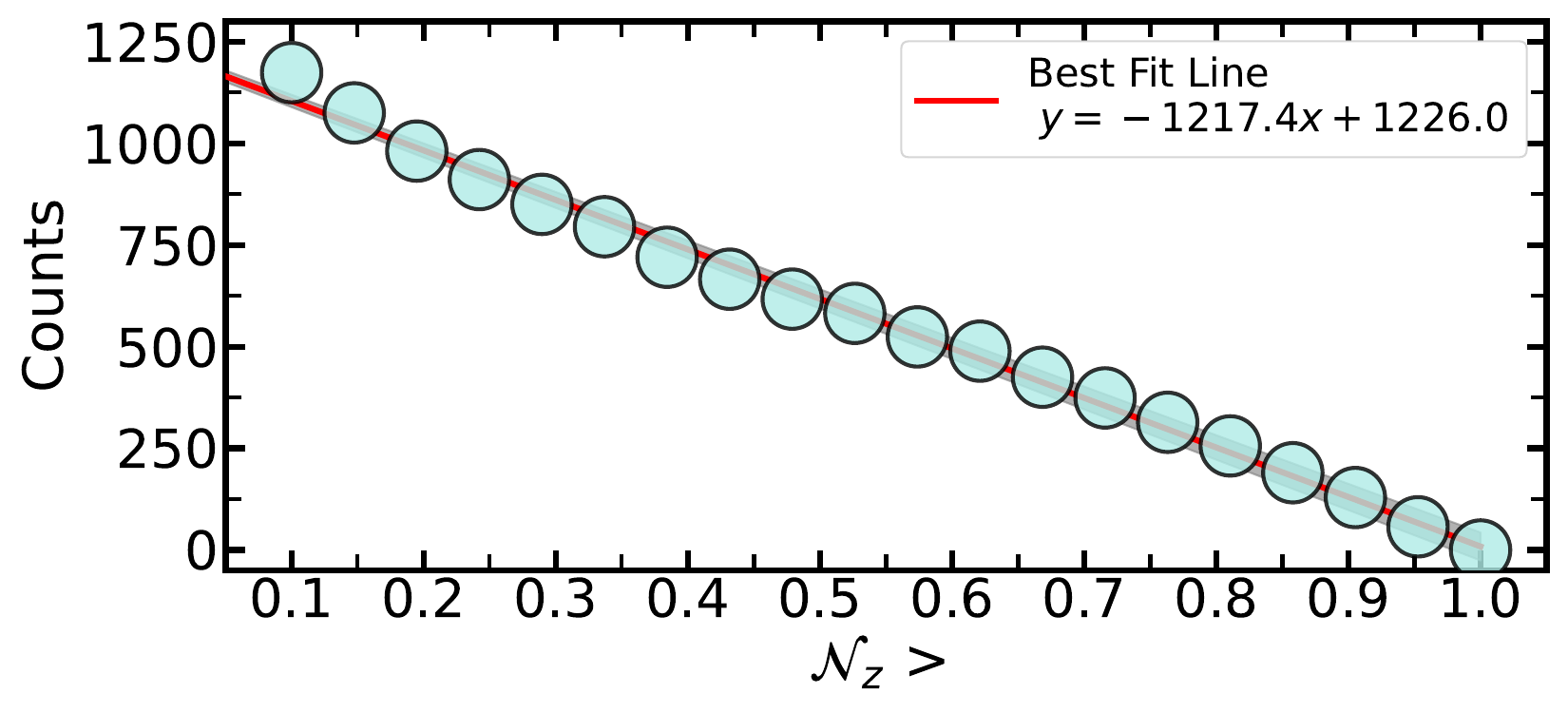}
    \caption{Total Number of galaxy pairs detected by applying different $\mathcal{N}_z$ threshold in all eight JWST fields.  The x-axis shows the threshold at which $\mathcal{N}_z$ is greater than a certain value, while the y-axis give the number of galaxies retrieved at that level. In general, the higher the value of $\mathcal{N}_z$, the more likely that a galaxy pair is real.   }

    \label{fig: N_z counts}
\end{figure*}

In this appendix, we present a catalog and cutout images of all merger close pairs detected with a probability ($\mathcal{N}_z$) higher than 0.90 and physical projection separation $10 - 50$ kpc, across all eight JWST fields: CEERS, JADES GOODS-S, NEP-TDF, NGDEEP, GLASS, El-Gordo, SMACS-0723, and MACS-0416, in \autoref{tab: close-pair catalog},  \autoref{fig: combined_first}, and \autoref{fig: comobined_second}. We also show the number of pairs selected with $\mathcal{N}_z$ greater than a certain threshold in \autoref{fig: N_z counts}. 

For each close-pair cutout graph, a 0.32 arcsecond aperture is drawn in red on each galaxy, with the redshift labeled at the top left of each aperture. We employ the Nakajima template \citep{Nakajima2022} for photometric AGN detection, marking the identified AGN-host galaxies in orange. A notable example is the close pair system, JADES GOODS-S 32751 and 32927, which have been observed with NIRSpec as part of the JADES project and with the F444W grism by FRESCO \citep{oesch2023jwst}. This system provides compelling evidence of a dual AGN merger and is a strong Lyman-$\alpha$ emitter. The detailed analysis of this system will be presented in a forthcoming paper by Li et al. (2024).

\FloatBarrier

\begin{table*}
\centering
\caption{Merger catalogs for galaxy pairs with $\mathcal{N}_z > 0.90$ and projection separations $10$ - $50$ kpc.}
\label{tab: close-pair catalog}
\begin{tabular}{c|cccc|cccc|cc}
\hline\hline 
\rule{0pt}{10.0pt} Field & Galaxy ID & RA Galaxy & DEC Galaxy & $z_\mathrm{phot}$ Galaxy & Paired ID & RA Paired & DEC Paired & $z_\mathrm{phot}$ Paired & dphys & $\mathcal{N}_z$ \\
& & (deg) & (deg) & & & (deg) & (deg) & & (kpc) & \\[2.0pt]
\hline
\rule{0pt}{10.0pt}CEERSP1 & 7910 & 214.982680 & 52.958951 & $5.20^{+0.04}_{-0.04}$ & 8221 & 214.981681 & 52.959053 & $5.19^{+0.03}_{-0.04}$ & 13.9 & 0.95 \\[2.5pt]
CEERSP3 & 6153 & 214.818854 & 52.856957 & $5.17^{+0.04}_{-0.04}$ & 6210 & 214.821281 & 52.858430 & $5.17^{+0.04}_{-0.05}$ & 47.3 & 1.00 \\[2.5pt]
CEERSP3 & 2774 & 214.822826 & 52.877943 & $8.30^{+0.19}_{-0.43}$ & 3498 & 214.825371 & 52.875729 & $8.27^{+0.30}_{-0.20}$ & 46.7 & 0.96 \\[2.5pt]
CEERSP3 & 2519 & 214.829189 & 52.883835 & $6.60^{+0.04}_{-0.24}$ & 2668 & 214.832181 & 52.885089 & $6.61^{+0.04}_{-0.17}$ & 43.8 & 0.95 \\[2.5pt]
CEERSP3 & 367 & 214.809560 & 52.881730 & $6.17^{+0.29}_{-0.18}$ & 876 & 214.812683 & 52.881541 & $6.25^{+0.18}_{-0.43}$ & 39.1 & 0.93 \\[2.5pt]
CEERSP4 & 2346 & 214.732548 & 52.735757 & $5.27^{+0.16}_{-0.33}$ & 2761 & 214.730042 & 52.735883 & $5.31^{+0.20}_{-0.27}$ & 34.2 & 0.95 \\[2.5pt]
CEERSP5 & 7926 & 214.941896 & 52.908949 & $5.39^{+0.16}_{-0.21}$ & 8230 & 214.944609 & 52.909779 & $5.44^{+0.16}_{-0.27}$ & 40.8 & 0.93 \\[2.5pt]
CEERSP6 & 1066 & 214.799211 & 52.815300 & $6.74^{+0.03}_{-0.03}$ & 1143 & 214.800184 & 52.814516 & $6.76^{+0.02}_{-0.04}$ & 19.3 & 0.96 \\[2.5pt]
CEERSP6 & 8835 & 214.846167 & 52.809366 & $7.29^{+0.17}_{-0.14}$ & 9614 & 214.847201 & 52.808346 & $7.37^{+0.10}_{-0.17}$ & 22.4 & 0.94 \\[2.5pt]
CEERSP7 & 8658 & 215.064295 & 52.938334 & $7.00^{+0.36}_{-0.38}$ & 8872 & 215.066323 & 52.937390 & $6.81^{+0.55}_{-0.12}$ & 30.0 & 0.98 \\[2.5pt]
CEERSP7 & 5660 & 215.070933 & 52.931833 & $5.69^{+0.26}_{-0.03}$ & 6202 & 215.068493 & 52.933263 & $5.74^{+0.19}_{-0.16}$ & 44.3 & 0.96 \\[2.5pt]
CEERSP8 & 3129 & 215.009319 & 52.875170 & $5.85^{+0.25}_{-0.09}$ & 3189 & 215.008545 & 52.874990 & $5.88^{+0.19}_{-0.16}$ & 10.7 & 0.99 \\[2.5pt]
CEERSP8 & 2827 & 215.007688 & 52.874106 & $5.83^{+0.25}_{-0.19}$ & 3189 & 215.008545 & 52.874990 & $5.88^{+0.19}_{-0.16}$ & 21.8 & 0.95 \\[2.5pt]
CEERSP8 & 2827 & 215.007688 & 52.874106 & $5.83^{+0.25}_{-0.19}$ & 3129 & 215.009319 & 52.875170 & $5.85^{+0.25}_{-0.09}$ & 30.9 & 0.93 \\[2.5pt]
CEERSP8 & 719 & 215.000358 & 52.855319 & $5.34^{+0.17}_{-0.23}$ & 845 & 214.999462 & 52.855449 & $5.36^{+0.19}_{-0.47}$ & 12.4 & 0.93 \\[2.5pt]
CEERSP8 & 8323 & 215.020056 & 52.910248 & $5.16^{+0.14}_{-0.12}$ & 9328 & 215.021006 & 52.910354 & $5.14^{+0.05}_{-0.07}$ & 13.3 & 0.92 \\[2.5pt]
CEERSP8 & 3129 & 215.009319 & 52.875170 & $5.85^{+0.25}_{-0.09}$ & 3443 & 215.006051 & 52.874802 & $5.94^{+0.33}_{-0.12}$ & 42.6 & 0.91 \\[2.5pt]
CEERSP9 & 1030 & 214.897539 & 52.788058 & $6.69^{+0.03}_{-0.59}$ & 1209 & 214.899339 & 52.790211 & $6.67^{+0.03}_{-0.05}$ & 47.7 & 0.98 \\[2.5pt]
CEERSP9 & 5959 & 214.943871 & 52.849344 & $5.16^{+0.03}_{-0.03}$ & 6273 & 214.941915 & 52.849884 & $5.16^{+0.05}_{-0.03}$ & 29.6 & 0.96 \\[2.5pt]
CEERSP9 & 6255 & 214.949979 & 52.855776 & $5.48^{+0.12}_{-0.21}$ & 6435 & 214.949113 & 52.856052 & $5.43^{+0.19}_{-0.13}$ & 13.1 & 0.95 \\[2.5pt]
CEERSP9 & 5283 & 214.874841 & 52.797010 & $6.80^{+0.06}_{-0.03}$ & 5581 & 214.875057 & 52.798800 & $6.81^{+1.07}_{-0.04}$ & 35.1 & 0.93 \\[2.5pt]
CEERSP9 & 5957 & 214.943725 & 52.849224 & $5.16^{+0.03}_{-0.03}$ & 6273 & 214.941915 & 52.849884 & $5.16^{+0.05}_{-0.03}$ & 29.1 & 0.92 \\[2.5pt]
CEERSP9 & 2774 & 214.963451 & 52.844896 & $5.98^{+0.28}_{-0.16}$ & 2783 & 214.965764 & 52.846560 & $5.92^{+0.28}_{-0.22}$ & 45.9 & 0.91 \\[2.5pt]
CEERSP10 & 7281 & 214.825163 & 52.779435 & $7.17^{+0.25}_{-0.10}$ & 7303 & 214.826919 & 52.780440 & $7.04^{+0.32}_{-0.05}$ & 27.9 & 0.94 \\[2.5pt]
CEERSP10 & 2600 & 214.868777 & 52.781066 & $4.89^{+0.10}_{-0.03}$ & 2902 & 214.867542 & 52.782023 & $4.88^{+0.01}_{-0.02}$ & 28.4 & 0.93 \\[2.5pt]
CEERSP10 & 5206 & 214.806862 & 52.753122 & $6.14^{+0.37}_{-0.46}$ & 5630 & 214.805488 & 52.754729 & $5.97^{+0.40}_{-0.14}$ & 37.9 & 0.93 \\[2.5pt]
JADES-Deep-GS & 32751 & 53.123009 & -27.796607 & $5.50^{+0.03}_{-0.03}$ & 32927 & 53.122473 & -27.796517 & $5.49^{+0.03}_{-0.03}$ & 10.7 & 1.00 \\[2.5pt]
JADES-Deep-GS & 18424 & 53.165775 & -27.784902 & $5.49^{+0.02}_{-0.03}$ & 18731 & 53.165012 & -27.784921 & $5.48^{+0.02}_{-0.03}$ & 14.9 & 1.00 \\[2.5pt]
JADES-Deep-GS & 35010 & 53.129241 & -27.783514 & $6.61^{+0.04}_{-0.04}$ & 36616 & 53.127127 & -27.782098 & $6.61^{+0.04}_{-0.04}$ & 46.7 & 1.00 \\[2.5pt]
JADES-Deep-GS & 31345 & 53.163316 & -27.736588 & $5.15^{+0.05}_{-0.05}$ & 31394 & 53.163945 & -27.735124 & $5.15^{+0.04}_{-0.03}$ & 35.7 & 0.99 \\[2.5pt]
JADES-Deep-GS & 31434 & 53.145541 & -27.765049 & $5.96^{+0.04}_{-0.04}$ & 32207 & 53.145021 & -27.762873 & $5.97^{+0.04}_{-0.03}$ & 46.9 & 0.99 \\[2.5pt]
JADES-Deep-GS & 36672 & 53.147331 & -27.751922 & $4.93^{+0.03}_{-0.03}$ & 36810 & 53.146916 & -27.752229 & $4.94^{+0.05}_{-0.03}$ & 11.2 & 0.99 \\[2.5pt]
JADES-Deep-GS & 25543 & 53.165707 & -27.757123 & $5.77^{+0.21}_{-0.10}$ & 25790 & 53.165795 & -27.755108 & $5.77^{+0.24}_{-0.11}$ & 43.3 & 0.98 \\[2.5pt]
JADES-Deep-GS & 33597 & 53.143980 & -27.758505 & $4.93^{+0.02}_{-0.03}$ & 36509 & 53.144828 & -27.759143 & $4.95^{+0.04}_{-0.03}$ & 22.9 & 0.98 \\[2.5pt]
JADES-Deep-GS & 22456 & 53.142274 & -27.806827 & $5.99^{+0.04}_{-0.06}$ & 22740 & 53.141697 & -27.807013 & $5.98^{+0.07}_{-0.07}$ & 11.4 & 0.98 \\[2.5pt]
JADES-Deep-GS & 5475 & 53.173496 & -27.825065 & $5.65^{+0.26}_{-0.01}$ & 5977 & 53.172864 & -27.824120 & $5.64^{+0.24}_{-0.03}$ & 23.9 & 0.98 \\[2.5pt]
JADES-Deep-GS & 11054 & 53.179285 & -27.792531 & $5.13^{+0.08}_{-0.05}$ & 11611 & 53.177523 & -27.793280 & $5.14^{+0.05}_{-0.05}$ & 39.5 & 0.97 \\[2.5pt]
JADES-Deep-GS & 15226 & 53.179755 & -27.774646 & $7.32^{+0.14}_{-0.10}$ & 15573 & 53.179537 & -27.773949 & $7.35^{+0.08}_{-0.12}$ & 13.5 & 0.97 \\[2.5pt]
JADES-Deep-GS & 9992 & 53.205694 & -27.753094 & $6.79^{+0.03}_{-0.03}$ & 11088 & 53.204222 & -27.751635 & $6.75^{+0.03}_{-0.03}$ & 38.4 & 0.97 \\[2.5pt]
JADES-Deep-GS & 22710 & 53.179627 & -27.745205 & $7.32^{+0.24}_{-0.08}$ & 23744 & 53.177210 & -27.744424 & $7.33^{+0.26}_{-0.06}$ & 42.6 & 0.97 \\[2.5pt]
JADES-Deep-GS & 17814 & 53.152721 & -27.808763 & $5.59^{+0.29}_{-0.00}$ & 18093 & 53.152046 & -27.808281 & $5.75^{+0.14}_{-0.15}$ & 16.6 & 0.97 \\[2.5pt]
JADES-Deep-GS & 32855 & 53.128192 & -27.787682 & $5.52^{+0.02}_{-0.03}$ & 32970 & 53.127170 & -27.788994 & $5.52^{+0.04}_{-0.03}$ & 35.1 & 0.97 \\[2.5pt]
JADES-Deep-GS & 33690 & 53.113478 & -27.795956 & $5.84^{+0.12}_{-0.13}$ & 33928 & 53.112674 & -27.796954 & $5.81^{+0.12}_{-0.13}$ & 26.2 & 0.97 \\[2.5pt]
JADES-Deep-GS & 24169 & 53.142956 & -27.798279 & $5.95^{+0.12}_{-0.15}$ & 24376 & 53.141888 & -27.800248 & $5.94^{+0.13}_{-0.24}$ & 46.2 & 0.97 \\[2.5pt]
JADES-Deep-GS & 36109 & 53.146673 & -27.752343 & $4.90^{+0.25}_{-0.07}$ & 36566 & 53.147042 & -27.751124 & $4.90^{+0.25}_{-0.02}$ & 29.5 & 0.96 \\[2.5pt]
\hline
\end{tabular}
\end{table*}
\begin{table*}
\centering
\begin{tabular}{c|cccc|cccc|cc}
\hline\hline 
\rule{0pt}{10.0pt} Field & Galaxy ID & RA Galaxy & DEC Galaxy & $z_\mathrm{phot}$ Galaxy & Paired ID & RA Paired & DEC Paired & $z_\mathrm{phot}$ Paired & dphys & $\mathcal{N}_z$ \\
& & (deg) & (deg) & & & (deg) & (deg) & & (kpc) & \\[2.0pt]
\hline
\rule{0pt}{10.0pt}JADES-Deep-GS & 24334 & 53.159470 & -27.771861 & $5.98^{+0.04}_{-0.11}$ & 25126 & 53.157055 & -27.772713 & $5.99^{+0.03}_{-0.17}$ & 48.4 & 0.96 \\[2.5pt]
JADES-Deep-GS & 18093 & 53.152046 & -27.808281 & $5.75^{+0.14}_{-0.15}$ & 18157 & 53.152883 & -27.806950 & $5.65^{+0.25}_{-0.21}$ & 33.0 & 0.95 \\[2.5pt]
JADES-Deep-GS & 35399 & 53.147371 & -27.749282 & $5.93^{+0.04}_{-0.18}$ & 36234 & 53.146150 & -27.749148 & $5.89^{+0.06}_{-0.09}$ & 23.1 & 0.95 \\[2.5pt]
JADES-Deep-GS & 12938 & 53.159047 & -27.818244 & $6.70^{+1.18}_{-0.08}$ & 13206 & 53.159039 & -27.817245 & $6.72^{+0.08}_{-0.08}$ & 19.7 & 0.95 \\[2.5pt]
JADES-Deep-GS & 17814 & 53.152721 & -27.808763 & $5.59^{+0.29}_{-0.00}$ & 18157 & 53.152883 & -27.806950 & $5.65^{+0.25}_{-0.21}$ & 39.6 & 0.95 \\[2.5pt]
JADES-Deep-GS & 15880 & 53.181486 & -27.769501 & $7.39^{+0.09}_{-0.14}$ & 15965 & 53.180119 & -27.771435 & $7.40^{+0.13}_{-0.12}$ & 42.5 & 0.94 \\[2.5pt]
JADES-Deep-GS & 21938 & 53.146256 & -27.802605 & $5.56^{+0.40}_{-0.01}$ & 22450 & 53.146417 & -27.800545 & $5.58^{+0.31}_{-0.02}$ & 45.2 & 0.94 \\[2.5pt]
JADES-Deep-GS & 36472 & 53.144838 & -27.751824 & $5.93^{+0.05}_{-0.21}$ & 36488 & 53.145560 & -27.751419 & $5.86^{+0.13}_{-0.17}$ & 16.1 & 0.94 \\[2.5pt]
JADES-Deep-GS & 35460 & 53.129543 & -27.777363 & $5.99^{+0.13}_{-0.12}$ & 36809 & 53.129930 & -27.778410 & $5.98^{+0.06}_{-0.21}$ & 23.2 & 0.94 \\[2.5pt]
JADES-Deep-GS & 15880 & 53.181486 & -27.769501 & $7.39^{+0.09}_{-0.14}$ & 15226 & 53.179755 & -27.774646 & $7.32^{+-7.32}_{-7.32}$ & 50.0 & 0.94 \\[2.5pt]
JADES-Deep-GS & 24313 & 53.140781 & -27.802173 & $5.95^{+0.05}_{-0.05}$ & 24602 & 53.141590 & -27.799983 & $5.97^{+0.04}_{-0.09}$ & 48.6 & 0.94 \\[2.5pt]
JADES-Deep-GS & 32310 & 53.153868 & -27.748166 & $6.48^{+0.11}_{-0.25}$ & 34004 & 53.152194 & -27.747403 & $6.45^{+0.08}_{-0.30}$ & 33.6 & 0.93 \\[2.5pt]
JADES-Deep-GS & 35238 & 53.140788 & -27.754309 & $5.73^{+0.19}_{-0.12}$ & 35940 & 53.142824 & -27.753893 & $5.62^{+0.25}_{-0.03}$ & 40.1 & 0.93 \\[2.5pt]
JADES-Deep-GS & 32983 & 53.124320 & -27.793421 & $6.01^{+0.03}_{-0.05}$ & 34370 & 53.122757 & -27.794645 & $6.01^{+0.06}_{-0.10}$ & 38.8 & 0.93 \\[2.5pt]
JADES-Deep-GS & 25052 & 53.127523 & -27.821078 & $5.03^{+0.03}_{-0.06}$ & 25624 & 53.127254 & -27.819147 & $5.04^{+0.07}_{-0.09}$ & 44.9 & 0.93 \\[2.5pt]
JADES-Deep-GS & 36103 & 53.132120 & -27.779131 & $5.95^{+0.04}_{-0.24}$ & 36809 & 53.129930 & -27.778410 & $5.98^{+0.06}_{-0.21}$ & 43.6 & 0.92 \\[2.5pt]
JADES-Deep-GS & 13392 & 53.167761 & -27.802318 & $6.04^{+0.04}_{-0.15}$ & 13456 & 53.167303 & -27.802868 & $6.00^{+0.03}_{-0.12}$ & 14.3 & 0.92 \\[2.5pt]
JADES-Deep-GS & 3607 & 53.206324 & -27.775727 & $5.93^{+0.11}_{-0.21}$ & 5262 & 53.204993 & -27.774029 & $5.94^{+0.04}_{-0.21}$ & 43.7 & 0.92 \\[2.5pt]
JADES-Deep-GS & 17226 & 53.154770 & -27.806513 & $6.21^{+0.08}_{-0.30}$ & 18579 & 53.152537 & -27.805958 & $6.19^{+0.17}_{-0.13}$ & 42.4 & 0.92 \\[2.5pt]
JADES-Deep-GS & 13116 & 53.188570 & -27.769387 & $5.82^{+0.12}_{-0.17}$ & 13377 & 53.186978 & -27.769892 & $5.82^{+0.17}_{-0.16}$ & 32.0 & 0.92 \\[2.5pt]
JADES-Deep-GS & 11867 & 53.191302 & -27.769902 & $5.96^{+0.06}_{-0.30}$ & 13013 & 53.189390 & -27.768425 & $5.90^{+0.06}_{-0.21}$ & 47.5 & 0.92 \\[2.5pt]
JADES-Deep-GS & 11067 & 53.168930 & -27.809406 & $5.95^{+0.04}_{-0.09}$ & 11102 & 53.168961 & -27.808741 & $5.98^{+0.04}_{-0.03}$ & 14.0 & 0.92 \\[2.5pt]
JADES-Deep-GS & 1711 & 53.209191 & -27.781479 & $4.89^{+0.05}_{-0.03}$ & 2114 & 53.207801 & -27.782372 & $4.90^{+0.21}_{-0.04}$ & 35.5 & 0.92 \\[2.5pt]
JADES-Deep-GS & 25055 & 53.153146 & -27.779160 & $5.97^{+0.13}_{-0.20}$ & 25526 & 53.152310 & -27.779072 & $6.07^{+0.04}_{-0.27}$ & 15.6 & 0.91 \\[2.5pt]
JADES-Deep-GS & 34534 & 53.112492 & -27.806222 & $5.99^{+0.06}_{-0.16}$ & 35140 & 53.110450 & -27.807030 & $6.00^{+0.03}_{-0.08}$ & 41.6 & 0.91 \\[2.5pt]
JADES-Deep-GS & 13013 & 53.189390 & -27.768425 & $5.90^{+0.06}_{-0.21}$ & 13116 & 53.188570 & -27.769387 & $5.82^{+0.12}_{-0.17}$ & 25.7 & 0.91 \\[2.5pt]
JADES-Deep-GS & 24376 & 53.141888 & -27.800248 & $5.94^{+0.13}_{-0.24}$ & 24551 & 53.142054 & -27.799387 & $5.94^{+0.06}_{-0.26}$ & 18.5 & 0.90 \\[2.5pt]
JADES-Deep-GS & 5769 & 53.166013 & -27.836107 & $10.52^{+0.25}_{-0.30}$ & 6235 & 53.165934 & -27.834233 & $10.63^{+0.31}_{-0.32}$ & 27.6 & 0.90 \\[2.5pt]
NEP-1 & 3263 & 260.733438 & 65.814259 & $6.13^{+0.17}_{-0.15}$ & 3657 & 260.735079 & 65.813488 & $6.13^{+0.16}_{-0.19}$ & 21.3 & 1.00 \\[2.5pt]
NEP-1 & 10530 & 260.772216 & 65.815156 & $6.03^{+0.14}_{-0.14}$ & 11046 & 260.774367 & 65.814862 & $6.02^{+0.12}_{-0.16}$ & 19.5 & 0.99 \\[2.5pt]
NEP-1 & 3592 & 260.706948 & 65.778486 & $6.15^{+0.13}_{-0.26}$ & 3750 & 260.708079 & 65.778979 & $6.10^{+0.18}_{-0.26}$ & 14.1 & 0.98 \\[2.5pt]
NEP-1 & 12099 & 260.714443 & 65.732351 & $5.78^{+0.16}_{-0.20}$ & 13036 & 260.719272 & 65.733133 & $5.74^{+0.20}_{-0.33}$ & 45.9 & 0.97 \\[2.5pt]
NEP-1 & 3263 & 260.733438 & 65.814259 & $6.13^{+0.17}_{-0.15}$ & 3365 & 260.734648 & 65.814805 & $6.22^{+0.12}_{-0.19}$ & 15.3 & 0.92 \\[2.5pt]
NEP-1 & 17063 & 260.734114 & 65.724736 & $5.89^{+0.12}_{-0.19}$ & 17944 & 260.739396 & 65.724244 & $5.93^{+0.16}_{-0.27}$ & 47.2 & 0.90 \\[2.5pt]
NEP-1 & 3365 & 260.734648 & 65.814805 & $6.22^{+0.12}_{-0.19}$ & 3657 & 260.735079 & 65.813488 & $6.13^{+0.16}_{-0.19}$ & 27.5 & 0.90 \\[2.5pt]
NEP-2 & 3792 & 260.873113 & 65.819014 & $6.20^{+0.21}_{-0.10}$ & 4883 & 260.873197 & 65.820684 & $6.26^{+0.12}_{-0.17}$ & 34.4 & 0.97 \\[2.5pt]
NEP-2 & 5395 & 260.865132 & 65.822733 & $6.98^{+0.06}_{-0.06}$ & 6437 & 260.866433 & 65.824823 & $6.99^{+0.10}_{-0.06}$ & 41.6 & 0.97 \\[2.5pt]
NEP-2 & 16454 & 260.807954 & 65.851888 & $6.93^{+0.20}_{-0.04}$ & 16457 & 260.812533 & 65.850840 & $6.92^{+0.11}_{-0.08}$ & 41.6 & 0.98 \\[2.5pt]
NEP-2 & 3094 & 260.851258 & 65.819816 & $5.99^{+0.15}_{-0.11}$ & 3406 & 260.853085 & 65.820139 & $5.96^{+0.15}_{-0.13}$ & 17.2 & 0.95 \\[2.5pt]
NEP-2 & 4086 & 260.767199 & 65.832102 & $5.17^{+0.06}_{-0.05}$ & 4100 & 260.766350 & 65.831768 & $5.17^{+0.04}_{-0.03}$ & 11.0 & 0.95 \\[2.5pt]
NEP-2 & 778 & 260.852378 & 65.814826 & $4.86^{+0.02}_{-0.12}$ & 1092 & 260.849386 & 65.815862 & $4.88^{+0.16}_{-0.10}$ & 37.6 & 0.94 \\[2.5pt]
NEP-2 & 1092 & 260.849386 & 65.815862 & $4.88^{+0.16}_{-0.10}$ & 1897 & 260.847586 & 65.817860 & $4.88^{+0.02}_{-0.02}$ & 49.9 & 0.94 \\[2.5pt]
NEP-2 & 4986 & 260.793908 & 65.830838 & $6.11^{+0.15}_{-0.16}$ & 5578 & 260.791291 & 65.831796 & $6.18^{+0.15}_{-0.25}$ & 29.8 & 0.93 \\[2.5pt]
NEP-2 & 15406 & 260.813870 & 65.850411 & $6.95^{+0.07}_{-0.04}$ & 16457 & 260.812533 & 65.850840 & $6.92^{+0.11}_{-0.08}$ & 13.5 & 0.93 \\[2.5pt]
NEP-2 & 9077 & 260.904667 & 65.825844 & $6.51^{+0.04}_{-0.32}$ & 9446 & 260.904706 & 65.826616 & $6.54^{+0.08}_{-0.30}$ & 15.5 & 0.92 \\[2.5pt]
\hline
\end{tabular}
\end{table*}
\begin{table*}
\centering
\begin{tabular}{c|cccc|cccc|cc}
\hline\hline 
\rule{0pt}{10.0pt} Field & Galaxy ID & RA Galaxy & DEC Galaxy & $z_\mathrm{phot}$ Galaxy & Paired ID & RA Paired & DEC Paired & $z_\mathrm{phot}$ Paired & dphys & $\mathcal{N}_z$ \\
& & (deg) & (deg) & & & (deg) & (deg) & & (kpc) & \\[2.0pt]
\hline
\rule{0pt}{10.0pt}NEP-2 & 458 & 260.961956 & 65.801143 & $6.42^{+0.10}_{-0.28}$ & 1479 & 260.966317 & 65.802746 & $6.43^{+0.13}_{-0.17}$ & 48.6 & 0.92 \\[2.5pt]
NEP-2 & 11232 & 260.812774 & 65.841356 & $5.23^{+0.25}_{-0.14}$ & 11423 & 260.810606 & 65.842099 & $5.19^{+0.28}_{-0.15}$ & 26.2 & 0.92 \\[2.5pt]
NEP-2 & 15564 & 260.807764 & 65.851006 & $6.92^{+0.36}_{-0.06}$ & 16457 & 260.812533 & 65.850840 & $6.92^{+0.11}_{-0.08}$ & 37.9 & 0.92 \\[2.5pt]
NEP-2 & 3792 & 260.873113 & 65.819014 & $6.20^{+0.21}_{-0.10}$ & 6365 & 260.878823 & 65.823235 & $6.16^{+-6.16}_{-6.16}$ & 49.8 & 0.91 \\[2.5pt]
NEP-2 & 15564 & 260.807764 & 65.851006 & $6.92^{+0.36}_{-0.06}$ & 16454 & 260.807954 & 65.851888 & $6.93^{+0.20}_{-0.04}$ & 17.1 & 0.90 \\[2.5pt]
NEP-3 & 1588 & 260.719973 & 65.888473 & $4.90^{+0.29}_{-0.01}$ & 1945 & 260.717975 & 65.887612 & $4.92^{+0.36}_{-0.05}$ & 27.7 & 0.93 \\[2.5pt]
NEP-3 & 3834 & 260.697960 & 65.873946 & $6.29^{+0.12}_{-0.15}$ & 4887 & 260.692599 & 65.873214 & $6.28^{+0.22}_{-0.25}$ & 47.4 & 0.93 \\[2.5pt]
NEP-4 & 11140 & 260.453023 & 65.820756 & $6.43^{+0.09}_{-0.22}$ & 11432 & 260.453387 & 65.820155 & $6.42^{+0.09}_{-0.27}$ & 12.5 & 0.98 \\[2.5pt]
NEP-4 & 12652 & 260.564772 & 65.804474 & $8.30^{+0.13}_{-0.12}$ & 13194 & 260.569008 & 65.803007 & $8.34^{+0.13}_{-0.20}$ & 39.3 & 0.96 \\[2.5pt]
NEP-4 & 10831 & 260.452343 & 65.821472 & $6.45^{+0.07}_{-0.18}$ & 11140 & 260.453023 & 65.820756 & $6.43^{+0.09}_{-0.22}$ & 15.5 & 0.95 \\[2.5pt]
NEP-4 & 8359 & 260.452210 & 65.825792 & $6.47^{+0.08}_{-0.14}$ & 9422 & 260.453744 & 65.824248 & $6.51^{+0.04}_{-0.26}$ & 33.5 & 0.95 \\[2.5pt]
NEP-4 & 3835 & 260.440177 & 65.836390 & $5.96^{+0.11}_{-0.16}$ & 4675 & 260.439310 & 65.834567 & $5.98^{+0.15}_{-0.17}$ & 39.2 & 0.95 \\[2.5pt]
NEP-4 & 787 & 260.514082 & 65.834185 & $19.55^{+0.72}_{-0.60}$ & 1592 & 260.501382 & 65.834120 & $5.15^{+-5.15}_{-5.15}$ & 24.1 & 0.94 \\[2.5pt]
NEP-4 & 13448 & 260.616281 & 65.796962 & $5.99^{+0.22}_{-0.16}$ & 14015 & 260.617649 & 65.795564 & $5.97^{+0.14}_{-0.14}$ & 31.7 & 0.93 \\[2.5pt]
NEP-4 & 10831 & 260.452343 & 65.821472 & $6.45^{+0.07}_{-0.18}$ & 11432 & 260.453387 & 65.820155 & $6.42^{+0.09}_{-0.27}$ & 28.0 & 0.91 \\[2.5pt]
NEP-4 & 994 & 260.512970 & 65.833906 & $19.53^{+0.48}_{-1.70}$ & 1592 & 260.501382 & 65.834120 & $5.15^{+-5.15}_{-5.15}$ & 22.1 & 0.91 \\[2.5pt]
NEP-4 & 12496 & 260.568470 & 65.804629 & $6.29^{+0.14}_{-0.15}$ & 12569 & 260.564696 & 65.804812 & $6.34^{+0.13}_{-0.12}$ & 31.8 & 0.91 \\[2.5pt]
\hline\hline
\end{tabular}
\end{table*}

\FloatBarrier


\bsp	
\label{lastpage}
\end{document}